\documentclass{aa}  

\usepackage{graphicx}
\usepackage{txfonts}
\newcommand{\Rs}{\ensuremath{R_{\odot}}}
\newcommand{\Ms}{\ensuremath{M_{\odot}}}

\newcommand{\Rt}{\ensuremath{R_t}}
\newcommand{\eg}{{\it e.g.}}
\newcommand{\cf}{{\it c.f.~}}
\newcommand{\ie}{{\it i.e.}}

\newcommand{\beq}{\begin{equation}}
\newcommand{\eeq}{\end{equation}}

\newcommand{\mzamsa}{\ensuremath{M_{\rm ZAMS,1}}}
\newcommand{\mzamsb}{\ensuremath{M_{\rm ZAMS,2}}}

\newcommand{\kmps}{\ensuremath{{\rm~km~s}^{-1}}}

\newcommand{\mcl}{\ensuremath{M_{cl}}}
\newcommand{\mproto}{\ensuremath{m_{\rm proto}}}
\newcommand{\rh}{\ensuremath{r_h}}
\newcommand{\rhstar}{\ensuremath{r_{h,\ast}}}

\newcommand{\rhbhbound}{\ensuremath{r_{\rm h,BH,bound}}}

\newcommand{\nbhbound}{\ensuremath{N_{\rm BH,bound}}}
\newcommand{\mbhbound}{\ensuremath{M_{\rm BH,bound}}}

\newcommand{\nnsbound}{\ensuremath{N_{\rm NS,bound}}}

\newcommand{\nbseven}{{\tt NBODY7~}}
\newcommand{\nbsix}{{\tt NBODY6 }}
\newcommand{\nbpp}{{\tt NBODY6++GPU }}
\newcommand{\nbsixseven}{{\tt NBODY6/7 }}
\newcommand{\bse}{{\tt BSE }}
\newcommand{\sse}{{\tt SSE }}
\newcommand{\ssebse}{{\tt SSE/BSE }}
\newcommand{\mobse}{{\tt MOBSE }}
\newcommand{\mocca}{{\tt MOCCA }}
\newcommand{\cmc}{{\tt CMC }}
\newcommand{\startrack}{{\tt StarTrack }}

\newcommand{\fbin}{\ensuremath{f_{\rm bin}}}

\newcommand{\fobin}{\ensuremath{f_{\rm Obin}}}

\newcommand{\mlwind}{{\tt MLWIND}}
\newcommand{\hrdiag}{{\tt HRDIAG}}
\newcommand{\kick}{{\tt KICK}}
\newcommand{\kw}{{\tt KW}}
\newcommand{\nsflag}{{\tt nsflag}}
\newcommand{\psflag}{{\tt psflag}}
\newcommand{\ecflag}{{\tt ecflag}}
\newcommand{\mfb}{\ensuremath{m_{\rm fb}}}
\newcommand{\fbtot}{\ensuremath{M_{\rm fb}}}
\newcommand{\fbfac}{\ensuremath{f_{\rm fb}}}
\newcommand{\ptsone}{{\tt pts1}}
\newcommand{\ptstwo}{{\tt pts2}}
\newcommand{\ptsthree}{{\tt pts3}}
\newcommand{\signs}{\ensuremath{\sigma_{\rm NS}}}
\newcommand{\mbh}{\ensuremath{m_{\rm BH}}}
\newcommand{\mns}{\ensuremath{m_{\rm NS}}}
\newcommand{\mnsav}{\ensuremath{\langle m_{\rm NS} \rangle}}
\newcommand{\mecsns}{\ensuremath{m_{\rm ECS,NS}}}
\newcommand{\mrem}{\ensuremath{m_{\rm rem}}}
\newcommand{\mco}{\ensuremath{m_{\rm CO}}}
\newcommand{\meff}{\ensuremath{m_{\rm eff}}}
\newcommand{\vkick}{{\rm v}_{\rm kick}}
\newcommand{\vkickns}{{\rm v}_{\rm kick,NS}}
\newcommand{\vesc}{{\rm v}_{\rm esc}}
\newcommand{\kconv}{\ensuremath{k_{\rm conv}}}
\newcommand{\kcol}{\ensuremath{k_{\rm col}}}
\newcommand{\fretbhm}{\ensuremath{f_{\rm ret,BH,m}}}
\newcommand{\fretbhn}{\ensuremath{f_{\rm ret,BH,n}}}
\newcommand{\fbhm}{\ensuremath{F_{\rm BH,m}}}
\newcommand{\fbhn}{\ensuremath{F_{\rm BH,n}}}
\newcommand{\fmrg}{\ensuremath{f_{\rm mrg}}}
\newcommand{\ace}{\ensuremath{\alpha_{\rm CE}}}

\begin{document} 
\title{{\tt BSE} versus {\tt StarTrack}: implementations of new wind, remnant-formation,
and natal-kick schemes in {\tt NBODY7} and their astrophysical consequences}
\titlerunning{{\tt BSE} versus {\tt StarTrack}: new ingredients and their astrophysical implications}
\author{S. Banerjee\thanks{Corresponding author. E-mail: sambaran@astro.uni-bonn.de}
        \inst{1,2}
	\and
	K. Belczynski\inst{3}
	\and
	C. L. Fryer\inst{4}
	\and
	P. Berczik\inst{5,6,7}
	\and
	J. R. Hurley\inst{8}
	\and
	R. Spurzem\inst{5,6,9}\thanks{Research Fellow of the Frankfurt Institute for Advanced Study}
	\and
	L. Wang\inst{10,11}
       }
\institute{
	Helmholtz-Instituts f\"ur Strahlen- und Kernphysik,
        Nussallee 14-16, D-53115 Bonn, Germany
	\and
	Argelander-Institut f\"ur Astronomie,
        Auf dem H\"ugel 71, D-53121, Bonn, Germany
	\and
	Nicolaus Copernicus Astronomical Centre of the Polish Academy of Sciences,
	ul. Bartycka 18,
	00-716 Warszawa, Poland
	\and
	Center for Theoretical Astrophysics, Los Alamos National Laboratory,
        P.O. Box 1663, Los Alamos, NM 87545, U.S.A.
	\and
        National Astronomical Observatories and Key Laboratory of Computational Astrophysics, Chinese Academy of Sciences,
        20A Datun Rd., Chaoyang District, Beijing 100101, China
	\and
	Astronomisches Rechen-Institut am Zentrum  f\"ur  Astronomie  der Universit\"at Heidelberg,
	M\"onchhofstra{\ss}e 12-14, 69120 Heidelberg, Germany
	\and
	Main Astronomical Observatory, National Academy of Sciences of Ukraine,
        27 Akademika Zabolotnoho St., 03143 Kyiv, Ukraine
        \and
	Swinburne University of Technology, 
        Hawthorn VIC 3122, Australia 
	\and
	Kavli Institute for Astronomy and Astrophysics, Peking University,
	5 Yi He Yuan Road, Haidian District, Beijing 100871, China
	\and
	Department of Astronomy, School of Science, The University of Tokyo,
	7-3-1 Hongo, Bunkyo-ku, Tokyo, 113-0033, Japan
	\and
        RIKEN Center for Computational Science,
	7-1-26 Minatojima-minami-machi, Chuo-ku, Kobe, Hyogo 650-0047, Japan
	}
\abstract{The masses of stellar-remnant black holes (BH), as a result of their formation
via massive single- and binary-stellar evolution, is of high interest in this era of
gravitational-wave detection from binary black hole (BBH) and binary
neutron star (BNS) merger events.}
{In this work, we present new developments in the
state-of-the-art N-body evolution program \nbseven in regards to its stellar-remnant formation
and related schemes. We demonstrate that the newly-implemented stellar-wind and
remnant-formation schemes in the \nbseven code’s stellar-evolutionary sector or \bse,
such as the ``rapid'' and the ``delayed'' supernova (SN) schemes along with
an implementation of pulsational-pair-instability and pair-instability supernova (PPSN/PSN),
now produces neutron star (NS) and BH masses that agree nearly perfectly, over
large ranges of zero-age-main-sequence (ZAMS) mass and metallicity, with those
from the widely-recognized \startrack population-synthesis program.
We also demonstrate the new, recipe-based implementations of various, widely-debated mechanisms
of natal kicks on NSs and BHs such as the ``convection-asymmetry-driven'', ``collapse-asymmetry-driven'',
and ``neutrino-emission-driven'' kicks, in addition to a fully consistent implementation of the
standard, fallback-dependent, momentum-conserving natal kick.}
{All the above newly-implemented schemes are also shared with the
standalone versions of \sse and \bse. All these demonstrations are performed with both the
updated standalone \bse and the updated $\nbseven/\bse$.}
{When convolved with stellar and primordial-binary populations as observed
in young massive clusters,
such remnant-formation and natal-kick mechanisms crucially determine the accumulated
number, mass, and mass distribution
of the BHs retaining in young massive, open, and globular clusters, which would become available for
long-term dynamical processing.}
{Among other interesting conclusions, we find that although the newer delayed-SN remnant formation
model gives birth to the largest number (mass) of BHs, the older remnant-formation schemes cause the
largest number (mass) of BHs to survive in clusters, when SN material fallback on to the BHs is incorporated.
The SN material fallback also causes the convection-asymmetry-driven SN kick to effectively
retain similar number and mass of BHs in clusters as for the standard, momentum-conserving kick.  
The collapse-asymmetry-driven SN kick would cause nearly all BHs to retain in clusters
irrespective of its mass, the remnant formation model, and the metallicity,
whereas the inference of a large population
of BHs in GCs would potentially rule out the neutrino-driven SN kick mechanism. 
Pre-SN mergers of massive primordial binaries would potentially cause BH masses to
deviate from the theoretical, single-star ZAMS mass-remnant mass relation unless
a substantial, up to $\approx40$\%, of the total merging stellar mass is lost during a merger process.
In particular, such mergers, at low
metallicities, have the potential to produce low-spinning BHs within the PSN mass gap that can be
retained in a stellar cluster and be available for subsequent dynamical interactions.
The new remnant-formation modelling reassures that young massive and open clusters would potentially
contribute to the dynamical BBH merger detection rate to a similar extent as their more
massive globular-cluster counterparts, as recent studies indicate.
}
\keywords{Stars: black holes --- Stars: massive --- Stars: mass-loss --- Stars: kinematics and dynamics
--- supernovae: general --- Methods: numerical}
\maketitle
\section{Introduction}\label{intro}

The formation mechanisms of double compact object binaries undergoing general-relativistic (hereafter GR)
inspiral and their occurrence rates has always been a topic of diverse interest \citep[\eg,][]{Abadie_2010}
and, naturally, it
has become a focal concern since the first direct detection of gravitational waves
(hereafter GW) from a stellar-origin binary-black hole (BBH) inspiral and merger by the Laser Interferometer
Gravitational-wave Observatory \citep[LIGO;][]{2016PhRvL.116f1102A}. Among the most important
ingredients in theoretical modelling and population-synthesis of BBH mergers are
the natal masses and kicks of stellar-remnant black holes (hereafter BH) --- this, as well, applies to
neutron stars (hereafter NS) while modelling NS-containing binaries.
In fact, the importance of realistic BH mass estimates became clear right after the 
very first LIGO detection of $\approx30\Ms$ BHs \citep{2016ApJ...818L..22A,Belczynski_2016}.

In fact, as long as the proto-remnant mass, $\mproto$, in a supernova (hereafter SN) explosion and the amount of
material fallback onto it during the SN depend on the carbon-oxygen (CO) core mass and on the pre-supernova
stellar mass \citep{Fryer_1999,Hurley_2000,Fryer_2001,Fryer_2012}, the final remnant mass would depend on the
entire life cycle of the progenitor star. In such cases, starting from a given zero-age main sequence (ZAMS) mass,
the sequentially-dependent evolutions of the star's radius, luminosity, mass loss via wind, and total mass
would determine the final remnant mass. This is typically the case for the formation of relatively heavier
($\gtrsim1.4\Ms$) NSs and BHs. For example, by applying ``alternative'', theoretical and semi-empirical wind recipes in
the main sequence and evolved stages of O- and B-type progenitors, \citet{Belczynski_2010} have
demonstrated that $\gtrsim30\Ms$ BHs, as observed by the LIGO-Virgo
\citep{2016PhRvL.116f1102A,Abbott_GW170104,Abbott_GW170814},
can easily form from progenitors having dwarf galaxy-
and globular cluster (hereafter GC)-like metallicities, which masses are impossible to achieve
with the ``standard'' \citet{Hurley_2000} wind prescription even at a very low metallicity.

The formation of compact binaries through both field binary evolution
\citep[\eg,][]{Belczynski_2002,Voss_2003,Belczynski_2010c,Belczynski_2016,Mandel_2017a,Mapelli_2017,Giacobbo_2018}
and dynamical interactions in stellar clusters
\citep[\eg,][]{Banerjee_2010,Ziosi_2014,Morscher_2015,Rodriguez_2016,Mapelli_2016,Wang_2016,Askar_2016,Askar_2018,Banerjee_2017,Banerjee_2017b,Banerjee_2018,Park_2017,Rodriguez_2018}
are now widely studied. In both formation channels, stellar wind, BH masses, and natal kicks
play a role.
A few widely-visible stellar and binary population synthesis programs such as
\startrack \citep[][hereafter B08]{Belczynski_2008}
and \mobse \citep{Giacobbo_2018,DiCarlo_2019}
have now adopted what can be called the state-of-the-art in stellar-wind and remnant-formation prescriptions
\citep{Belczynski_2010,Fryer_2012,Belczynski_2016a,2017arXiv170607053B}.
On the other hand, much simpler and crude prescriptions are often used 
in works that explore the dynamical-formation channel, especially,  
those using the publicly-available and widely used direct N-body codes
$\nbsix$ \citep{2003gnbs.book.....A,Nitadori_2012},
$\nbseven$ \citep{2012MNRAS.422..841A}, and
$\nbpp$ \citep{Wang_2015}.
Note that such studies of the dynamical-formation channel, nevertheless, incorporate adequate or near-adequate treatment
of stellar content, Newtonian dynamical interactions, and their GR counterparts (given that both the
field-population-synthesis and the dynamical codes presently resort to similar, parametrized recipes for
treating the tidally-interacting and the mass-transferring binaries' internal evolution; \citealt{Hurley_2002}).
This, in turn, often makes model ingredients such as BH retention and mass distribution,
that critically influence a cluster's evolution and compact-binary formation, questionable in an
otherwise realistic simulation. Note, in particular, that the BH population that is retained
in a cluster following the BHs' birth and which then becomes available for dynamical processing,
influences the formation of both BH- and NS-containing compact binaries \citep{Banerjee_2017b},
alongside governing the cluster's long-term evolution \citep{2016arXiv160300884C,Banerjee_2017,ArcaSedda_2018}.

This work, for the first time, introduces three major upgrades to the
widely-used direct N-body evolutionary program \nbseven \citep{2012MNRAS.422..841A}, a direct descendant
of \nbsix \citep{2003gnbs.book.....A,Nitadori_2012}: (a) the semi-empirical stellar wind prescriptions as in 
\citet[][hereafter B10]{Belczynski_2010}, (b) remnant formation and material fallback according to the
``rapid'' and the ``delayed'' SN models of
\citet[][hereafter F12]{Fryer_2012}, incorporating the occurrences of
pair-instability supernova (PSN) and pulsation pair-instability
supernova (PPSN) according to the conditions of \citet[][hereafter B16]{Belczynski_2016a},
and (c) an explicit modulation of the BHs' and the
NSs' natal kicks based on the fallback fraction during their formation.
The stellar- and binary-evolution drivers of \nbsixseven
(as well of the Monte Carlo-based star cluster-evolution codes \mocca;
\citealt{Hypki_2013,Giersz_2013} and \cmc; \citealt{Joshi_2000})
are the well-known, fitting formulae and recipe based (as also for, \eg, \startrack) \sse \citep{Hurley_2000} and \bse
\citep{Hurley_2002} programs
\footnote{Since \sse is a part of \bse, \bse will imply both of them hereafter
in this paper. Similarly \nbseven, that is utilized in this work, will, hereafter,
imply both \nbsix and \nbseven since both adopt \ssebse.}. 
After the introduction of the above updates (a), (b), and (c), \bse and \nbseven
are now similarly rich and diverse in essentially all stellar- and binary-evolutionary aspects as 
modern population-synthesis codes such as \startrack and \mobse. This would not only enable 
the ``\nbsixseven community'' to do more realistic computations but also allow for
direct, cross-community comparisons of results.

Note that the above-mentioned upgrades (a) and (c) are partly available in the current, public version
of \nbseven but the implementations do not produce outcomes fully consistent with \startrack
\citep[see, \eg,][]{Banerjee_2017}, as it
should since \bse and \startrack utilize the same underlying fitting formulae for the stellar-structural
parameters \citep{Hurley_2000}.
In particular, the ``maximum BH mass effect'' \citep{Belczynski_2010} is absent in that
implementation.

In this work, we validate the above new updates of $\bse$ by comparing its outcomes,
for wide ranges of ZAMS mass and metallicity, with those from $\startrack$. We note
that detailed comparisons among several binary-population-synthesis codes, in the context
of the formation of white dwarf (hereafter WD) containing binaries, have recently
been conducted by \citet{Toonen_2014}. In this work, we will, however, focus
mostly on NS and BH outcomes since the above upgrades mostly affect SN outcomes.

In Secs.~\ref{newbse}, \ref{newwind}, and \ref{newrem},
we describe our new updates to $\bse$. In Sec.~\ref{comp1},
we make detailed comparisons of the remnant outcomes of the updated $\bse$
with those from $\startrack$. In Secs.~\ref{nbcode}, \ref{stdkick}, and \ref{altkick},
we describe how the updated $\bse$ is adapted to the direct N-body
integration code $\nbseven$, the new implementation of the explicit use
of SN fallback fraction in determining SN natal kicks, and
the new implementations of alternative natal kick models.
In Sec.~\ref{primbin}, we discuss how the single-stellar initial mass-final mass
relations get affected by the presence of an observationally-motivated
population of massive primordial binaries in a dynamically-active environment as in
young massive and open clusters. In Sec.~\ref{comp2}, we make a preliminary
comparison between (isolated) massive binary evolutions from the updated
$\bse$ and $\startrack$. In Sec.~\ref{summary}, we summarize our results
and indicate forthcoming studies along this line.

\section{The updated \bse: stellar-wind and remnant-formation schemes}\label{newbse}

In the following, the new implementations in the standalone \bse are described
and the outcomes are compared with those of \startrack.  

\subsection{The stellar-wind prescription}\label{newwind}

The recipe for stellar wind follows the semi-empirical prescriptions of B10,
which is currently considered as the state-of-the-art. In the subroutine \mlwind,
these prescriptions are implemented in such a way that they are activated \emph{individually} in the following
sequence of priority when the star is evolved until the second asymptotic giant branch (AGB)
\citep[\bse stellar type $\kw\leq6$;][]{Hurley_2000}:
(1) metallicity($Z$)-independent luminous-blue-variable (LBV) wind \citep[Eqn.~8 of B10;][]{Humphreys_1994},
(2) $Z$-dependent \citet{Vink_2001} winds (Eqns.~6 \& 7 of B10), 
(3) $Z$-dependent \citet{Nieu_1990} wind (Eqn.~3 of B10),
(4) the \citet{Hurley_2000} treatments for lower-mass/colder stars (\citealt{Kudritzki_1978} wind
for the giant branch and beyond and \citealt{Vassiliadis_1993} wind for the AGB; Eqns.~1 \& 2 of
B10 respectively). For naked-Helium and more evolved stars ($\kw>6$), only the $Z$-dependent
Wolf-Rayet (WR) wind \citep[Eqn.~9 of B10;][]{Hamann_1998,Vink_2005} is applied.
The conditions for activating the various winds remain the same as in B10. 
The key difference
between this new $\mlwind$ function and that of the original $\bse$
(or of the currently-public version of $\nbseven$) is that
in the original (older) codes the \citet{Nieu_1990} and the \citet{Kudritzki_1978} 
winds are applied for low-mass and high-mass stars.
In new $\mlwind$, for massive, hot stars, \citet{Vink_2001} winds are applied instead.
As it turns out, this
makes a substantial difference in the resultant wind mass loss and hence
in the remnant mass, especially for sub-solar metallicities.

\subsection{The remnant-mass prescriptions}\label{newrem}

The remnant-mass model is a very important ingredient of the stellar-evolution
modelling in $\nbseven$ and $\bse$, that determines the masses of the NSs and the
BHs formed as the final product of stellar (and binary; see Sec.~\ref{primbin})
evolution. In other words, such a prescription, together with the underlying
stellar-structure model and mass-loss recipes (see \citealt{Hurley_2000} and
Secs.~\ref{intro} \& \ref{newwind}), determines the ``initial-final''
relation of stellar evolution. Although the final remnant mass can
be expected to be an overall increasing function of the initial ZAMS
mass, such initial-final curves often posses intricate excursions
depending on the details of the remnant-mass prescription,
the wind mass loss prescription, and the underlying
stellar structure (see, \eg, B10, F12; below).

The subroutine \hrdiag~in \nbseven (which is directly ported also
into the standalone \bse after omitting the
$100\Ms$ limit from $\bse$/{\tt star}), that derives the remnant mass from the
CO core mass in the AGB or the naked-Helium phases (in addition to dealing with the
earlier evolutionary phases), adequately incorporates 
the older remnant-mass prescriptions provided by
\citet{Belczynski_2002}, B08, and \citet{Eldridge_2004}. The subroutine
is now enhanced by incorporating the remnant-mass prescriptions
of F12 for the rapid and the delayed SN (Eqns.~15-17 and Eqns.~18-20 of F12 respectively)
\footnote{Formula for ${\rm a_1}$ in Eq.~16 (rapid explosions) of F12 
should be: ${\rm a_1} = 0.25 - \frac{1.275}{M - M_{\rm proto}}$}. 
Moreover, the PPSN and PSN conditions, as described in B16 (see references therein), 
are implemented in the \hrdiag~routine. In the present version of the routine,
any of the five remnant-mass schemes can be chosen
with a switch ($\nsflag=1$, 2, 3, 4, 5 for \citealt{Belczynski_2002}, B08, F12-rapid, F12-delayed,
and \citealt{Eldridge_2004} recipes respectively) and the PPSN/PSN mass cutoff can
be (de)activated independently ($\psflag=0$ or 1).
The fallback amount and fraction onto a BH, $\mfb$ and $\fbfac$ respectively, are calculated for each
case according to their definitions in F12. As in previous studies,
a 10\% neutrino mass loss is assumed to obtain the gravitational mass of
the remnant from its baryonic mass ($=\mproto + \mfb$) in the case of a BH formation
and Eqn.~13 of F12 \citep{Lattimer_1989,Timmes_1996}
to account for the neutrino loss during an NS formation. The already-available
electron-capture-supernova\citep[ECS;][]{Podsiadlowski_2004}-NS formation scheme in $\nbseven/\hrdiag$,
which is analogous to such scheme in B08 producing the characteristic
$\mecsns=1.26\Ms$ NSs, is kept intact (activated with
$\ecflag=1$). Hereafter in this text, `NS' will imply a regular NS produced
through core-collapse SN and an ECS product will be denoted by `ECS-NS'.

Both the new $\mlwind$ and $\hrdiag$ routines, as described above, are now shared
between the standalone \bse and \nbseven.

All the amendments in the standalone $\bse$ code, to implement
the new ingredients as described above
and also in Secs.~\ref{stdkick} and \ref{altkick}, are elaborated
in Sec.~\ref{changes}. Note that this updated $\bse$,
which is used in the rest of this work,
retains the same binary-interaction physics (\ie, treatments of tidal evolution,
mass transfer, common envelope evolution, and mergers) as described in
\citet{Hurley_2002} along with its subsequent amendments as available
in the public version of $\bse$.

\subsection{Comparison with \startrack}\label{comp1}

Fig.~\ref{fig:cmp1} demonstrates the agreement in ZAMS mass-remnant mass
relation between the updated \bse and \startrack
for the B08\footnote{The ZAMS mass-remnant mass relations for the
B08 case correspond to those of B10 which adopts the B08 remnant-mass
model along with the newer wind recipes as described in Sec.~\ref{newwind}
of this paper. This same wind model accompanies all the remnant-mass models
considered in this work.},
F12-delayed, and F12-rapid remnant-mass models without PPSN/PSN
and for $Z=0.0002$, 0.006, and 0.02. Fig.~\ref{fig:cmp_psn} demonstrates
similar agreements, for the F12-rapid case and the metallicities indicated
therein, including the PPSN/PSN recipes of B16. The middle panel of
Fig.~\ref{fig:cmp_psn} exhibits the PSN ``mass gap'' from \bse
for $Z=0.0001$ and that
beyond this gap ($\gtrsim260\Ms$ along the ZAMS axis) the BH formation resumes.

The bottom panel of Fig.~\ref{fig:cmp_psn} shows such ZAMS mass-remnant mass
relations (for the F12-rapid remnant mass model including B16-PPSN/PSN)
up to very large ZAMS masses, beyond the PSN mass gap, for a wide range of $Z$ (colour coding).
It also demonstrates the excellent agreement between \bse (dashed lines) and
\startrack (solid lines) ZAMS mass-remnant mass curves for such large
ZAMS masses. For each $Z$, the extent of the PSN mass gap (\ie, both
the ZAMS mass at which PSN commences and the ZAMS mass, BH mass point at which BH formation
resumes) is also in agreement. Note that with increasing $Z$, the
PSN mass gap diminishes. The BH formation resumes whenever the mass of
the He core,
in the evolved phase (AGB and beyond) of the parent star,
exceeds $135\Ms$ (see B16; \citealt{Woosley_2017}). However, at low $Z$, depending
on the total wind mass loss, an amount
of H-envelope retains (\ie, the star is of AGB type) so that the least massive pre-SN star,
beyond the PSN mass gap,
that would directly collapse to BH (see B16; for a pre-SN
star of such a large mass, the direct collapse condition
is satisfied irrespective of the remnant-mass model)
may well exceed $135\Ms$. For solar-like $Z$, the
much stronger winds eliminate the
H-envelope nearly completely (\ie, the pre-remnant star is
of naked Helium type) so that the pre-SN (pre-collapse)
star, just beyond the PSN mass gap, is of $\approx135\Ms$
which, given the 10\% neutrino mass loss (see B16; Sec.~\ref{newrem}),
results in an $\approx120\Ms$ BH. Hence, the overall PSN BH mass gap,
as obtained from $\bse/\startrack$, ranges between $\approx40\Ms-\approx120\Ms$
as indicated in Fig.~\ref{fig:cmp_psn} (bottom panel; black, horizontal lines) which
limits closely agree with those of \citet{Woosley_2017}.

Fig.~\ref{fig:cmp_ext} individually shows the
close agreements between \bse and \startrack ZAMS mass-remnant mass
curves, until $300\Ms$ ZAMS mass, for the F12-rapid remnant mass model
without (top panels) and with (bottom panels) PPSN/PSN and for $Z=0.02$,
0.002, and 0.0002. The ``clipping'' of the BH mass at $40.5\Ms$, due
to PPSN, is apparent in the $Z=0.0002$ column.

It is worth considering such comparisons between initial mass-final mass
relations also for lower ZAMS masses which are WD progenitors. 
Fig.~\ref{fig:cmp_wd} demonstrates the agreement between the
initial mass-final mass relations from $\bse$ and $\startrack$
over the ZAMS mass range of $1\Ms-20\Ms$. As evident,
the outcomes from $\bse$ and $\startrack$ agree nearly perfectly
beginning from low-mass WDs to massive NSs.

\subsubsection{The time-step parameters in \bse}\label{tstep}

Note that such near-perfect agreement of remnant masses between
\bse and \startrack depends on the choice of the
stellar-evolutionary time step parameters $\ptsone$, $\ptstwo$, and
$\ptsthree$ \citep{Hurley_2000} for the main sequence (MS) and the evolved phases
respectively. These \bse input parameters ($\ptsone$, $\ptstwo$, and $\ptsthree$ respectively)
are essentially fractions of stellar lifetimes in the main sequence,
sub-giant, and more evolved phases that are taken as stellar-evolutionary
time steps in the respective evolutionary stages. For a given underlying
stellar structure \citep{Hurley_2000}, the accuracy and hence the
convergence of \bse's stellar-evolutionary calculation would be compromised
if the time step parameters are chosen to be too large (on the other hand,
choosing too small values would result in impractically long computing time
and as well numerical difficulties).

This is demonstrated
in Fig.~\ref{fig:tstep}, where the B08 remnant-formation scheme
(that is already available in the to-date-public $\nbseven/\hrdiag$) is applied.
As can be seen, the spurious spikes in the 
\bse ZAMS mass-remnant mass curves (thin lines), that appear in
the curves' ``maximum-mass'' part (see B10) especially at
low $Z$,
decline with decreasing time-step parameters. In particular,
$\ptsone=0.05$, $\ptstwo=0.01$, $\ptsthree=0.02$, as defaulted
in $\bse$, still produce considerable spikes (see top-right panel of Fig.~\ref{fig:tstep})
for the lower $Z$s. 
The combination of $\ptsone=0.001$, $\ptstwo=0.01$, $\ptsthree=0.02$
is found to be optimal between speed and convergence over
a wide range of $Z$ (third row, right panel  
of Fig.~\ref{fig:tstep}) which is what is chosen in the \bse and \nbseven
computations in Figs.~\ref{fig:cmp1}, \ref{fig:cmp_psn}, \ref{fig:cmp_ext}
and the rest of the figures in this work. The final panel of
Fig.~\ref{fig:tstep} demonstrates that despite taking very small  
time-step parameters, the currently-public version of $\nbseven/\mlwind$
produces ZAMS mass-remnant mass relations that overshoot
their \startrack counterparts significantly, especially for low $Z$ and
$\gtrsim100\Ms$ ZAMS masses. In particular, the ``saturation'' effect
in BH mass is completely absent unlike the cases with the new \mlwind~(with
the same implementation of the B08 remnant scheme in \hrdiag~) and
\startrack whose outcomes agree nearly perfectly, as demonstrated above.

\section{The adoption of the updated \bse in \nbseven: remnant natal kicks}\label{nbcode}

Although the new $\mlwind$ and $\hrdiag$ routines can be readily shared between the
standalone \bse and \nbseven, the main stellar- and binary-evolution engines in the two codes
are implemented in different ways, which may produce different remnant masses.
Fig.~\ref{fig:cmp_nb} demonstrates as good agreements between the ZAMS mass-remnant mass
relations, over a range of $Z$ and remnant-mass prescriptions,
when the new $\mlwind$ and $\hrdiag$ are adopted in \nbseven.
The \nbseven data in Fig.~\ref{fig:cmp_nb} are obtained whilst evolving model clusters
of initial mass $\mcl(0)\approx5.0\times10^4\Ms$ ($N(0)\approx85$K),
initially-\citet{Plummer_1911} profile with half-mass radius $\rh(0)\approx2$ pc,
having the standard stellar initial mass function \citep[IMF;][]{Kroupa_2001}, and
with all stars initially single. With the
choices of the \bse time-step parameters suggested
in Sec.~\ref{tstep} (which needs to be parametrized
in the routine {\tt trdot} in \nbseven as opposed to specifying them as
input parameters in the standalone \bse), such runs practically do not slow down
during the evolution of the most massive remnant progenitors (the first $\approx20$ Myr),
since the primary bottleneck on computing time (up to a given
physical time) still comes from the direct N-body integration.

\subsection{The retention of black holes in stellar clusters: standard,
fallback-controlled natal kicks}\label{stdkick}

In the present context, the retention of stellar remnants, especially of BHs, in stellar clusters
(young massive clusters, open clusters, and GCs) after their birth is widely debated.
How many and what masses of BHs remain gravitationally bound to their parent cluster after
their birth through a core-collapse SN, depending on their natal kicks,
is instrumental in determining their long-term population evolution in the cluster,
their impact on the structural and internal-kinematic evolution of the cluster
\citep[\eg;][]{Kremer_2018,Askar_2018},
and the nature of their dynamical pairing and GR merger
\citep[\eg;][]{Banerjee_2017,Farr_2017,Rodriguez_2018,Samsing_2018}.
The BH natal kicks are also very important for the formation of BBHs and
BH-star systems and their coalescences through field binary evolution
\citep[\eg,][]{Belczynski_2016,Mandel_2017a}. However, BH natal kicks
are, to date, poorly constrained and understood from both observational and
theoretical point of view
\citep{Willems_2005,Fragos_2009,Repetto_2012,Repetto_2015,Mandel_2016,Belczynski_2016b,Repetto_2017}. 
The single Galactic-field NSs (the vast majority of which would be the
products of core-collapse SN), on the other hand, are observationally inferred
to have large natal kick magnitudes, distributed according to
a Maxwellian with 1-dimensional dispersion $\signs\approx265\kmps$
\citep[average speed of $\approx 420 \kmps$;][]{Hobbs_2005}.

A commonly-used model \citep[as in, \eg,][]{Belczynski_2008,Giacobbo_2018}
for core-collapse-SN natal kick magnitude, $\vkick$,
is to assume NS-like kicks also
for BHs \citep[see, \eg,][]{Repetto_2012,Janka_2013,Repetto_2015} but which are scaled down linearly
with increasing material-fallback fraction, $\fbfac$ (see Sec.~\ref{newrem}),
so that for $\fbfac=1$ (\ie, a failed SN or direct collapse) the natal kick is necessarily zero:
\beq
\vkick=\vkickns(1-\fbfac),
\label{eq:vkick_std}
\eeq
where $\vkickns$ is chosen randomly from a Maxwellian of $\signs=265\kmps$. For the
model computations presented in this subsection, both BHs and
NSs are treated with this kick scheme, except for the ECS-NSs which are given zero or
$\sim {\rm~few}\kmps$ natal kicks \citep{Podsiadlowski_2004,Gessner_2018}.

The $\vkick$, according to Eqn.~\ref{eq:vkick_std}, is applied in \nbseven
by transporting the $\fbfac$, as computed in $\hrdiag$ (see Sec.~\ref{newrem}),
to the standard version of the \nbseven's subroutine $\kick$ via a dedicated {\tt common block}.
The $\kick$ routine already includes an elaborate algorithm for generating
an isotropic distribution of velocity vectors with their magnitudes, $\vkick$s, chosen
from a Maxwellian distribution of a specified dispersion. As in the original $\nbseven/\kick$ routine,
the ECS-NSs are distinguished based on their $\mecsns=1.26\Ms$, which are subjected
to lowered-dispersion or zero $\vkick$s and are exempted from the above
fallback treatment (the ECS engine is assumed to produce a full explosion).
Keeping in mind that the newly-planted
remnant-formation schemes in $\hrdiag$ such as F12-rapid (Sec.~\ref{newrem})
produce non-monotonic ZAMS mass-NS mass
relations and, in particular, sub-Chandrasekhar NSs (see, \eg, Fig.~\ref{fig:kcomp_bhdist}),
a very narrow mass window around $\mecsns=1.26\Ms$ is allowed for the ECS-NS treatment
(unlike in the default $\nbseven/\kick$ where the ECS-NS treatment
was invoked for $\mns\leq1.28\Ms$, since the ECS-NSs were anyway the distinctly
least massive NSs produced according to the \citealt{Belczynski_2002} and B08
prescriptions). Analogous updates are now implemented also in the
standalone \bse's kick treatment which would then accordingly modify
a binary's response to an SN.

Fig.~\ref{fig:bhmass_cmp1} shows the early (up to 20 Myr) evolution of the total mass, $\mbhbound$ (top-left panel),
and number, $\nbhbound$ (bottom-left panel), of the BHs
bound within the $\mcl(0)=5.0\times10^4\Ms$ (initially all single stars)
model cluster (see Sec.~\ref{nbcode}) for the various remnant-formation schemes (\ie, B08, F12-delayed, F12-rapid,
F12-rapid+B16-PPSN/PSN; see Sec.~\ref{newrem}),
for this standard, fallback-controlled kick prescription, as given by Eqn.~\ref{eq:vkick_std},
when adopted in \nbseven as above ($Z=0.0001$ assumed)
\footnote{Hereafter, the subscript `bound' indicates quantities measured within the cluster's tidal radius,
$\Rt\approx90$ pc,
as is customary. Most of the BHs within $\Rt$ are indeed gravitationally bound to the cluster but,
at a given time, a few of them may be on their way to escape the cluster.
The number of the latter depends on the BHs' $\vkick$ and on the escape speed from the cluster
($\vesc\approx40\kmps$ here).}.
The differences in the $\nbhbound-t$ curves between the remnant-formation cases arise due to   
the differences in $\fbfac$ which quantity determines the $\vkick$ of the BH (or NS).  
The differences in the $\mbhbound-t$ curves additionally arise due to the differences in the BHs'
masses (for a given ZAMS mass and $Z$) between the various remnant-formation cases.
Note that in all of the remnant-formation prescriptions considered here, all BHs and the
heaviest NSs form with a nonzero $\fbfac$. In the F12-rapid/delayed schemes, all NSs form
with a small amount of fallback (fallback mass, $\fbtot>0.2\Ms$; see F12). However, for all
NSs, $\fbfac$ is small ($\lesssim 10^{-2}$) and the corresponding $\vkick$ is typically large 
enough (but see Sec.~\ref{altkick}) to eject them from a young massive, open, or globular cluster.
The only NSs that would retain in clusters following their birth would be the ECS-NSs that
receive small natal kicks (see above), in the present (and the following; Sec.~\ref{altkick}) kick
scheme(s).

As can be seen, in terms of retained BHs (but see below), the B08 remnant-mass model produces more BHs
(in both mass and number) than any other remnant model, due to the wider ``direct-collapse hump'' in
its ZAMS mass-remnant mass relations for lower
ZAMS masses ($\sim20-40\Ms$; see Fig.~\ref{fig:cmp1} and B10), and hence expands a cluster more
(see the bottom-right panel of Fig.~\ref{fig:bhmass_cmp1}; the expansion beyond $\sim50$ Myr
is driven mainly by the centrally-segregated BH subsystem). For the F12-rapid case, having PPSN/PSN or not
makes a little difference in practice (as long as bulk properties are concerned,
the PPSN truncation has important implications for interpreting BBH merger masses;
see, \eg, B16, \citealt{Mandel_2017}):
due to the standard IMF in the model cluster, there are only a few stars undergoing PPSN/PSN.
The dips in the $\mbhbound-t$ and $\nbhbound-t$ curves for the F12-rapid cases
(top-left and bottom-left panels of Fig.~\ref{fig:bhmass_cmp1}) are due to a significant
number of BHs receiving
partial fallback, occurring in the dip between the two direct-collapse humps of
the corresponding ZAMS-remnant curves (see Fig.~\ref{fig:cmp_nb}),
which escape marginally, taking time. For massive GCs and nuclear clusters,
whose escape speeds, $\vesc$, are much higher ($\vesc\gtrsim100\kmps$), this feature would vanish.
The BH subsystems, for all cases, nevertheless sink and concentrate similarly for all
the remnant-formation cases (top-right panel of Fig.~\ref{fig:bhmass_cmp1}).

Fig.~\ref{fig:bhmass_cmp2} shows the BH mass distributions for the four remnant-formation scenarios
as indicated in the panels' legends,
for $Z=0.0001$. On
each panel, both the mass distribution with which the BHs are born (steel blue histogram)
and the mass distribution of the BHs retaining at $t\approx20$ Myr (blue histogram), in the
$\mcl(0)=5.0\times10^4\Ms$ cluster (standard IMF; initially single-only stars),
are shown. At a time such as $t\approx20$ Myr, which is well after the completion
of the BH formation (the last BH forms at $\approx 11$ Myr), all the unbound
BHs (due to $\vkick>\vesc$) have escaped through the tidal radius but the remaining,
bound BHs are still in the midst of segregating towards the
cluster's center (Fig.~\ref{fig:bhmass_cmp1}, top-right panel) and their dynamical processing
is yet to begin. Therefore, the blue histograms fairly represent the mass distributions
of the BHs which become available for long-term dynamical processing, for the various
remnant-formation schemes (corresponding to $\mcl(0)\approx5.0\times10^4\Ms$ with
$\vesc\approx40\kmps$, standard IMF, $Z=0.0001$). 

In fact, as seen in Fig.~\ref{fig:bhmass_cmp2}, the F12-delayed case inherently produces the
largest number of BHs from stellar evolution (as opposed to the resulting retained BHs where
B08 wins; see above),
due the largest proto-remnant core + fallback masses (see F12) resulting in the lowest BH-(and NS-)formation
threshold with respect to the core-collapsing progenitor mass,
but it also produces the largest number of escapees due to its smallest direct-collapse hump,
ultimately resulting in the smallest number of retainers (see Fig.~\ref{fig:bhmass_cmp1}).
The F12-rapid cases produce less escapees (more direct- and near direct-collapse BHs)
due to the ``double hump'' (see Fig.~\ref{fig:cmp_nb}).
The BH distribution is truncated at $\approx40\Ms$, as expected, when PPSN/PSN is included.

The discrepancy between the natal and the retained mass distribution,
as in Fig.~\ref{fig:bhmass_cmp2},
gives the number and the mass retention fraction of BHs in a cluster,
given the $Z$ and the remnant-formation model. Table~\ref{tab_retfrac} provides such
BH retention fractions, corresponding to
the standard, fallback-controlled natal kicks (Eqn.~\ref{eq:vkick_std}),
for the various remnant-formation models and $Z$s considered here. 

\subsection{The retention of black holes in stellar clusters: alternative natal-kick
prescriptions}\label{altkick}

While the fallback-modulated natal kick prescription has some observational and theoretical motivation
(see Sec.~\ref{stdkick} and references therein) and is often applied
in N-body and population-synthesis studies, it is far from any
robust constraint on observational or theoretical grounds, especially for BHs. Therefore,
there exists ample room for alternatives of such prescriptions, in particular,
for translating the NSs-like kicks to BHs that comprise a much wider remnant-mass range (see, \eg,
Fig.~\ref{fig:bhmass_cmp2}). From the
theoretical perspective, a key uncertainty lies in identifying the mechanism(s)
that generate spatial and directional asymmetries in the SNs' ejected (baryonic) material, which, in turn,
momentum-propel the remnant. On the other hand, the asymmetry in the
neutrino flux during an SN could alone be enough to kick an NS with the
typical observed speed \citep{Hobbs_2005}. If we assume \emph{one dominant
SN-kick mechanism} and take the observed NS natal kicks as the constraint, then
we are already left with several possibilities as formulated below.

\subsubsection{Convection-asymmetry-driven natal kick}\label{nkconv}

The convection-asymmetry-driven kick mechanism assumes that the
natal kicks are produced by asymmetries in the convection
within the collapsing SN core \citep{Scheck_2004,Scheck_2008,Fryer_2007}.
This is in the line of the standard kick
prescription but a more accurate prescription would be to boost the kick for systems with
longer convection durations: a lower-mode convection produces stronger kicks and it develops with time,
so the longer it takes to explode the stronger is the kick. In that case,

\beq
\left\lbrace
\begin{array}{lrl}
\vkick = &  \vkickns\frac{\mnsav}{\mrem}(1-\fbfac) & \mco\leq3.5\Ms\\ 
\vkick = & \kconv\vkickns\frac{\mnsav}{\mrem}(1-\fbfac) & \mco>3.5\Ms. 
\end{array}
\right.
\label{eq:vkick_conv}
\eeq

Here, \mnsav is a typical NS mass, \mrem is the remnant (NS or BH) mass, \mco is
the CO core mass, and \kconv is an efficiency factor (somewhere between 2-10)
with the theory that we are more likely to get a low-mode convection in
explosions that take longer to develop, \ie, those with larger $\mco$.

\subsubsection{Collapse-asymmetry-driven natal kick}\label{nkcollapse}

Another alternative ejecta kick mechanism would be the one produced when asymmetries in the
pre-collapse silicon and oxygen shells produce asymmetric explosions
\citep{Burrows_1996,Fryer_2004,Meakin_2006,Meakin_2007}. In contrast to the
convection-asymmetry mechanism (see above),
this mechanism would produce stronger kicks for shorter convective durations (smaller $\mco$)
since the asymmetries would be washed out by long periods of convection:

\beq
\left\lbrace
\begin{array}{lrl}
\vkick = &  \vkickns\frac{\mnsav}{\mrem}(1-\fbfac) & \mco\leq3.0\Ms\\ 
\vkick = & \kcol\vkickns\frac{\mnsav}{\mrem}(1-\fbfac) & \mco>3.0\Ms. 
\end{array}
\right.
\label{eq:vkick_col}
\eeq
Here, $\kcol\approx0.1$ is the suppression factor for larger $\mco$s. 

\subsubsection{Neutrino-driven natal kick}\label{nknu}

The neutrino mechanism \citep{Fuller_2003,Fryer_2006} would produce the kick
through asymmetric neutrino emission. In that case, a formulation would be

\beq
\vkick = \vkickns\frac{\min(\mrem,\meff)}{\mrem}.
\label{eq:vkick_nu}
\eeq

Here, $\meff$ ($5\Ms\lesssim\meff\lesssim10\Ms$) is the effective remnant mass
beyond which the neutrino emission would not increase significantly
with increasing mass of the SN core. Unlike the baryonic material,
neutrinos do not fall back on to the remnant to return a part (or whole)
of the ejecta momentum, despite the occurrence of such a fallback
(or of a failed SN). Hence the omission of the $(1-\fbfac)$ term
in Eqn.~\ref{eq:vkick_nu}.

\subsubsection{The dependence of BH retention on natal-kick mechanism}\label{nkmech}

It would be interesting to study the impact of these different natal-kick
prescriptions on the retention of BHs and NSs in a cluster and,
potentially, formulate signatures to observationally distinguish between
such cases. For this purpose, we continue to utilize
the $\mcl(0)\approx5.0\times10^4\Ms$, $\rh(0)\approx2$ pc model
cluster (initially single-only stars; Sec.~\ref{nbcode}) here, which would facilitate comparisons. 
For definiteness, $\kconv=5.0$ (Eqn.~\ref{eq:vkick_conv}), $\meff=7.0\Ms$ (Eqn.~\ref{eq:vkick_nu}),  
and $\mnsav=1.4\Ms$ are used. 

Fig.~\ref{fig:kcomp_num} shows the early time evolution of
$\mbhbound$, $\nbhbound$, and $\nnsbound$ in the cluster for the
different natal-kick prescriptions formulated in Secs.~\ref{stdkick}, \ref{nkconv},
\ref{nkcollapse}, and \ref{nknu},
as obtained from \nbseven computations ($Z=0.0001$). For implementing the
alternative kick recipes in $\nbseven/\kick$ and $\bse/\kick$, that can
be selected with a switch, {\tt KMECH}, in the $\kick$ routines,
the corresponding $\mco$ is also transported from $\hrdiag$ 
to $\kick$ via the dedicated {\tt common block} (see Sec.~\ref{stdkick}).

As seen, the standard and the convection-asymmetry-driven cases collect very similar
number and mass of BHs within the cluster while the
collapse-asymmetry-driven case collects more BHs. This is
expected since the latter kick model imparts much lower kicks
(\cf, Eqns.~\ref{eq:vkick_std}, \ref{eq:vkick_conv}, and \ref{eq:vkick_col})
onto the BHs and also onto the relatively massive NSs.
The neutrino mechanism does the worst in this regard with no BHs
retained in the cluster: if the observed high NS kicks are
indeed due to asymmetric neutrino emission alone then all BHs
(provided the BH natal kick mechanism is the same as for NSs)
would get substantially higher
kick despite significant fallback (the emitted neutrinos would not return
any momentum since they do not participate in fallback; see Sec.~\ref{nknu}).
If there is indeed, depending on dynamical age, a significant number ($\sim100$)
of stellar-mass BHs retaining and comprising a BH subsystem in several Galactic GCs
\citep[\eg,][]{Askar_2018,Kremer_2018,Kremer_2018b}, then, perhaps, the neutrino-driven
kick mechanism is disfavoured.

The $\nnsbound$ converges to a few after $\approx50$ Myr after which
ECS-NSs, due to their slow/zero $\vkick$ dispersion (see Sec.~\ref{stdkick}),
start to build up.
However, there is a significant number of slow-escaping NSs' buildup
around 10-20 Myr age (see Fig.~\ref{fig:kcomp_num}, bottom panel) which is
unique for the case of the collapse-asymmetry-driven kick that may be interesting for
radio observations in young massive clusters and for supporting or ruling-out
such a kick scenario. Note that the model clusters utilized in these
\nbseven runs have $\vesc\approx40\kmps$ which is 1/2 to 1/3 of typical
$\vesc$s in massive GCs, so that one would potentially retain some BHs
in massive, compact GCs in all the kick-mechanism cases anyway,
although the overall trends would be similar as in Fig.~\ref{fig:kcomp_num}.
A survey of such GC models exploring these different natal-kick mechanisms,
based on the Monte Carlo cluster evolution program \mocca, is underway
(Leveque, A., et al., in preparation).

Fig.~\ref{fig:kcomp_bhdist} (left panel) compares, analogously to Fig.~\ref{fig:bhmass_cmp2}
(see Sec.~\ref{stdkick}),
the mass spectra of the BHs retained in the parent cluster (at $t\approx20$ Myr), for the
different natal-kick prescriptions considered here ($Z=0.0001$ is assumed).
As above, the standard and the convection-asymmetry-driven mechanisms
produce similar retained BH mass spectra and nearly same BH retention in
terms of number and mass. BHs $\lesssim10\Ms$ escape the cluster at or shortly after birth.
In contrast, the collapse-symmetry-driven mechanism produces low natal kicks for all BHs
(see also Fig.~\ref{fig:kcomp_comass}) so that all of them retain in the cluster, following their
natal mass spectrum (the dashed, magenta histogram of Fig.~\ref{fig:kcomp_bhdist}).
Note that such complete retention of BHs for the collapse-asymmetry-driven
kick mechanism is due to the low kicks for $\mco>3\Ms$ (Eqn.~\ref{eq:vkick_col})
which is below the BH-formation threshold. 
Therefore, this is a generic feature and it holds true for all
remnant-mass schemes and for all metallicities.
In particular, as seen in Fig.~\ref{fig:kcomp_bhdist} (left panel),
$\lesssim10\Ms$ BHs would also be retained in GCs and open clusters;
this particular feature is unique to the collapse-asymmetry-driven kick mechanism which can,
therefore, be
used to support or rule out such a kick scenario. Note, in this context,
that the BH identified in the GC NGC 3201, through radial-velocity measurements,
has a minimum mass of $4\Ms$ \citep{Giesers_2018}.

The natal mass distribution of the NSs is also shown in
Fig.~\ref{fig:kcomp_bhdist} (right panel, black histogram). The prominent
sub-Chandrasekhar peak is a feature of the F12-rapid scheme applied,
convolved with the cluster's standard IMF -
these NSs are just the
($1\Ms$ proto-remnant $+ 0.2\Ms$ fallback)$\times0.9$ NSs (see F12).
As mentioned before (Sec.~\ref{stdkick}), only the ECS-NSs retain
in the cluster for longer term which is demonstrated 
in Fig.~\ref{fig:kcomp_bhdist} (right panel, blue histogram). For massive GCs,
some of the slow-escaping NSs would additionally retain in the collapse-asymmetry-driven-kick
case (see above; Fig.~\ref{fig:kcomp_bhdist}, right panel, gray histogram).

Fig.~\ref{fig:kcomp_comass} shows the $\vkick$s generated by \nbseven,
as a function of $\mco$ (left panel) and $\mrem$ (right panel),
for the different natal-kick recipes considered here (due to the logarithmic vertical
axis, direct-collapse BHs with $\fbfac=1$ and $\vkick=0$ are not shown in
these panels). Comparing with the present model clusters' typical
$\vesc$ (the solid, blue line), these panels clarify that
for such clusters the standard and the convection-asymmetry-driven
kick models would yield similar BH retention, the collapse-asymmetry-driven
case would retain all BHs, whereas the neutrino-driven case would eject nearly
all BHs. The $1.26\Ms$ ECS-NSs (see Sec.~\ref{stdkick}),
which are given $\vkick$s from a Maxwellian distribution of much
lowered dispersion (1-dimensional dispersion of $\sigma_{\rm ECSNS}\approx3\kmps$;
see, \eg, \citealt{Gessner_2018}), make up the vertical
stripe along the $\mrem$ axis (Fig.~\ref{fig:kcomp_comass}, right panel).
Most of these ECS-NSs retain in these model clusters due to
their low natal kicks. From Fig.~\ref{fig:kcomp_comass}, it
is also apparent that such conclusions concerning the retention of BHs and
NSs in clusters remain nearly unaltered for $10\kmps\lesssim\vesc\lesssim100\kmps$, \ie,
for young massive clusters, moderately-massive open clusters, and typical GCs.
The mass gap between NSs and BHs (Fig.~\ref{fig:kcomp_comass}, right panel) 
is a characteristic of the F12-rapid remnant scheme adopted in this example.

\section{The effect of massive primordial binaries}\label{primbin}

Until now, model clusters initially comprising only single stars are considered in
this work.
It would be worth looking at the case where a population of primordial
binaries is present in the model, especially among the BH-progenitor
stars. The merger among the members of a massive-stellar binary,
especially during their advanced (non-remnant) stellar-evolutionary stages,
can lead to a stellar structure (the relative masses of the CO-core, He-core,
and H-envelope) that is not achievable through
the evolution of a single star (even of the ZAMS mass that is equal to the sum
of the ZAMS masses of the members; see below, Fig.~\ref{fig:bhwbin2})
and hence can produce a (single) BH
that deviates significantly from the corresponding single-star ZAMS mass-remnant mass relation
(see, \eg, \citealt{Marchant_2016,Kruckow_2018,Spera_2019}).
Such merger events would, therefore, modify the natal and the retained
BH mass spectra which would, in turn, modify the BH-driven long-term
cluster evolution and the dynamical GR-merger events involving BHs.
Note that in the dynamically-active environment of a dense stellar
cluster, massive-stellar mergers can happen not only in the course of
a massive-binary evolution (leading to
Roche lobe overflow and common-envelope phases; see, \eg, \citealt{Ivanova_2013})
but also during close
binary-single and binary-binary interactions
(\eg, in-orbit stellar collisions during
Kozai oscillations in a dynamically-formed hierarchical
triple or during a resonant triple-interaction;
see, \eg, \citealt{Banerjee_2012,Leigh_2013}). Even the outcomes of the
massive-binary evolution channel can get significantly altered via close dynamical
encounters that would alter the binaries' orbital parameters.
In principle, in such an environment,
a given massive star can undergo multiple    
mergers until a BH is formed \citep[\eg,][]{Fujii_2013,Wang_2020}
or even a BH non-progenitor can grow
in mass to become a BH progenitor.
Finally, note that BH masses can get altered by not only such star-star
mergers but also other stellar-hydrodynamical processes such as envelope
ejection during a common-envelope (CE) phase and enhanced stellar wind
in a tidally-interacting binary; such processes can also be either an
outcome of massive-binary evolution or triggered dynamically (or both).

A detailed study of the influence of dynamical interactions on massive-binary
evolution is beyond the scope of the present work. To have an idea
of to what extent BH masses can get altered due to their
progenitors being binary members, we evolve with $\nbseven$
$\mcl(0)\approx5.0\times10^5\Ms$, $\rh(0)\approx2$ pc model
clusters as before but with all stars (also sampled from a standard
IMF) of ZAMS mass $\gtrsim16\Ms$ initially paired in binaries
among themselves. A high fraction of massive-stellar (O-star) binaries
($\fobin(0)=100$\% initially in these models) is indeed consistent with observations
of young massive clusters \citep{Sana_2011,Sana_2013}.
Motivated by such observations, these O-type-stellar binaries are
taken to initially follow
the orbital-period distribution of \citet{Sana_2011}
and a uniform mass-ratio distribution
(an O-star is paired only with another O-star, as typically observed, and the pairing among the
lower-mass stars is obtained separately; see below).
The orbital periods of the non-O-star ($\lesssim16\Ms$ ZAMS)
primordial binaries are taken to follow a
\citet{Duq_1991} distribution that represents a dynamically-processed binary population
\citep{Kroupa_1995a} and their mass-ratio distribution is also taken to be uniform.
The initial binary fraction among the non-O-type stars is taken to be small;
$\fbin(0)\approx5$\%. The initial eccentricities of the O-type stellar
binaries follow the \citet{Sana_2011} eccentricity distribution and those
for the rest of the binaries are drawn from the thermal eccentricity distribution
\citep{1987degc.book.....S}\footnote{We note the recent work by \citet{Geller_2019} questioning the
validity of the assumption of an initial thermal eccentricity distribution of
the primordial binaries. The sub-population of primordial binaries that directly
influences the present results is the O-star-binary population for which
the adopted (sub-thermal) initial \citet{Sana_2011} eccentricity distribution is motivated
by observations.}.
As explained in \citet{Banerjee_2017b}, such a scheme for including
primordial binaries provides a reasonable compromise between the economy of computing
and consistencies with observations.

The filled, black squares in Fig.~\ref{fig:bhwbin} (top row) show the remnant mass
versus ZAMS mass for the above model cluster with massive primordial binaries, when
evolved with $\nbseven$ that includes the new $\bse$ (Sec.~\ref{newbse}); for the
BH progenitors that have undergone a star-star merger before the BH formation, the ZAMS mass
of the primary (the more massive of the members participating in the star-star merger, at the
time of the merger) is plotted in the abscissa. Due to BH progenitors undergoing mergers 
in their evolved phases (and/or due to other stellar-hydrodynamical processes)
before the BH formation (see above)
\footnote{The BH masses are printed right \emph{at their formation} by arranging a
special output in the new $\kick$ routine (Secs.~\ref{stdkick}, \ref{altkick}),
as also for the $\nbseven$ data points in Fig.~\ref{fig:cmp_nb}.
Hence, any merger(s) or stellar-hydrodynamical process(es), influencing
an $\nbseven$ BH mass in Fig.~\ref{fig:bhwbin}, must have happened
before the BH formation. Otherwise,
the BH mass will necessarily lie on the ZAMS mass-remnant mass relation as in Fig.~\ref{fig:cmp_nb}.
In particular, the post-formation mass gain of a BH (due to, \eg, mass transfer, BH-star
collision) is not captured in Fig.~\ref{fig:bhwbin}.},
the resulting BH masses deviate significantly from the ZAMS mass-remnant mass relations
(which relations, by definition, are for single ZAMS stars) 
for the assumed remnant-formation scheme (F12-rapid with and without PPSN/PSN) and
metallicity ($Z=0.0001$). The majority of the BHs are over-massive, showing broader natal and retained 
BH mass distributions as contrasted with those from single stars only, for 
models with and without PPSN/PSN physics included (Fig.~\ref{fig:bhwbin}, second row;
the standard, fallback-controlled natal kick is assumed). 
Also, due to the formation of more massive BHs, the total \emph{mass} of the BHs
formed/retained in the cluster is similar to that of the
initially-single-star-only model clusters (Fig.~\ref{fig:bhwbin}, third row)
despite the BHs forming/retaining in a smaller \emph{number}
(due to the mergers; Fig.~\ref{fig:bhwbin}, fourth row).  
These dependencies, as demonstrated in Fig.~\ref{fig:bhwbin},
suggest that BH retention in a cluster and
the retained BHs' mass spectrum both depend on the BHs' progenitor stars'
binary properties. Table~\ref{tab_retfrac} demonstrates the somewhat
increased mass and number retention \emph{fractions} of BHs, at Z=0.0001,
in the presence of a massive primordial-binary population as
adopted here (\cf the top and the bottom sections of the table).
Note that although complex, dynamical-interaction-mediated channels
leading to star-star mergers (see above) among
members from different binaries can and do occur in these models,
most of these mergers are between the members of the same O-star primordial
binary occurring due to Roche lobe overflow (the overflow itself can,
of course, be triggered via dynamical perturbation of the binary's orbit)
\footnote{Some authors distinguish between ``coalescence'' occurring
due to mass transfer in a binary and ``merger'' due to an in-orbit or a
hyperbolic collision. In this paper, we will call the event a ``merger'',
irrespective of the channel, whenever two stars combine to become a single entity.}.

The BH masses in Fig.~\ref{fig:bhwbin} should, however, be taken with caution.
The BH masses depend on how the evolved stars are ``mixed'' at their merger
to compose the stellar collision product. This is rather simplistic in $\bse$
where a complete mixing is assumed to compute the stellar-evolutionary age
and hence the CO-core and He-core masses of the merged star. A simple recipe for
the ``dynamical'' mass loss during the merger process is assumed which results in
a mass loss of $\lesssim 15$\% of the total available
mass budget per merger \citep[\eg,][]{Glebbeek_2009} 
\footnote{In the version of \bse adopted
in \nbseven, a 30\% mass loss from the secondary (the less massive
merger participant) is assumed when
the kinetic energy of approach of the merging non-compact stars exceeds the
internal binding energy of the secondary, provided the stars geometrically
overlap at the merger and
provided they are in the main sequence. Merger among more evolved
stars and grazing encounters are even poorly understood/studied and no mass loss is associated
with such mergers in $\nbseven/\bse$. If one of the merging member
is an NS or a BH, an unstable Thorne-{\.Z}ytkow object is formed.
In such a case, the entire secondary is assumed to be dissipated in the case
of an NS core and half of the secondary mass is retained onto the BH in
the case of a BH core.}. In reality, a star-star merger is likely to
be a much more complex process whose study is beyond the scope
of the present work. Nevertheless, this study suggests that
the natal/retained BH mass function in clusters can potentially be
substantially modified and broadened, despite PPSN/PSN,
provided the clusters contain a high fraction of massive primordial
binaries at their young age, as observed in young massive and open
clusters.

This is due to the ``unusual'' structure obtained in  
merger products that contain a much more massive H-envelope when 
the He-core is below the PPSN threshold of $45\Ms$ (that see B16),
compared to that of a single-stellar-evolution product of the
same He-core mass.
The substantial H-envelope in the merger products results
in BHs (formed via direct collapse) well exceeding $50\Ms$ 
despite the sub-PPSN He-core. Note that the
BH mass surpasses the $\mbh=40.5\Ms$ PPSN cutoff (which mass would result
from the collapse of a $45\Ms$ He-core after subtracting for
the 10\% neutrino mass loss)
for ZAMS mass approaching $\approx100\Ms$ from below (resulting in the He-core approaching the PPSN
threshold of $45\Ms$)
also for pure single-star evolution, but only by a few $\Ms$ (see Figs.~\ref{fig:cmp_psn} \&
\ref{fig:cmp_nb}).  
Owing to much more massive H-envelope for similar He-core masses,
this surpassing happens to a much larger extent for the
BHs obtained from merger products that result from
the observationally-motivated massive binary population, 
as adopted here, in a dynamically-active environment.
However, this still retains the notion that
BBH mergers comprising one or both of the members having $\mbh\gtrsim40\Ms$
(the single-stellar PPSN cutoff limit)
are necessarily of dynamical origin \citep[see, \eg,][]{Rodriguez_2018}. This is because
after being formed from a star-star merger product,
a BH can continue to interact with stellar members and remnants
and participate in a GR merger only in a dynamically-active environment.
However, the occurrence of BBH mergers within the ``PSN mass gap'' in this way 
would not require low natal spins of (first-generation) BHs. On
the other hand, if stellar progenitors indeed produce
very low spin BHs irrespective of their own spin, internal structure, and
chemical composition \citep{Spruit_2002,Fuller_2019}, then, again,
in contrast to the widely-conceived notion of
second-generation BHs having high spins, such mass-gap BBH mergers
would exhibit a low effective spin parameter, provided
the first-generation, mass-gap BH merges with another first-generation BH.
See also \citet{Spera_2019} in this context.

On the other hand, when the He-core mass exceeds the PPSN threshold
of $45\Ms$ and until the end of the PSN gap is reached (He-core
mass of $135\Ms$; see Sec.~\ref{comp1}), the outcome becomes independent
of the mass of the H-envelope around it. Namely, when the He-core
mass lies between $45\Ms-65\Ms$ (the PPSN range) a BH of $40.5\Ms$ is born
via direct collapse as a result of PPSN and for He-core between
$65\Ms-135\Ms$ no remnant is left due to PSN (the BH formation
is resumed and the BH mass will
again depend on the H-envelope mass if the He-core exceeds $135\Ms$; see Sec.~\ref{comp1}
and Fig.~\ref{fig:cmp_psn}). Note that in the present \bse scheme
a stellar object is always treated in the same, single-star
way (see Sec.~\ref{newrem}) irrespective of it being a merger product
or not. Hence, even for a merger product with unusually-massive H-envelope,
a $40.5\Ms$
BH (no BH) is formed whenever the PPSN (PSN) condition is reached, such
conditions and the corresponding outcomes being dependent solely on
the He-core mass. This is why in Fig.~\ref{fig:bhwbin} (top row, right
panel, filled, black squares), there are only $40.5\Ms$ or no BHs
beyond the start of the horizontal, single-star PPSN line (at ZAMS mass $\approx100\Ms$
for $Z=0.0001$) even in the presence of massive primordial binaries:
a merger, which typically leads to a similarly- or
more-massive He-core (depending on the mass ratio and the
epochs of the merger members), will exclusively satisfy either the PPSN or the PSN
condition when a star of ZAMS mass $\gtrsim100\Ms$ is involved.
In this example, all but one of the PPSN derived BHs and PSN points,
beyond $100\Ms$ ZAMS, come from star-star merger products. 
For reference, Fig.~\ref{fig:bhwbin2} re-plots the top row of Fig.~\ref{fig:bhwbin}
with the sum of the ZAMS masses of the star-star mergers' members
along the abscissa, for the primordial-binary model. Here, of course,
the deviations from the single-star ZAMS mass-remnant mass
relations are less dramatic, especially for large combined ZAMS masses,
and not all BHs beyond $\gtrsim100\Ms$ combined ZAMS are formed
via PPSN.

Note that when a much more massive H-envelope,
compared to what would be available from single stellar evolution, 
is present due to, say, a merger as considered here, a PPSN may
shed only a part of the envelope resulting in a BH that
is significantly more massive than what the collapse of the He-core alone
(\ie, if the envelope is entirely expelled) would give.
The PPSN/PSN recipe of B16 (see above and Sec.~\ref{newrem}), as implemented
in the current \startrack and in the new \bse introduced
in this work and which is applied also for star-star merger products, assumes
complete stripping of the H-envelope, always resulting in a $40.5\Ms$ BH.
This is in line with the hydrodynamic study of PPSN/PSN
in single massive stars by \citet{Woosley_2017} where
a small remaining H-envelope results in BHs up to $\approx52\Ms$
as outcomes of PPSN in low-$Z$, massive stars.
More massive BHs would potentially result from merger products
where the PPSN is triggered in the presence of a substantially more massive
H-rich envelope. This would, in turn, diminish the PSN mass gap (see
Sec.~\ref{comp1} and references therein, Fig.~\ref{fig:cmp_psn}).
Also, due to the enhanced pressure from such an envelope,
PPSN may be triggered ``prematurely'' when the He-core is
of $\lesssim45\Ms$ resulting in a longer PPSN phase with a larger number of
mass loss episodes. At present, such stellar models are unavailable.
Although beyond the scope of the present study, a more elaborate
study of this regime will be presented elsewhere.

Of course, the extent of the exceeding of the single-star PPSN
cutoff depends on the amount of ``dynamical'' mass loss during the
star-star merger process. The physical process of merger among stars
and the associated mass loss is poorly understood;
especially so for massive stars beyond the main sequence.
At present, the mass loss from star-star mergers in $\nbseven/\bse$
is at most moderate, ranging from $\approx15$\% (energetic, equal-mass MS-MS mergers)
to zero (non-MS or low-K.E. mergers; see above), the
MS-MS merger mass loss being consistent with
basic hydrodynamic \citep[\eg,][]{Lombardi_2002} and
``sticky-star'' \citep[\eg,][]{Gaburov_2008} treatments of the process.
However, if the mass loss per merger is much higher, say,
comparable to the mass of the secondary in all mergers, 
then the H-envelope in the merger products would be
similarly massive as in single-stellar evolution.
In such a case, the remnant mass-ZAMS mass relations
and the PSN mass gaps would be much less affected
by massive-stellar mergers from primordial binaries.

Fig.~\ref{fig:bhwbin3} demonstrates this for the case of F12-rapid
remnant-formation including PPSN/PSN and low metallicity ($Z=0.0001$). The panels
in Fig.~\ref{fig:bhwbin3} show the remnant mass-ZAMS mass 
outcomes (black, filled squares) from different calculations
of the primordial-binary model with increasing
fraction, $\fmrg$, of the secondary's
(the less massive merger participant's) mass lost
during a merger: from $\fmrg=0.3$ in energetic MS-MS mergers only ($\nbseven/\bse$ default; left panel),
through $\fmrg=0.5$ in all star-star mergers (middle panel),
to $\fmrg=0.8$ in all star-star mergers (right panel)
\footnote{These enhanced $\fmrg$s are implemented
in \nbseven for demonstration purposes only.
It would, perhaps, be more realistic to have at least a mass condition
for the value of $\fmrg$ which will be implemented
in the near future.}
\footnote{The filled, black points won't always
map through the panels in Fig.~\ref{fig:bhwbin3} since varying $\fmrg$ alters
the local gravitational potential and hence the dynamical
interactions, so that a particular merger does not necessarily re-occur
in the different N-body computations and the sequence
of the mergers is also altered over the different runs.}.
Fig.~\ref{fig:bhwbin3} suggests that despite mergers
of massive primordial binaries, $\fmrg=0.5$ ($\lesssim25$\%
loss of the total primary + secondary mass at the time of
the merger; see \citealt{deMink_2013} in this context) already restores
the PPSN/PSN plateauing of BH mass as obtained
from the latest core-collapse calculations (maximum
BH mass of $\approx52\Ms$; \citealt{Woosley_2017})
and $\fmrg=0.8$ ($\lesssim40$\% loss of the total available
mass budget) nearly restores the entire single-star
remnant mass-ZAMS mass relation. In other words,
if the single-star remnant mass-ZAMS mass relation
is to be unaffected by massive primordial-binary mergers,
then a substantial, up to $\approx40$\%, mass loss would be necessary
in such mergers.

\subsection{Comparison with $\startrack$ binary evolution}\label{comp2}

In Sec.~\ref{comp1}, the comparison between the remnant outcomes of $\startrack$ and $\bse$
were limited to single stellar evolution.
In the present context, it would be worth comparing between the outcomes of
binary evolution with $\startrack$ and the updated $\bse$. Such a comparison
(as such, between any two binary-evolution programs)
is a more complex affair than comparing single-stellar evolution
due to the differences in the treatments of binary-interaction
physics, namely, tidal interaction, mass transfer, CE evolution and
as well in the treatments of star-star and star-remnant merger products.
The difficulty of such a comparison exercise is
augmented by the fact that these binary-interaction processes
are themselves poorly understood \citep{Ivanova_2013}.

At the moment, it is formidable to do a $\bse-\startrack$ comparison
while integrated within $\nbseven$ as done above for $\bse$ alone.
Integration of $\startrack$ within $\nbseven$ is far from
being straightforward partly due to the difference in
programming languages ({\tt C} in $\startrack$ versus {\tt Fortran}
in $\nbseven$ and $\bse$) and partly due to the rather
rigid structure of $\nbseven$'s subroutines that connect
$\bse$ to its N-body integration engine. A foreseeable
solution would be to use a bridging framework such as {\tt AMUSE}
\citep{PortegiesZwart2018} but, currently, neither $\nbseven$
nor $\startrack$ is adequately adapted for use in {\tt AMUSE}.

Therefore, to preliminarily assess the differences in the outcomes of
$\bse$ and $\startrack$ in the context of
remnant formation, which is the focus in this work, 
we evolve 12 selected massive binaries with updated $\bse$ and $\startrack$.
The individual binary parameters and the corresponding remnant
outcomes with $\bse$ and $\startrack$ are summarized in Table~\ref{tab_binevol}.
For both $\bse$ and $\startrack$ evolutions, metallicity $Z=0.0001$,
CE efficiency parameter $\ace=1.0$, F12-rapid+B16-PPSN/PSN remnant
model, and star-star and star-BH merger mass loss fraction of
$\fmrg=0.5$ of the secondary (less massive)
member are applied. The mass transfer is set to be
Eddington limited (Eddington limiting factor $=1.0$).

The main difference between these $\bse$ and $\startrack$ outcomes
is that $\startrack$ leads to more star-star mergers (especially
during CE) than $\bse$, leading to a single BH remnant. As evident
from Table~\ref{tab_binevol}, all but one binaries yield
just one remnant in \startrack (including one
no-remnant outcome due to PSN). In contrast, through $\bse$ evolution,
8 out of the 12 binaries lead to double remnants,
3 of which are BBHs and 2 of the BBHs merge within a short delay time.

The main cause for difference in binary-evolutionary outcomes between
$\bse$ and $\startrack$, in these examples, is due to differences
in the treatment of tidal energy loss among the interacting binary members.
In particular, as described in \citet{Belczynski_2008},
$\startrack$ employs 50 times stronger tidal energy loss
than $\bse$ \citep{Hurley_2002} for convective stellar envelopes and the energy dissipation
is weaker for radiative envelopes due to the use of tidal constants of \citet{Claret_2007}.
These differences cause the initial, post-SN,
and post-CE-ejection detached (and eccentric)
binaries to shrink (and circularize) at different rates, leading to
subsequent binary interactions at different stellar-evolutionary ages.
In particular, the more efficient tidal dissipation for convective envelopes would
lead to faster orbit shrinking (during spin-orbit synchronization),
leading to Roche lobe filling at a younger age when the member
stars have smaller radii, thus leading to mergers in more
of the systems. At the same time, for radiative envelopes,
the reduced tidal efficiency leads to mergers by slowing down the orbital circularization,
enabling Roche lobe overflow near the periastron at a
younger age when the stars have swelled less.

Interestingly, if one considers only the most massive remnant
produced from each binary evolution, they agree reasonably well among 
$\bse$ and $\startrack$ and very closely for 7 out of the 12 binaries. 
This is demonstrated in Fig.~\ref{fig:bevcomp}; see also Table~\ref{tab_binevol}.
The differences between $\startrack$ and $\bse$ remnant masses are
less visible in this way since, for
most of the evolutions, the mass of the heavier among the two remnants or
of the single remnant is determined mainly by the PPSN mass
ceiling and star-star and star-BH merger mass loss
which are taken to be identical for $\bse$ and $\startrack$,
in this exercise.

The difference in treatments of tidal interaction and also in
implementations and numerical approaches would lead to differences
in outcomes of low-mass interacting binaries as well.
Such a study is beyond the present scope; see \citet{Toonen_2014}
in this context.

\section{Summary and outlook}\label{summary}

The work presented above is summarized below where the next
steps are also motivated.

\begin{figure*}
\centering
\includegraphics[width=5.86cm,angle=0]{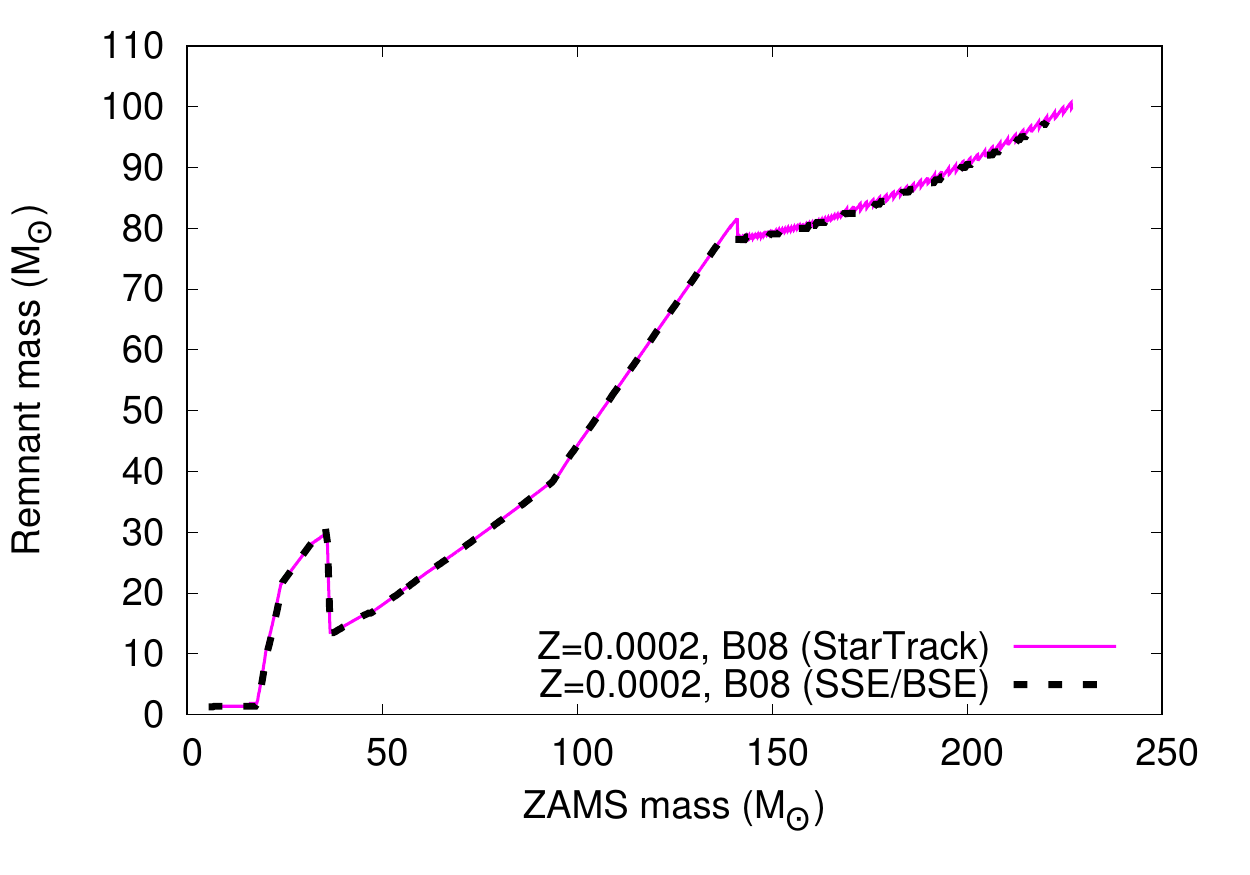}
\includegraphics[width=5.86cm,angle=0]{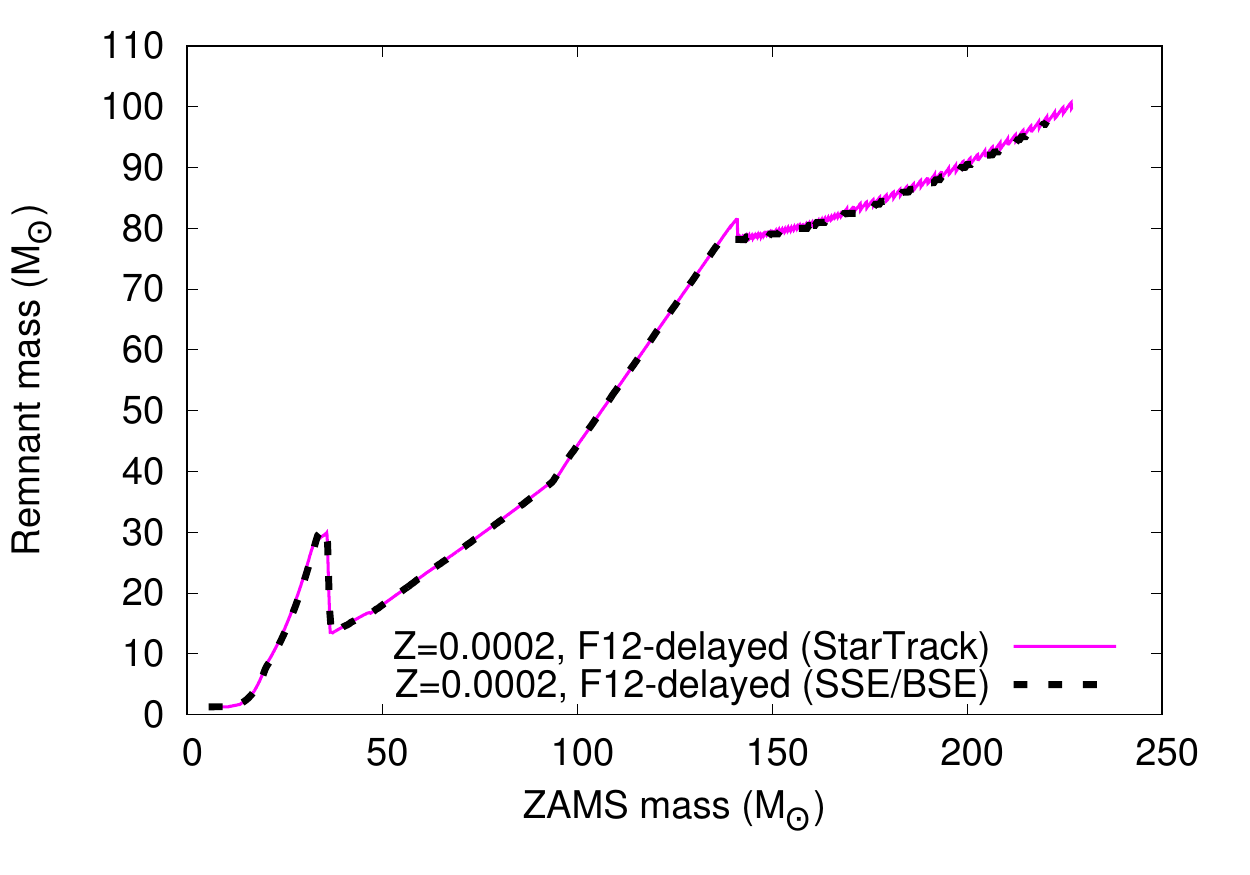}
\includegraphics[width=5.86cm,angle=0]{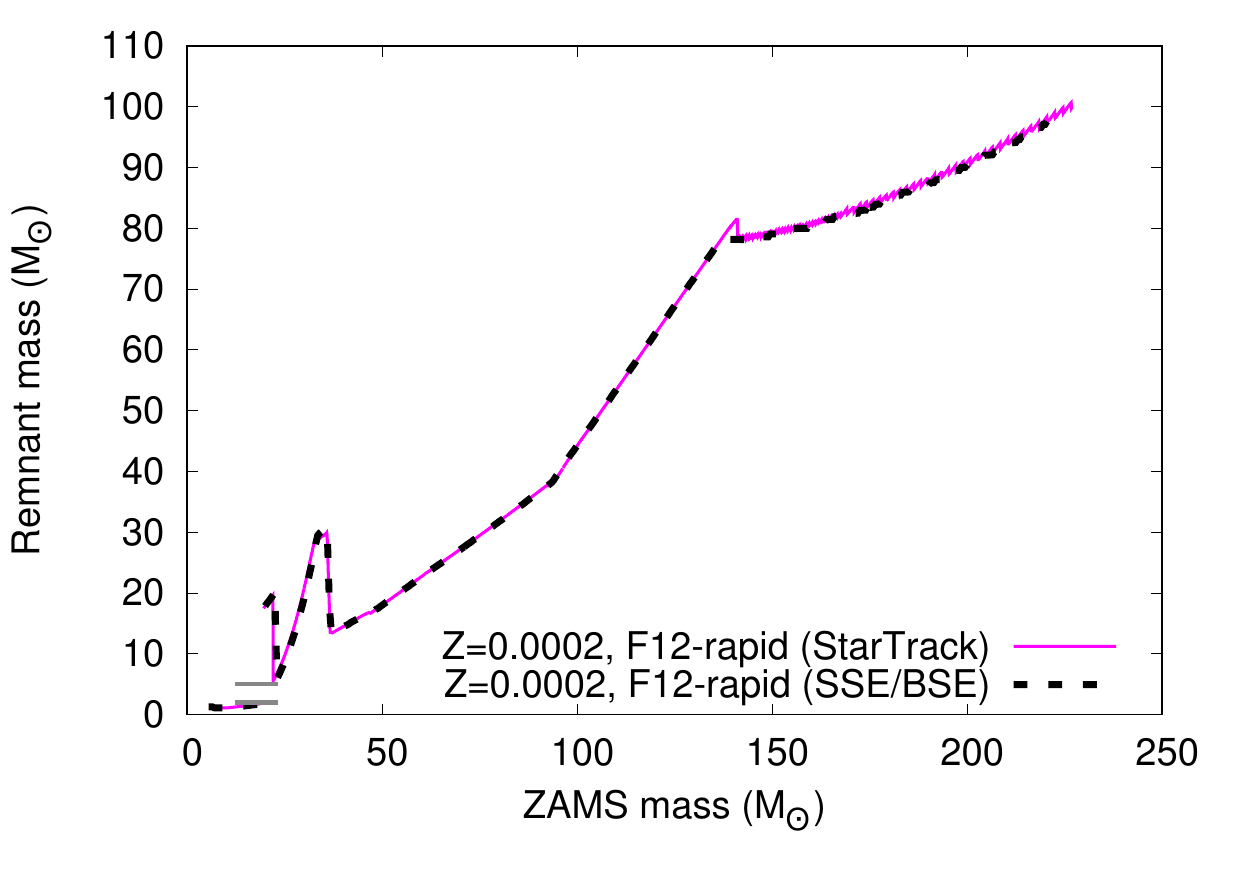}\\
\includegraphics[width=5.86cm,angle=0]{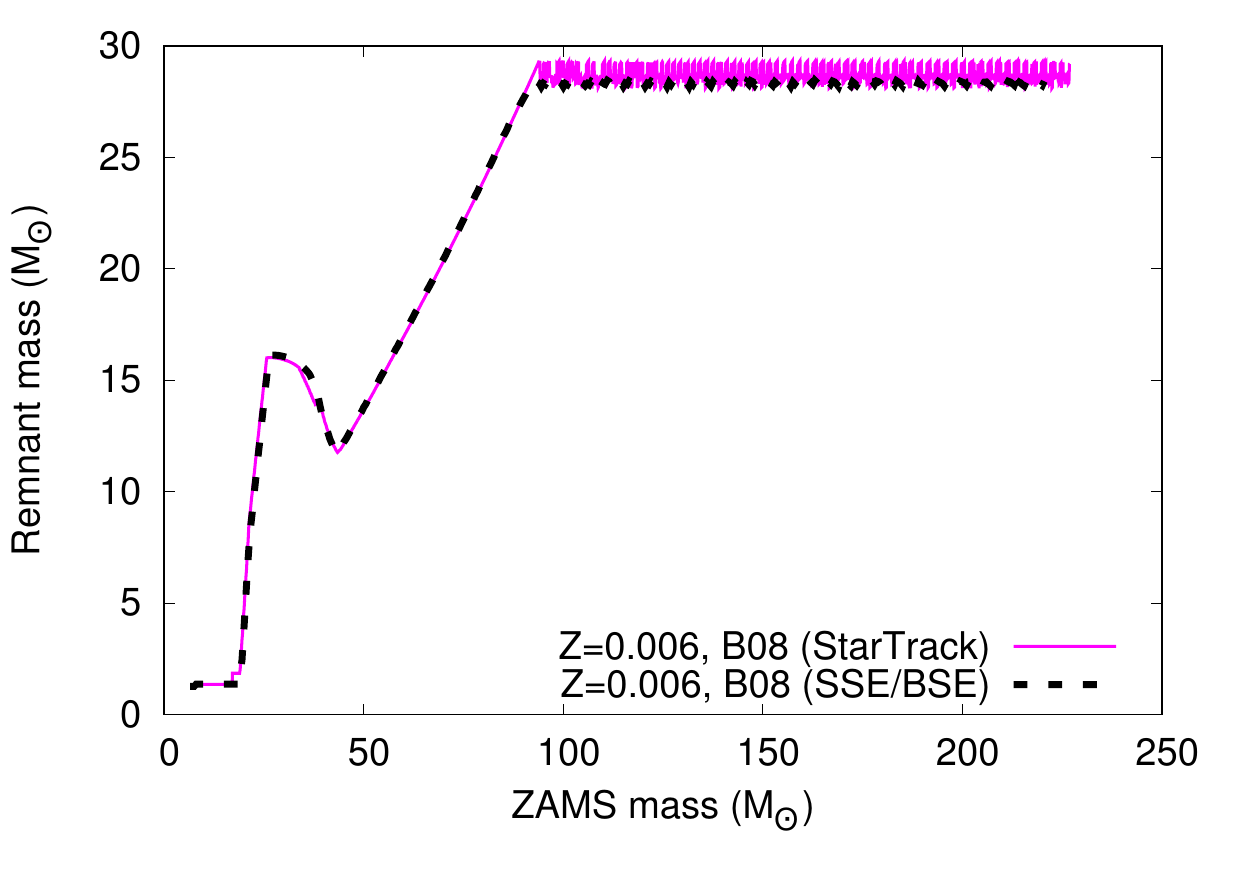}
\includegraphics[width=5.86cm,angle=0]{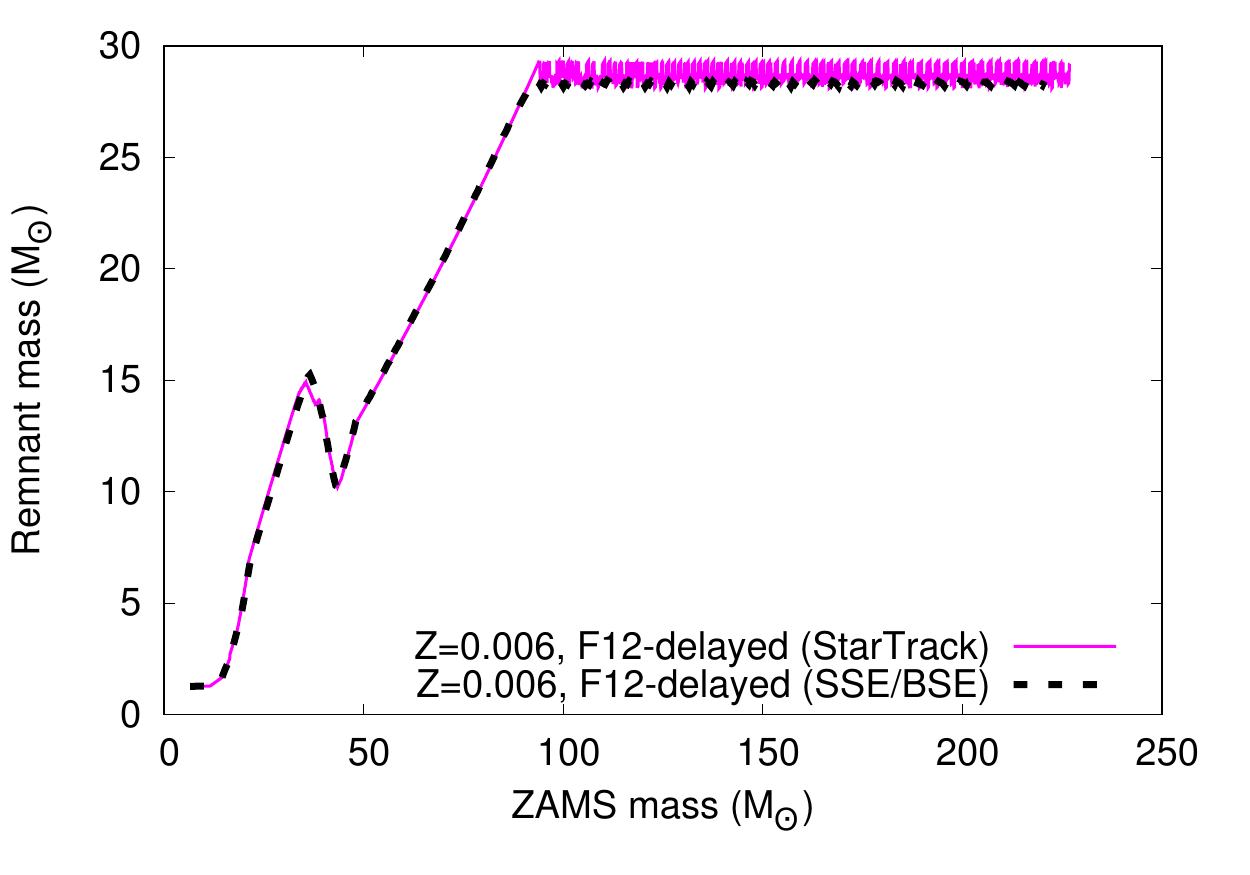}
\includegraphics[width=5.86cm,angle=0]{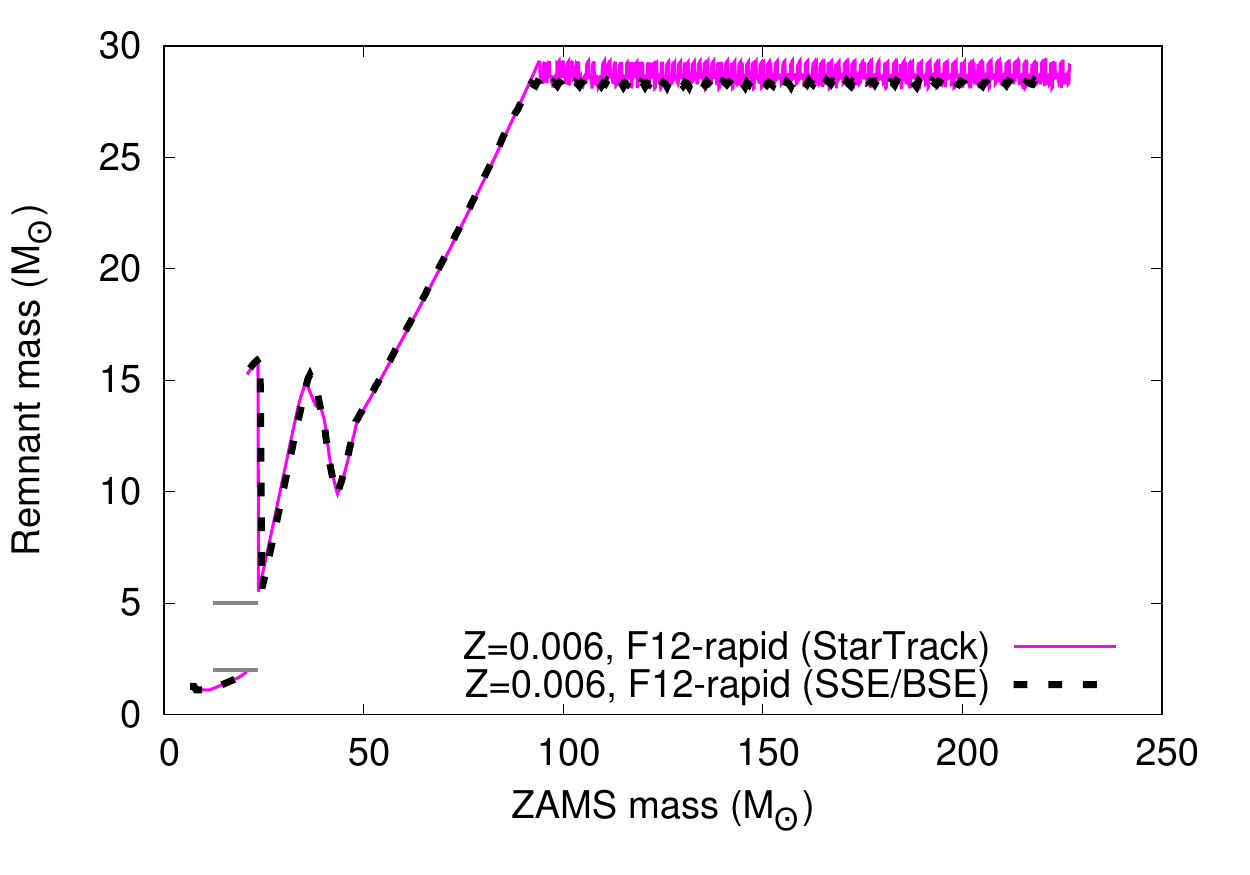}\\
\includegraphics[width=5.86cm,angle=0]{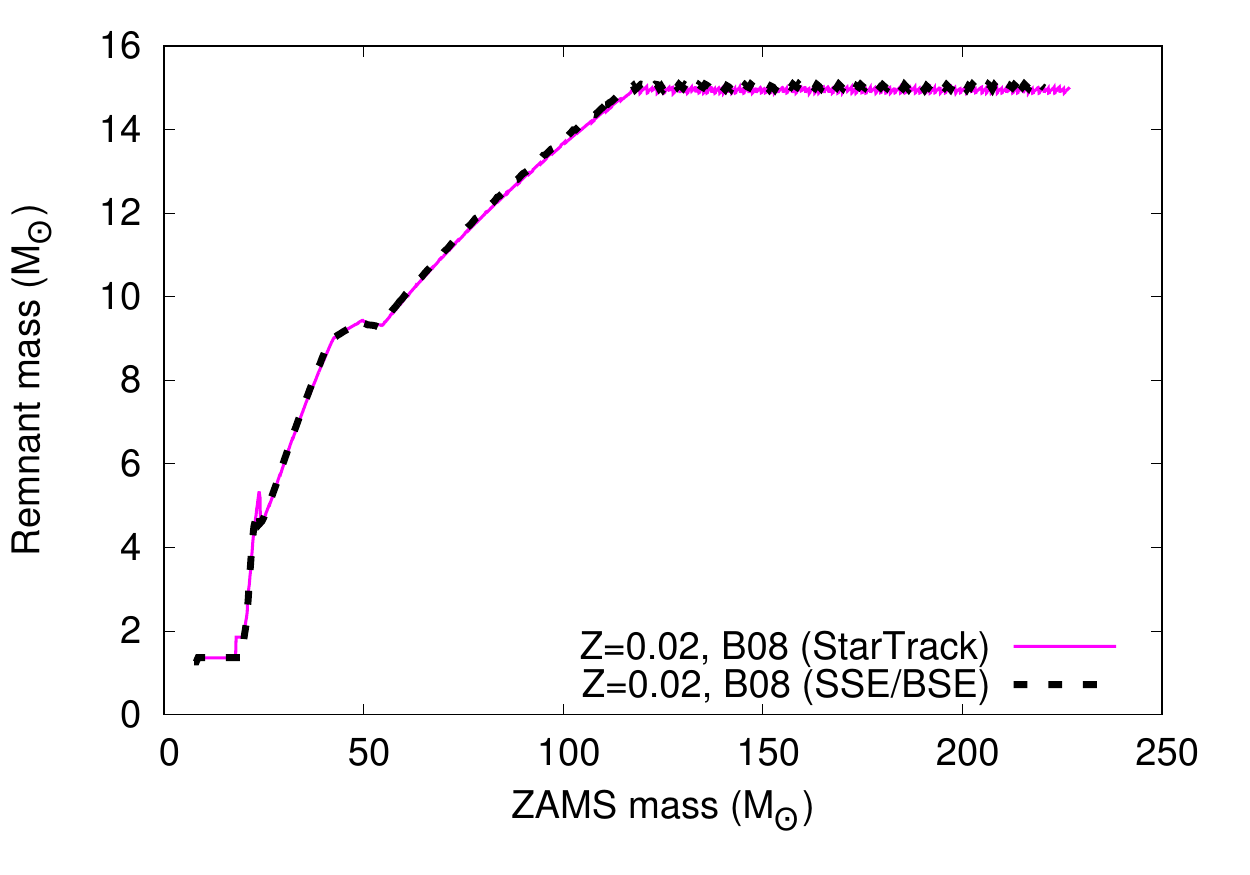}
\includegraphics[width=5.86cm,angle=0]{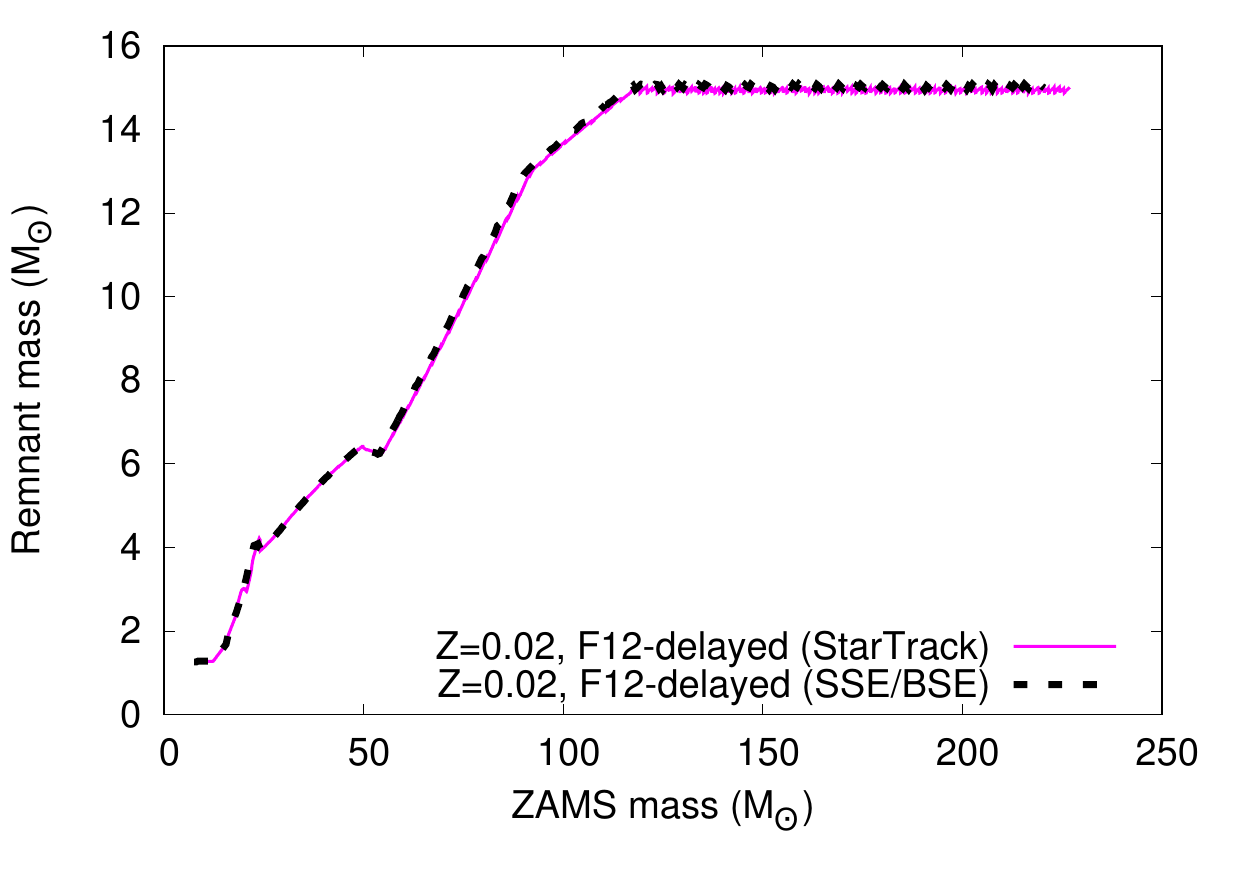}
\includegraphics[width=5.86cm,angle=0]{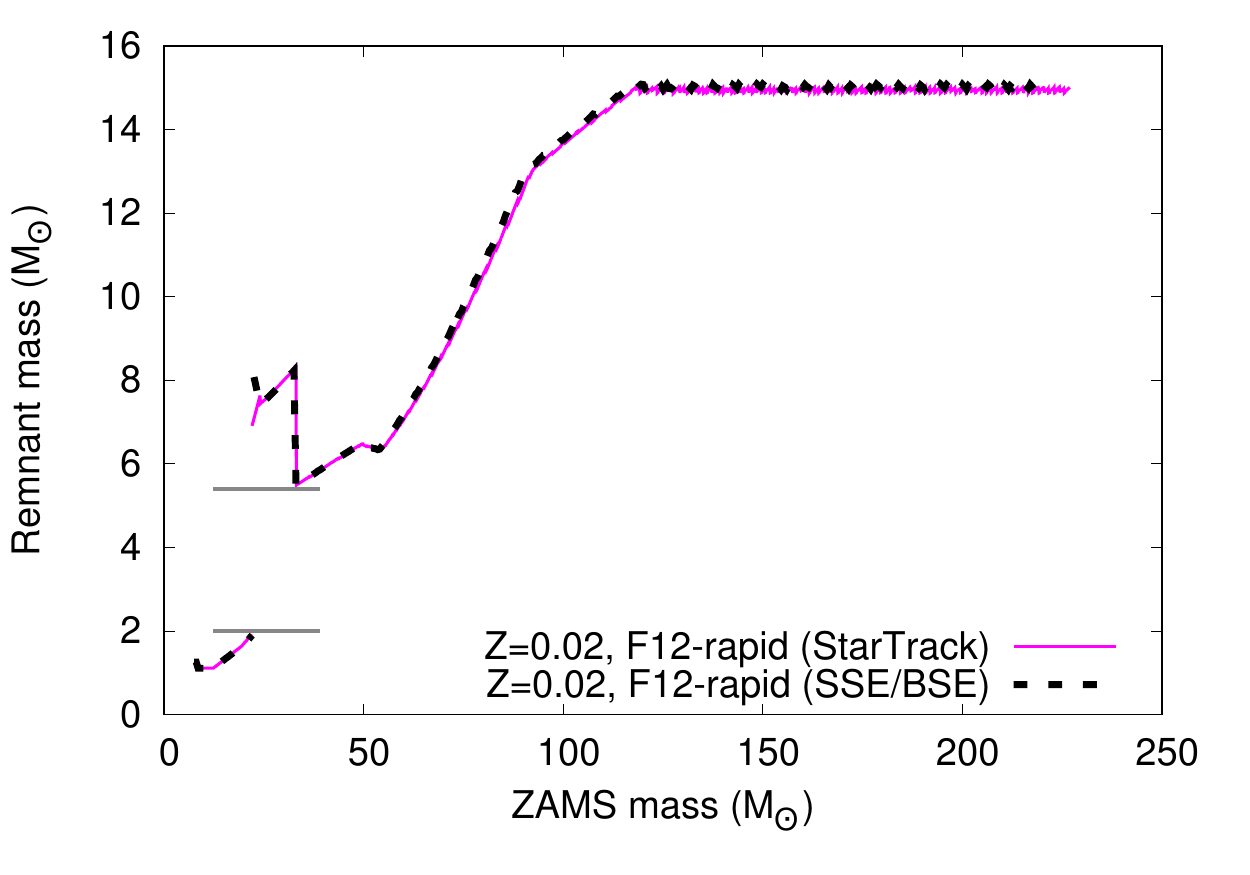}
\caption{Comparisons of the ZAMS mass-remnant mass relation between the updated \bse
(Secs.~\ref{newwind} \& \ref{newrem}) and \startrack. The comparison is done
for the cases of B08, F12-delayed, and F12-rapid remnant-mass models and for
the metallicities $Z=0.0002$, 0.006, and 0.02, as indicated in the legends.
In all cases in this figure, PPSN/PSN is disabled. 
In all panels, the black, dashed line is the outcome from the new \bse and the
blue, solid line is the corresponding \startrack result.
The mass gap
between NSs and BHs, that is characteristic of the F12-rapid remnant mass model,
is also indicated in the corresponding panels (the grey, horizontal lines at $\approx2.0\Ms$
and $\approx5.0\Ms$ in the F12-rapid panels).}
\label{fig:cmp1}
\end{figure*}

\begin{figure*}
\centering
\includegraphics[width=9.0cm]{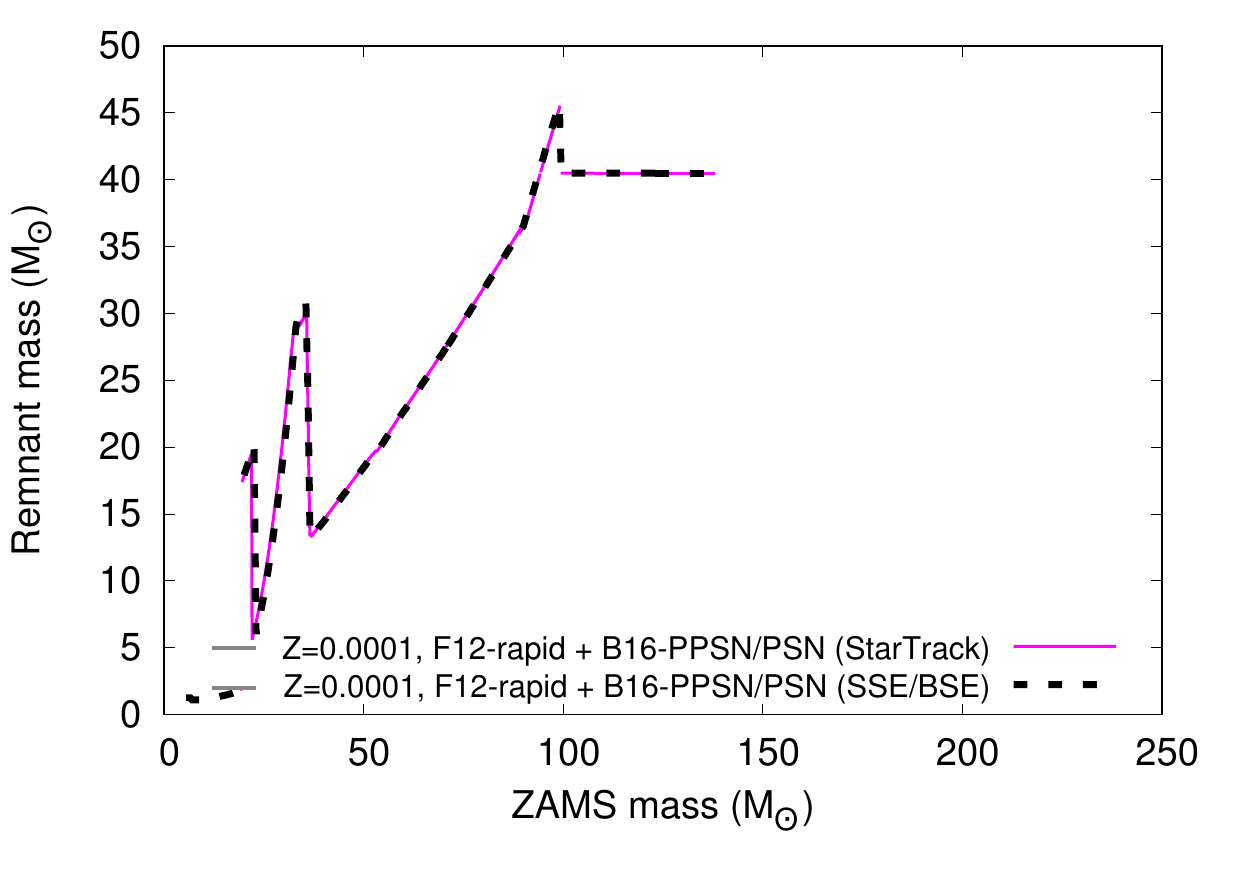}
\includegraphics[width=9.0cm]{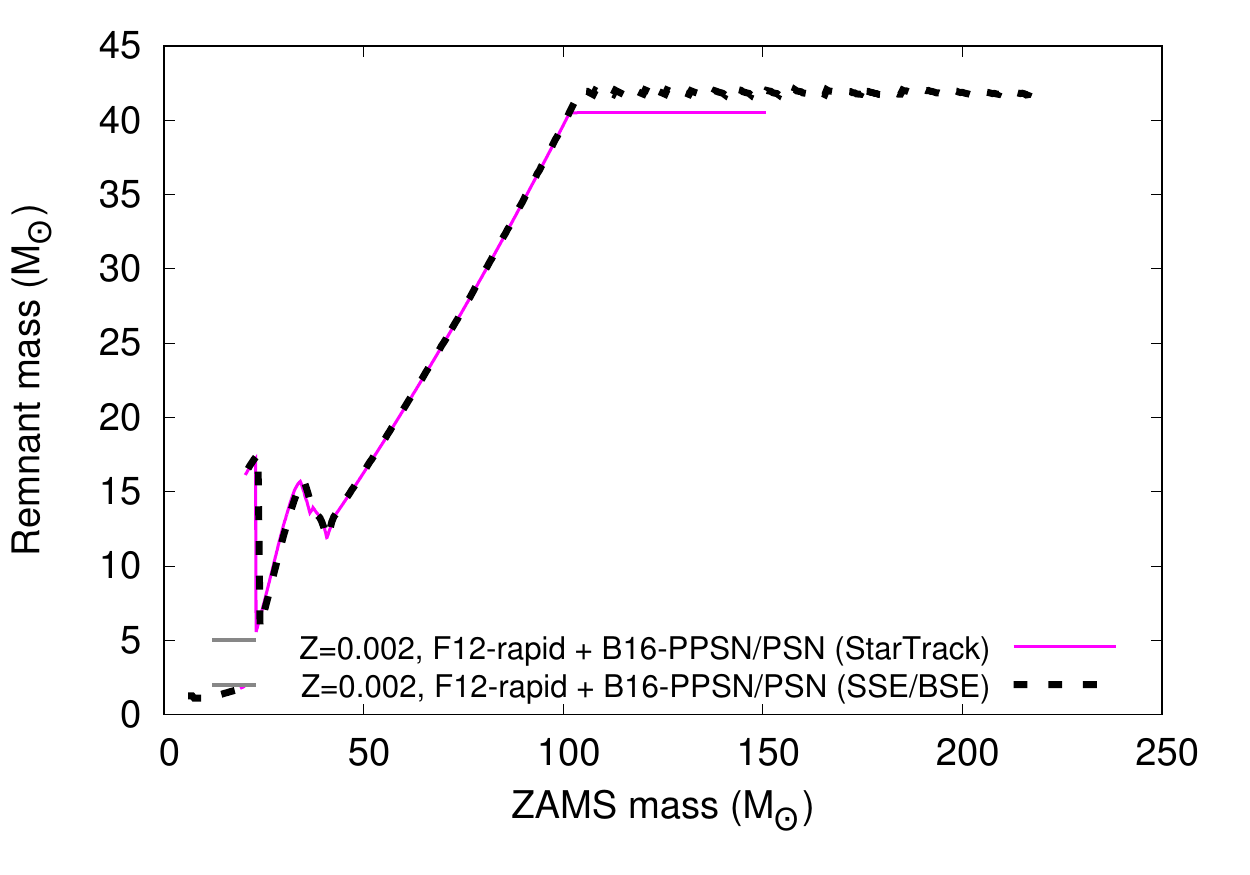}\\
\includegraphics[width=18.0cm]{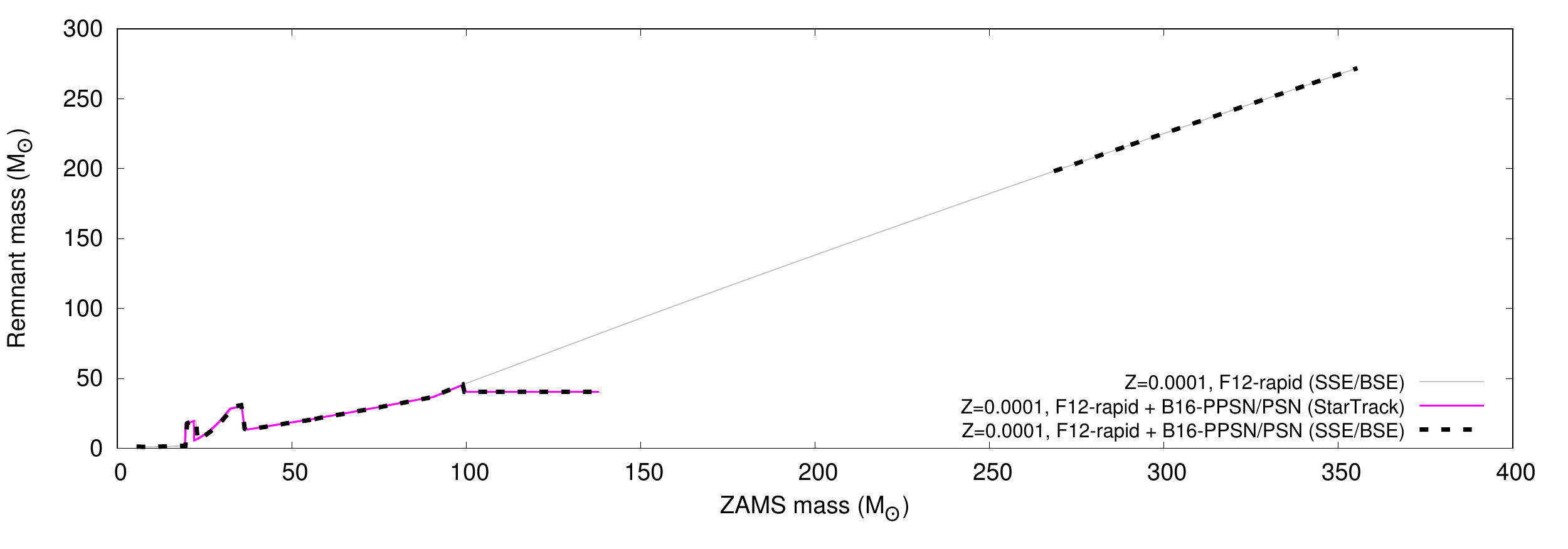}\\
\includegraphics[width=18.0cm]{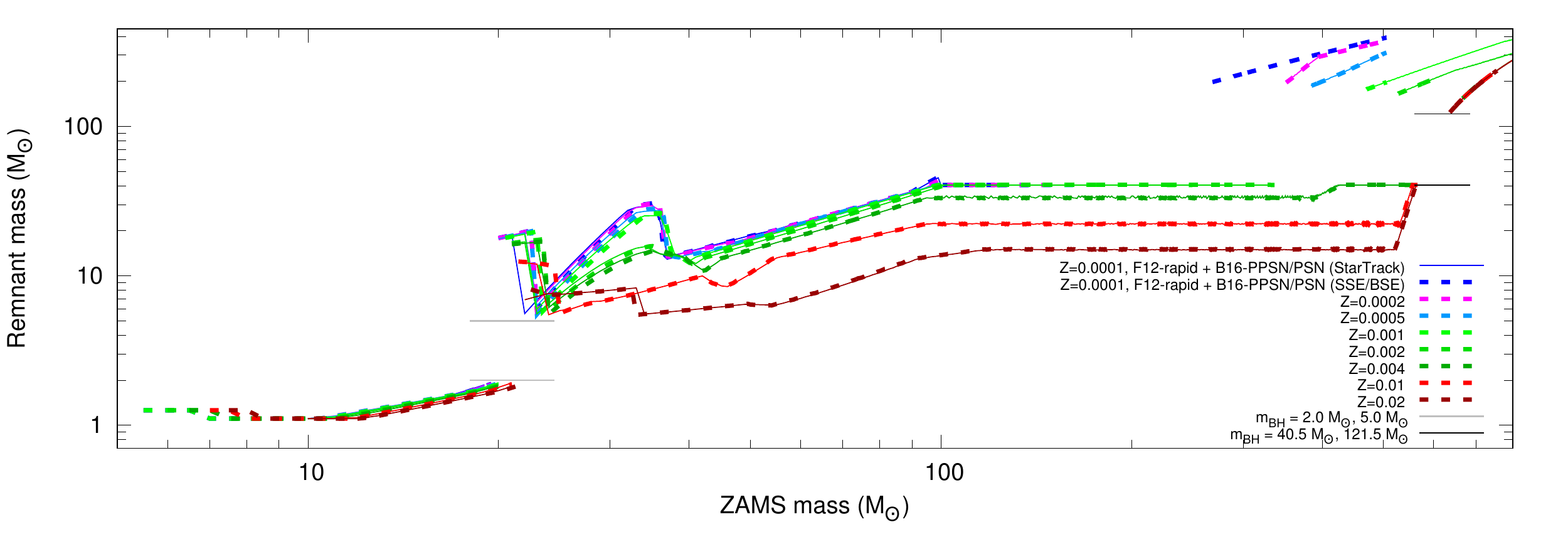}
\caption{Comparisons of the ZAMS mass-remnant mass relation between the updated \bse
and \startrack including PPSN/PSN (F12-rapid SN model) for $Z=0.0001$ and 0.002.
The line legends are the same as in Fig.~\ref{fig:cmp1}. The middle panel demonstrates
the PSN ``mass gap'', as obtained from the current \bse for $Z=0.0001$, by extending
the ZAMS mass beyond $300\Ms$. The lower panel compares between the ZAMS mass-remnant mass relations
as obtained from the updated \bse (thick, dashed lines) and \startrack
(thin, solid lines) up to very large ZAMS masses, extending
beyond the PSN ``mass gap'' (the black, horizontal lines at $40.5\Ms$ and
$121.5\Ms$), for a wide range of metallicities. The lower mass gap
between NSs and BHs, that is characteristic of the F12-rapid remnant mass model,
is also indicated in this panel (the grey, horizontal lines at $\approx2.0\Ms$
and $\approx5.0\Ms$).} 
\label{fig:cmp_psn}
\end{figure*}

\begin{figure*}
\centering
\includegraphics[width=5.86cm,angle=0]{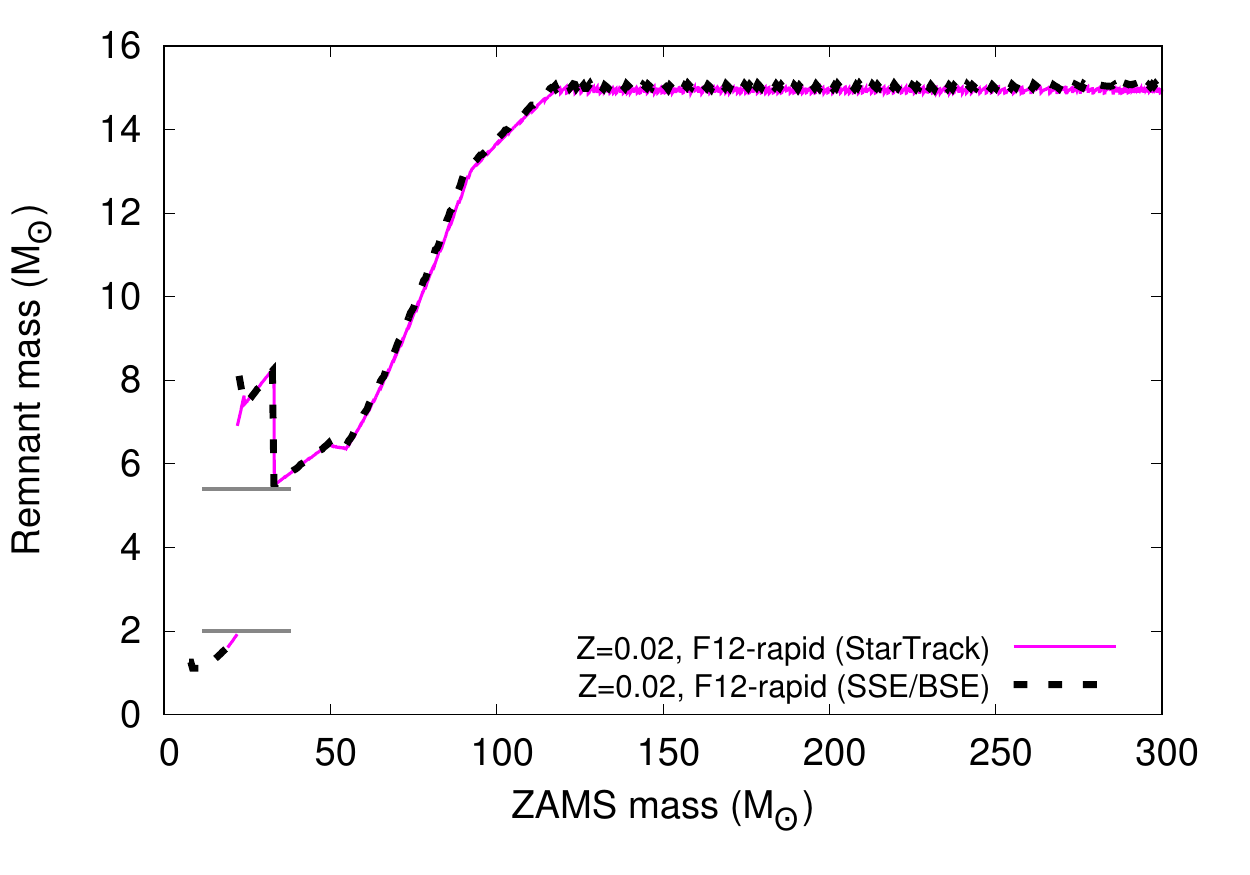}
\includegraphics[width=5.86cm,angle=0]{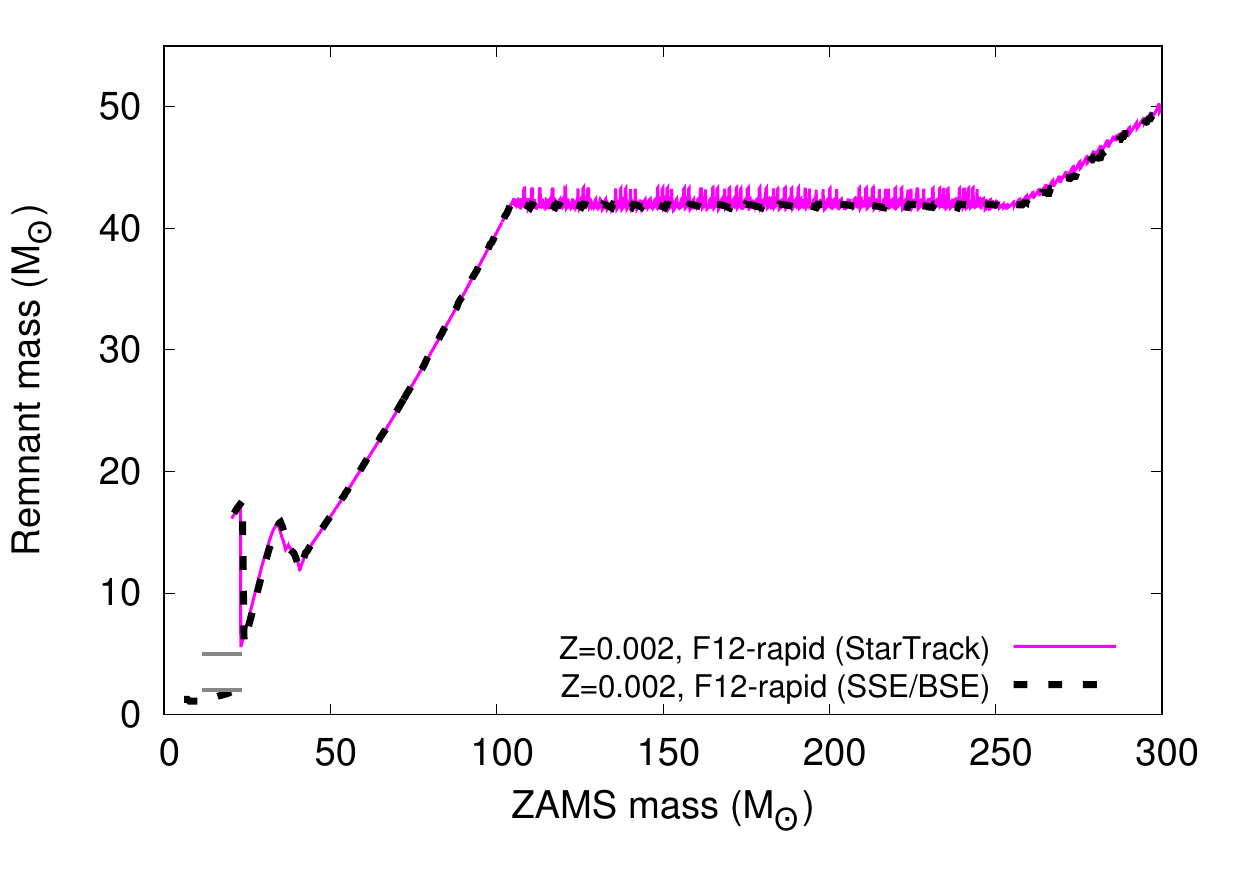}
\includegraphics[width=5.86cm,angle=0]{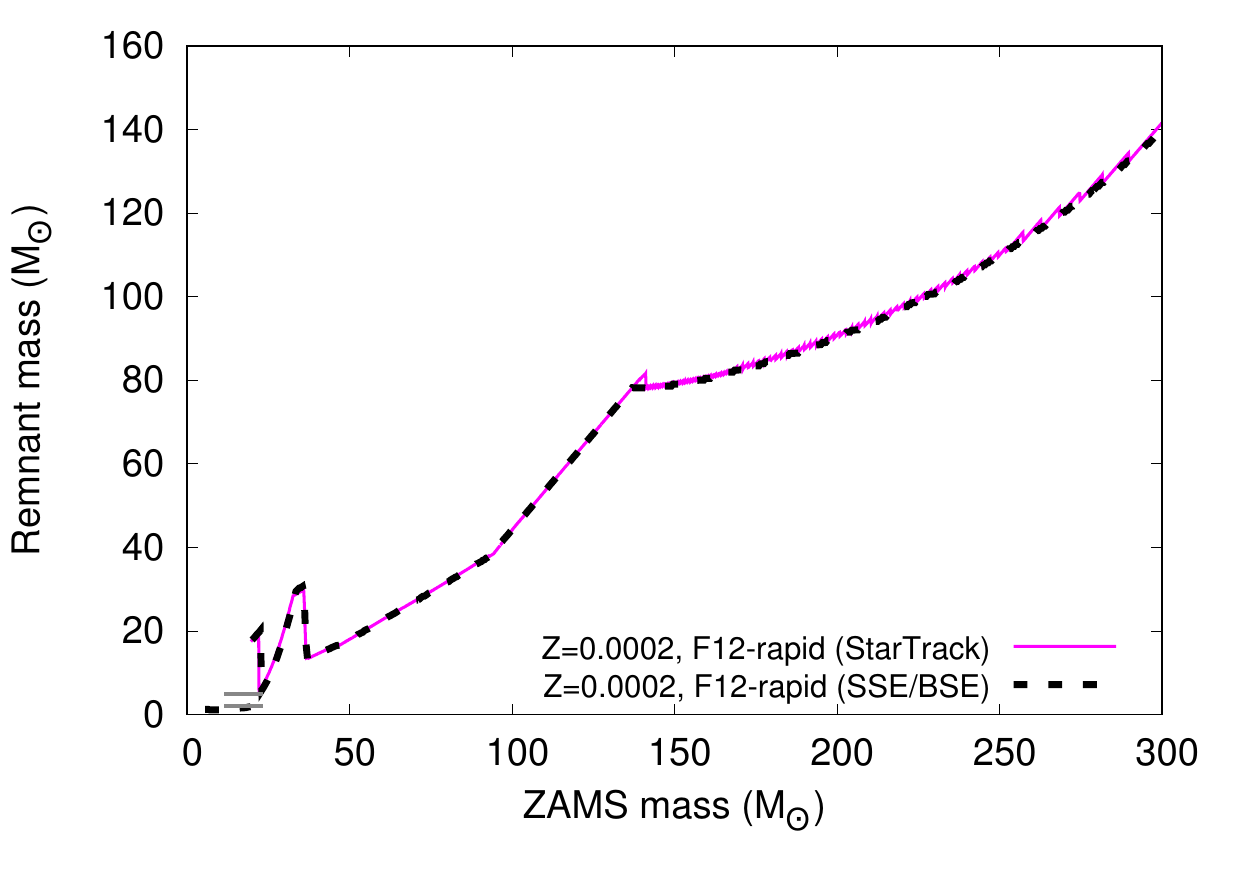}\\
\includegraphics[width=5.86cm,angle=0]{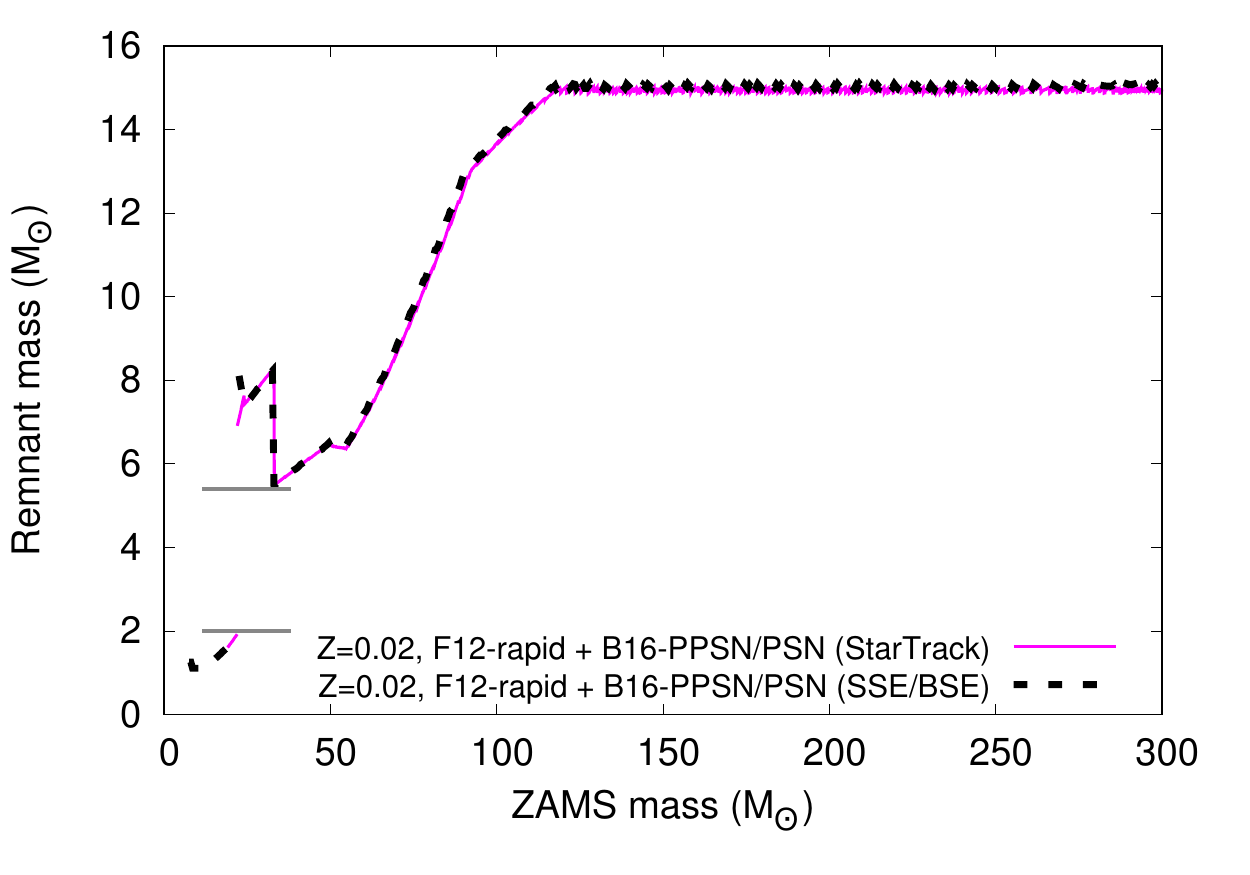}
\includegraphics[width=5.86cm,angle=0]{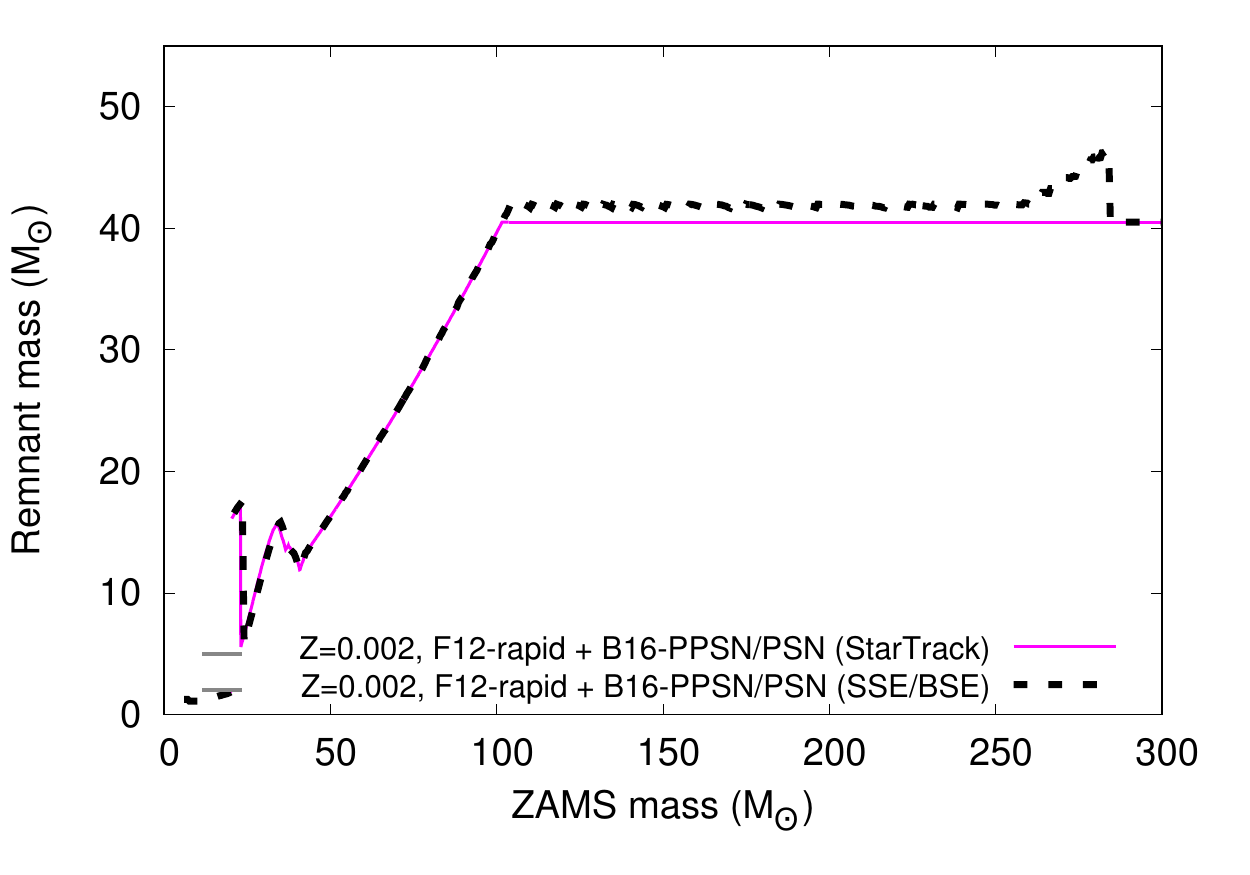}
\includegraphics[width=5.86cm,angle=0]{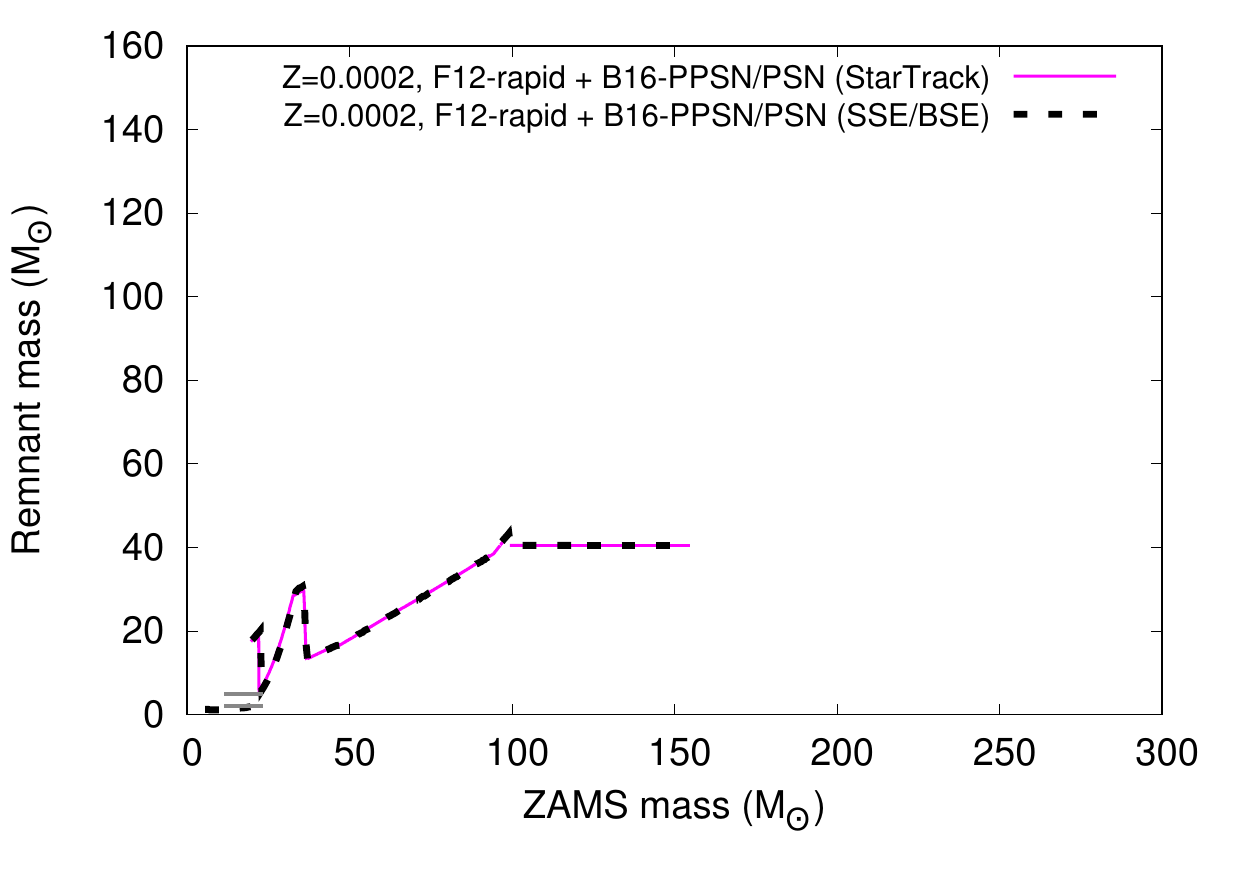}
\caption{Comparisons of the ZAMS mass-remnant mass relation between the updated \bse
and \startrack up to $\approx300\Ms$ ZAMS mass, excluding (top panels)
and including (bottom panels) PPSN/PSN (F12-rapid SN model)
for $Z=0.0002$, 0.002, 0.02. The line legends are the same as in Fig.~\ref{fig:cmp1}.
Note the significant ``clipping'' of BH mass (at $\approx40\Ms$) for
ZAMS masses $\gtrsim100\Ms$ at very
low metallicities such as $Z=0.0002$, when
PPSN/PSN is included (see also B16 and Sec.~\ref{newrem}).}
\label{fig:cmp_ext}
\end{figure*}

\begin{figure*}
\centering
\includegraphics[width=9.0cm]{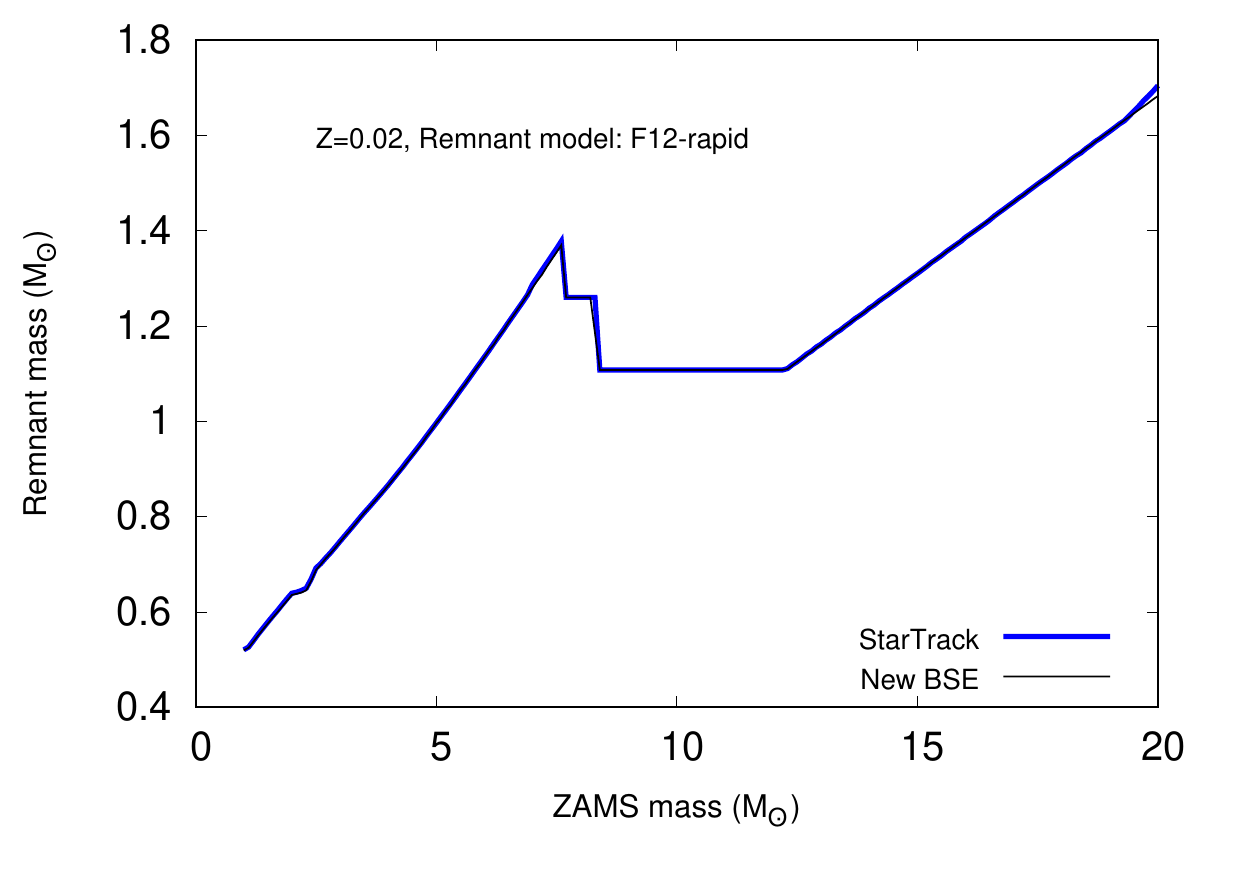}
\includegraphics[width=9.0cm]{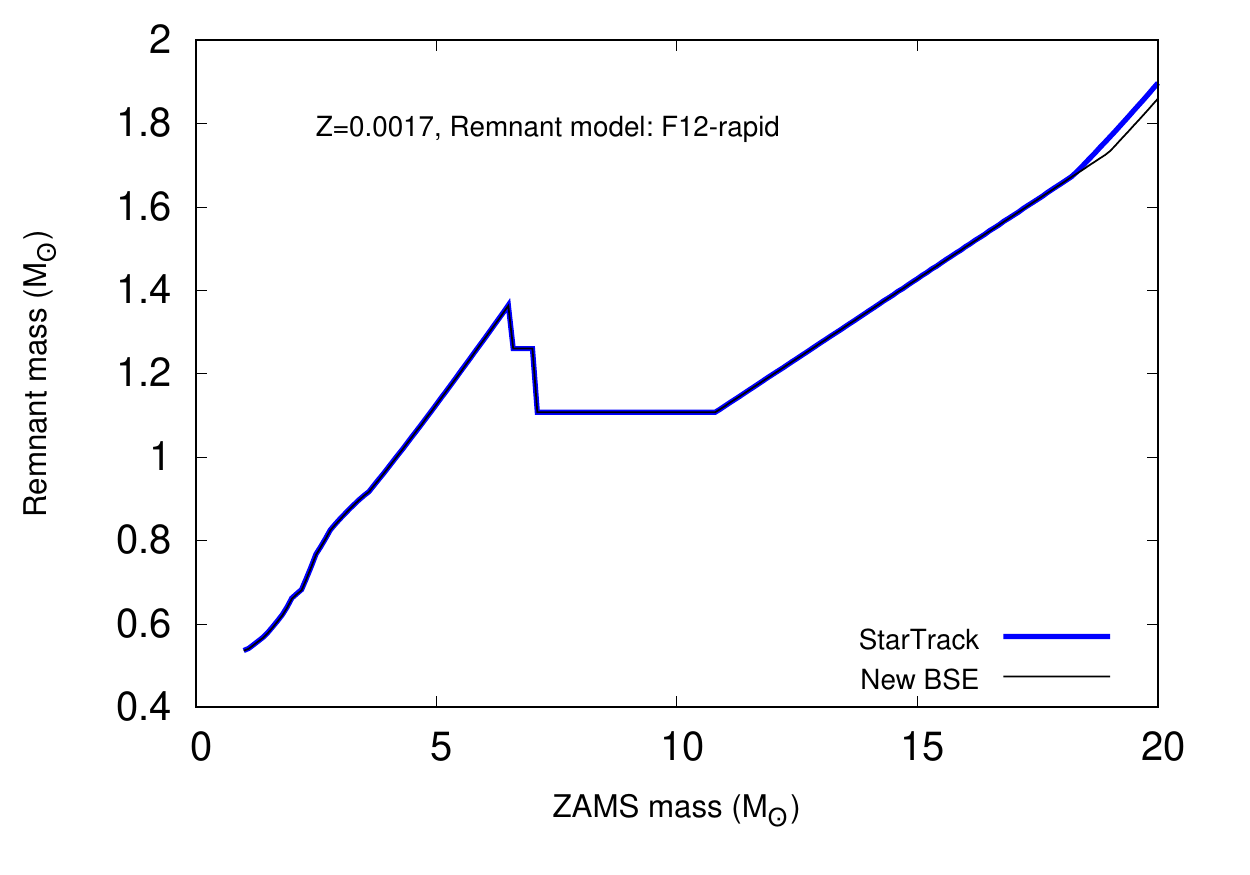}
\caption{Comparisons of the ZAMS mass-remnant mass relation between the updated \bse (thin, black line)
and \startrack (thick, blue line) for lower ZAMS masses ($1\Ms-20\Ms$) for F12-rapid remnant mass model.
Over this mass range, the switching from WD formation to NS formation is evident (NS formation
from the first drop of the curve to $1.26\Ms$). The comparison
is shown for the two metallicities $Z=0.02$ and $0.0017$.}
\label{fig:cmp_wd}
\end{figure*}

\begin{figure*}
\centering
\includegraphics[width=16.0cm]{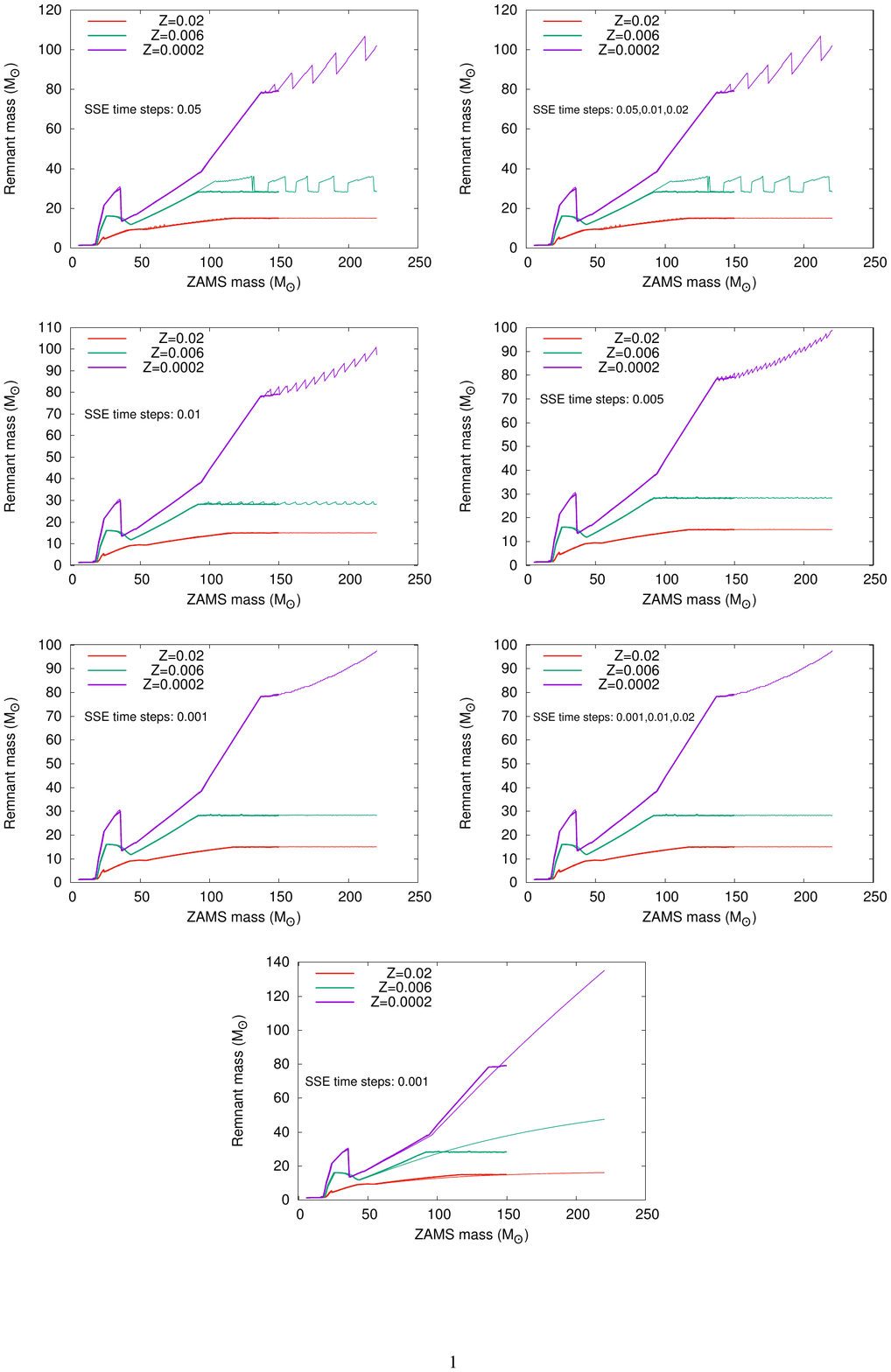}
\caption{{\bf Panels 1-6:} Comparisons of the ZAMS mass-remnant mass relation between the updated \bse (thin lines)
and \startrack (thick lines) for the B08 remnant-formation scheme and for $Z=0.0002$, $0.006$, $0.02$, for different
choices of the stellar-evolutionary time step parameters $\ptsone$, $\ptstwo$, and $\ptsthree$. The
values of these time-step parameters are indicated in the corresponding panels; only one value
is indicated when $\ptsone=\ptstwo=\ptsthree$, otherwise the respective values are indicated.
{\bf Panel 7:} The same comparison with $\ptsone=\ptstwo=\ptsthree=0.001$
when the default, to-date-public version of $\hrdiag/\mlwind$ is
utilized (see Sec.~\ref{tstep} for the details). The \startrack data in these panels are from B10.}
\label{fig:tstep}
\end{figure*}

\begin{figure*}
\centering
\includegraphics[width=8.0cm]{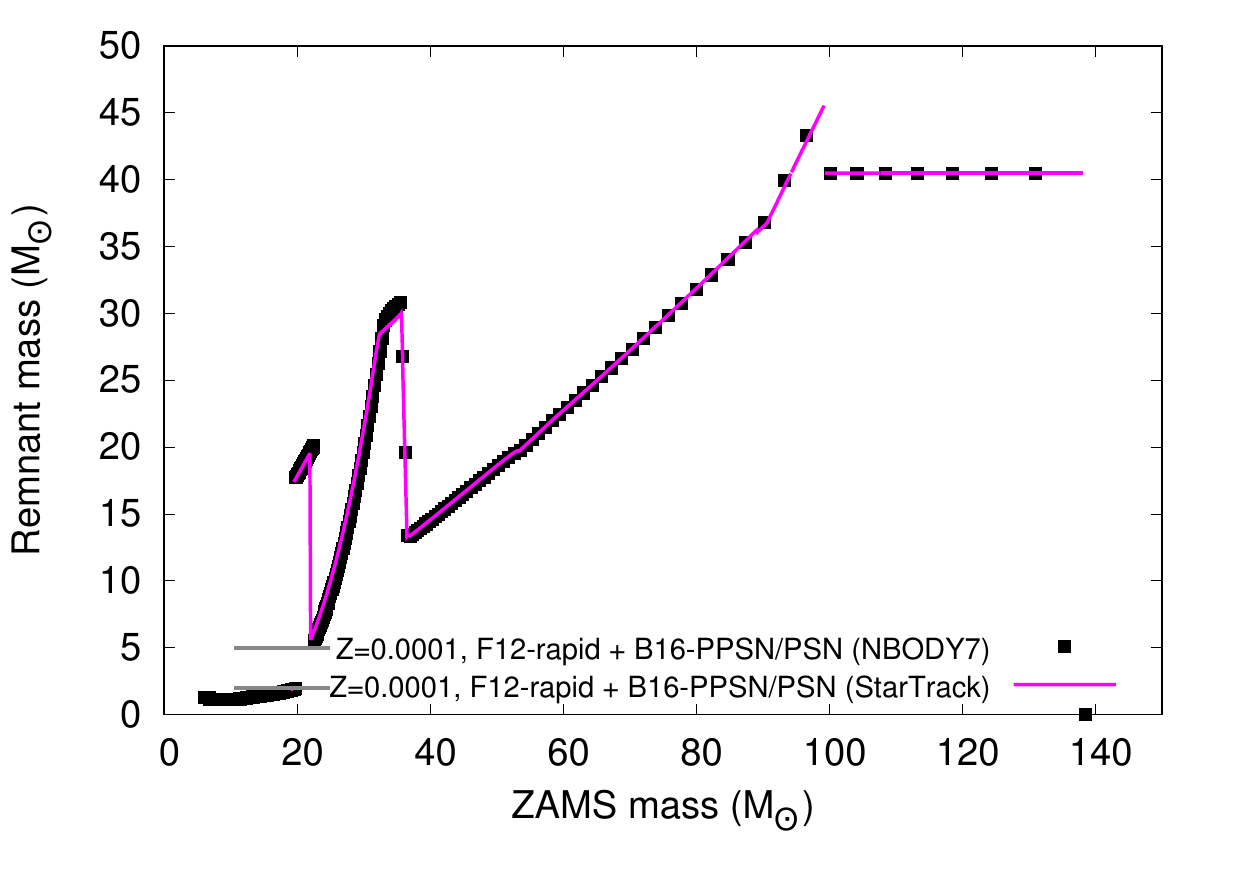}
\includegraphics[width=8.0cm]{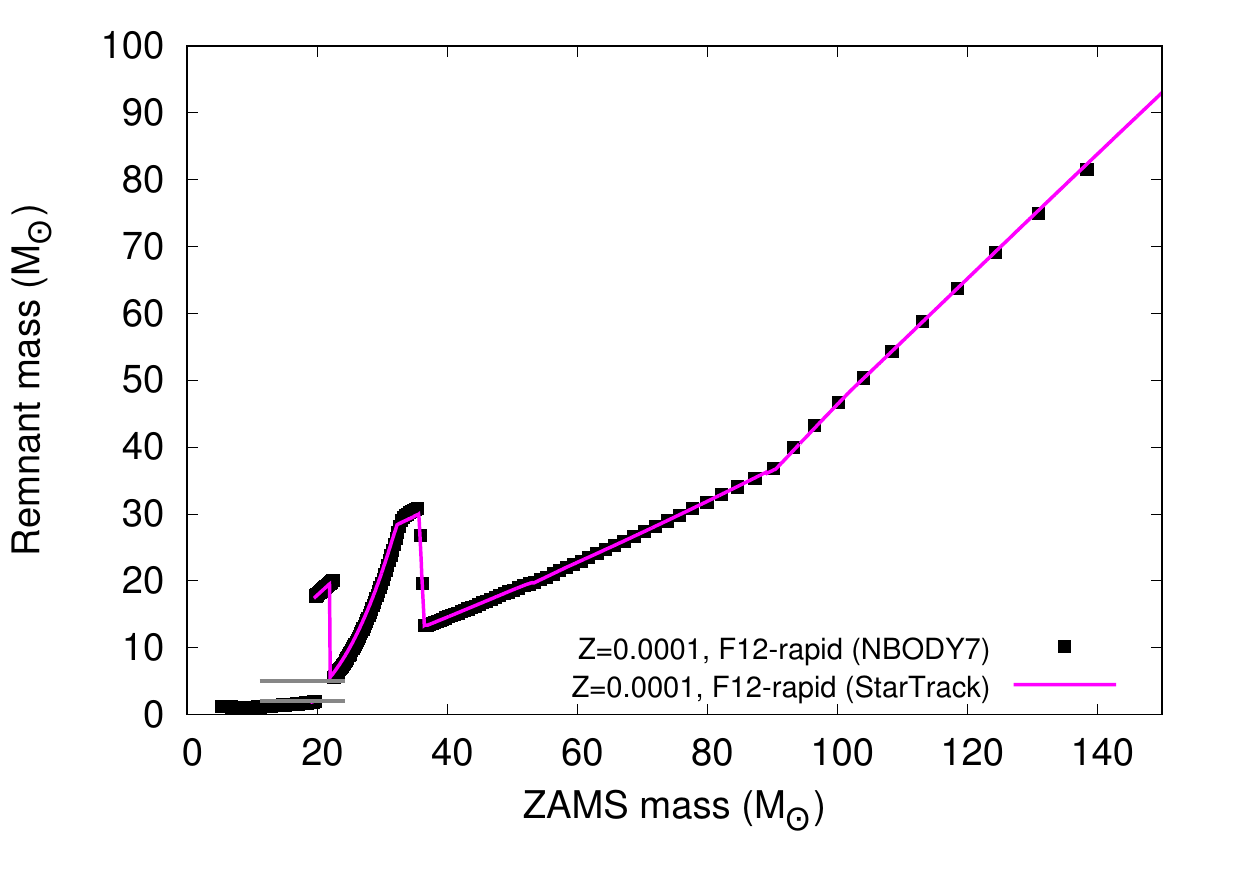}\\
\includegraphics[width=8.0cm]{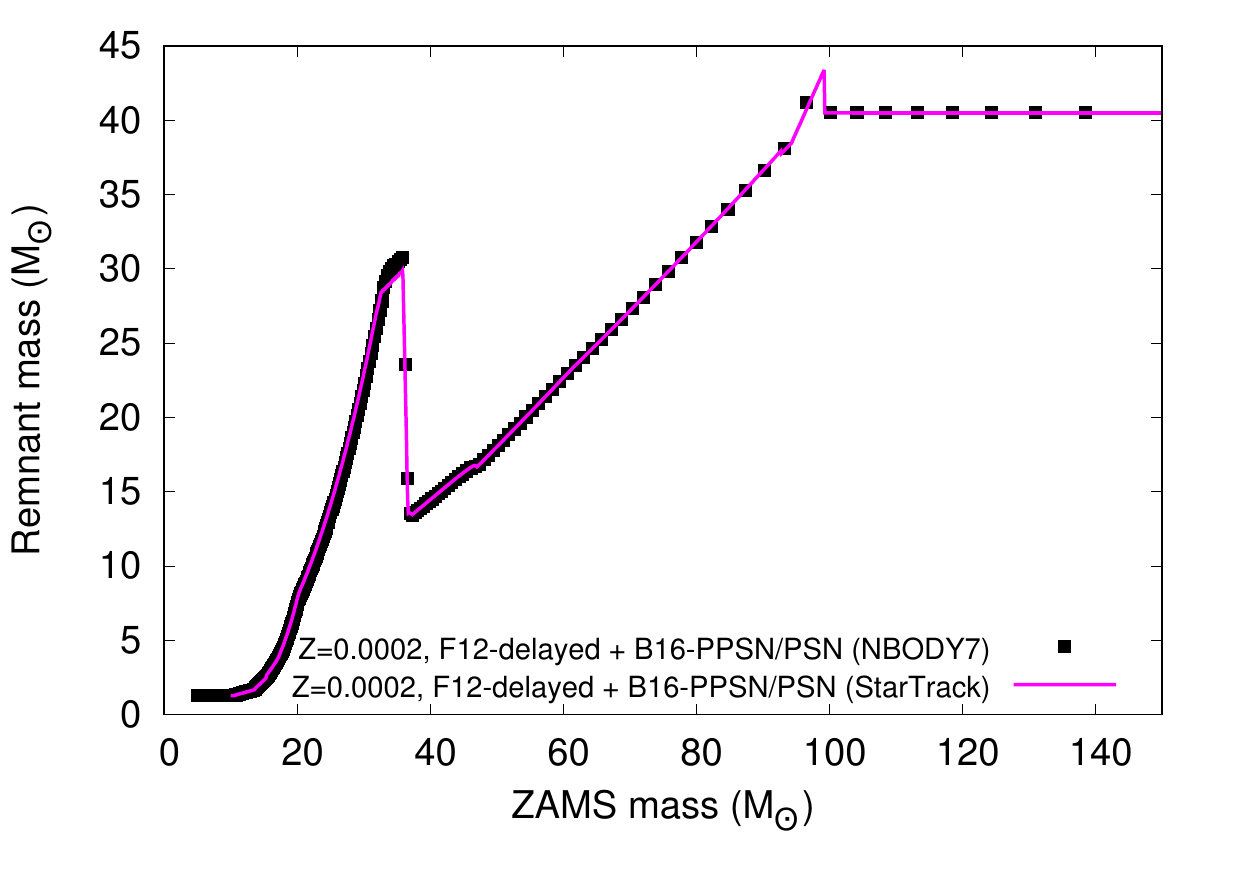}
\includegraphics[width=8.0cm]{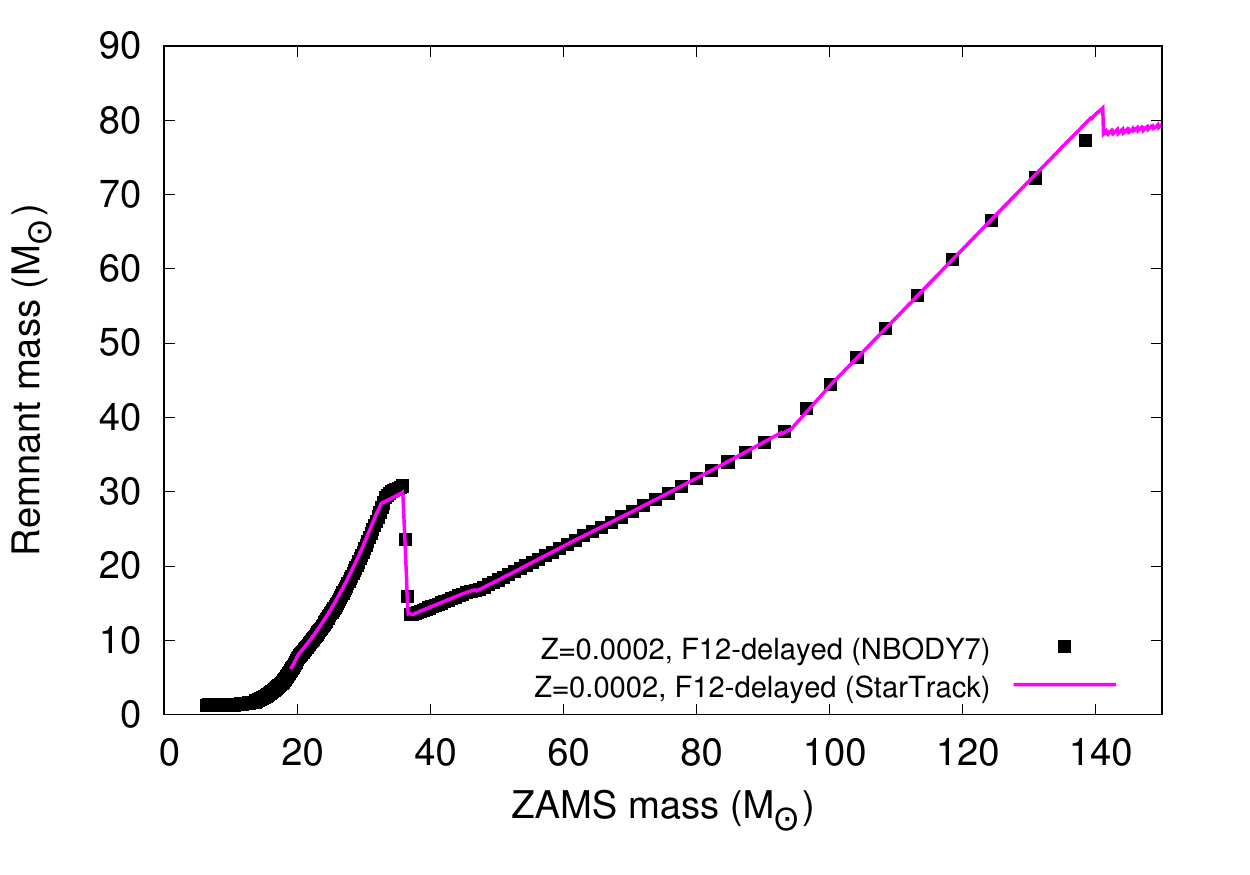}\\
\includegraphics[width=8.0cm]{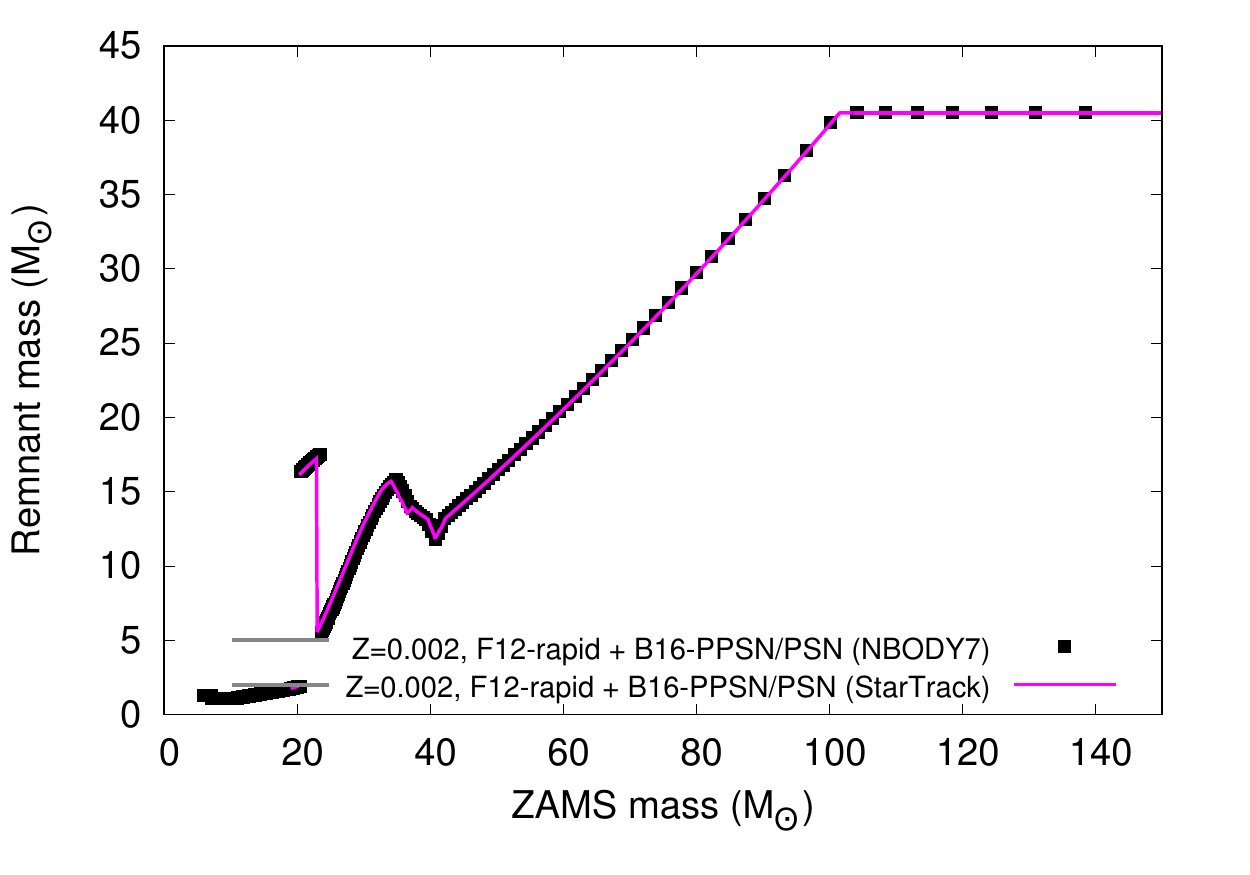}
\includegraphics[width=8.0cm]{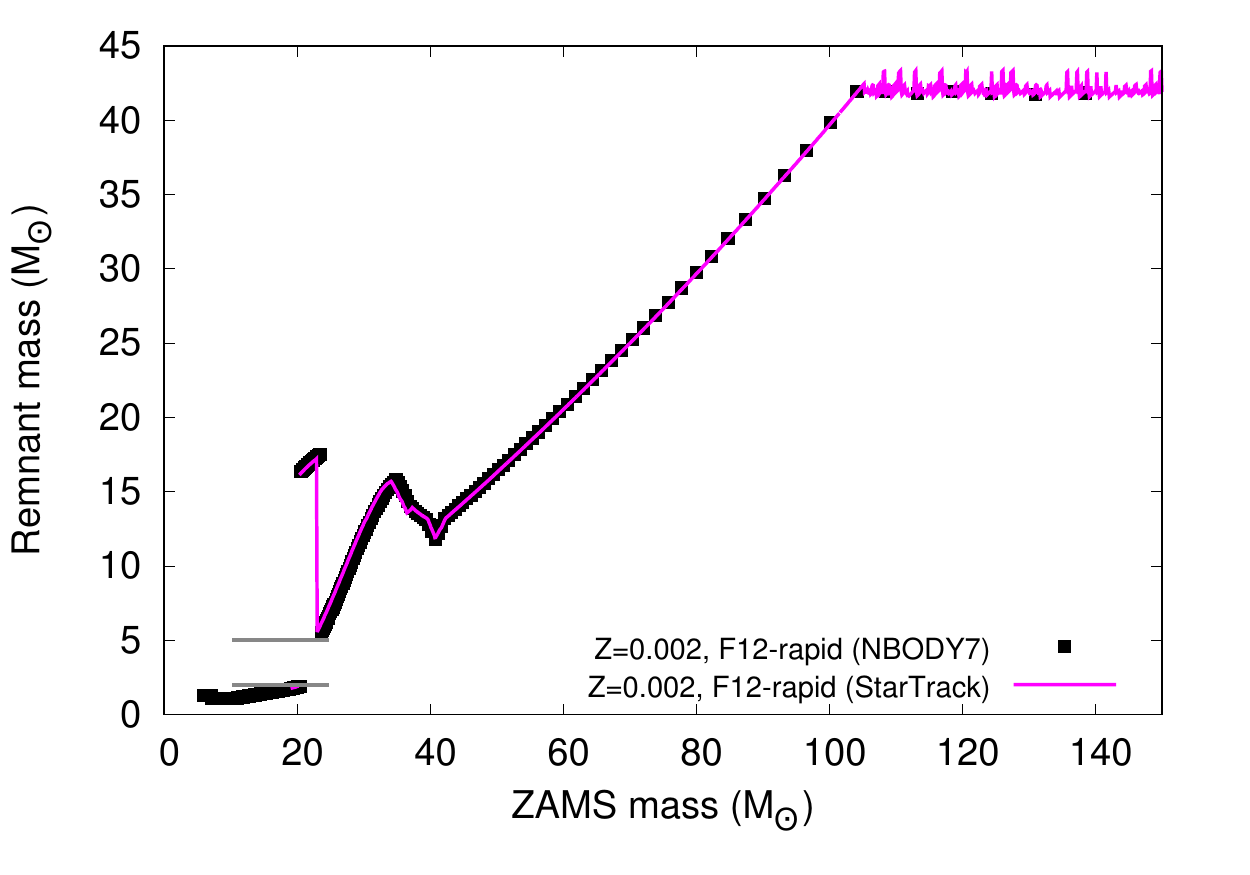}\\
\includegraphics[width=8.0cm]{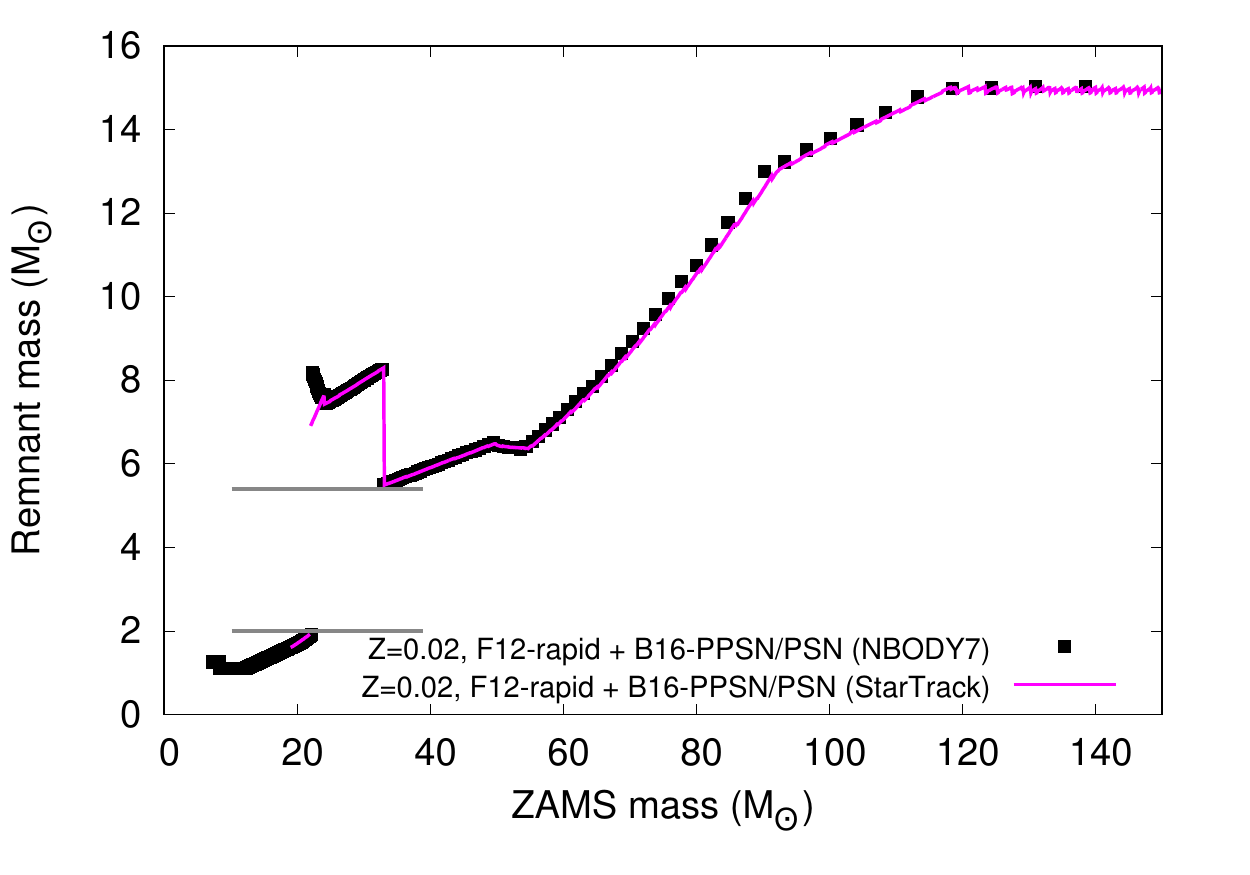}
\includegraphics[width=8.0cm]{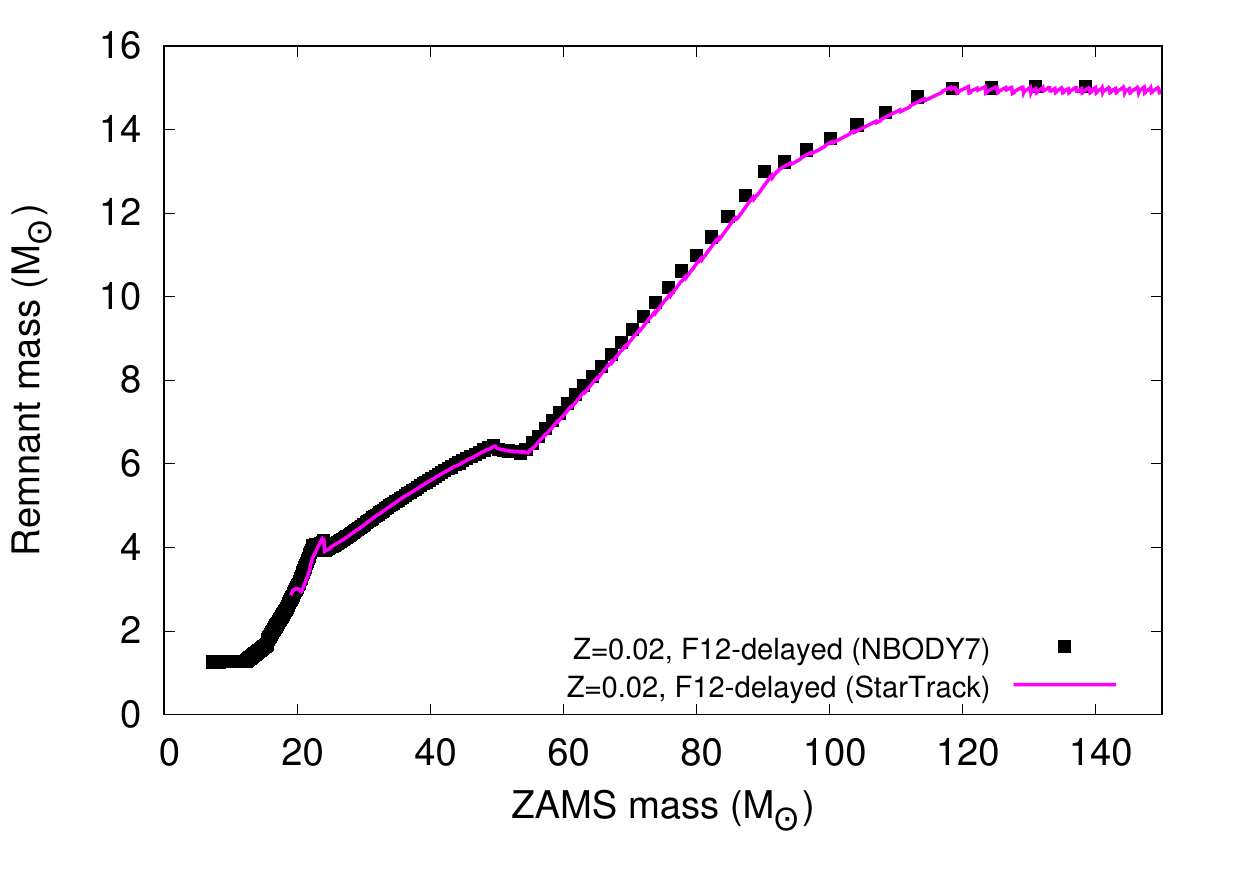}
\caption{Comparisons of the ZAMS mass-remnant mass relation between the updated \bse
from within \nbseven and \startrack, for different $Z$ and remnant-formation
prescriptions. Note that for $Z=0.002$ and $Z=0.02$, PPSN/PSN do not affect the ZAMS mass-remnant mass
relation since the He-core mass never reaches the PPSN/PSN threshold (see B16).
The mass gap between the NSs and BHs for the F12-rapid cases is indicated
with the grey horizontal lines, as in the previous figures.}
\label{fig:cmp_nb}
\end{figure*}

\begin{figure*}
\centering
\includegraphics[width=18.0cm]{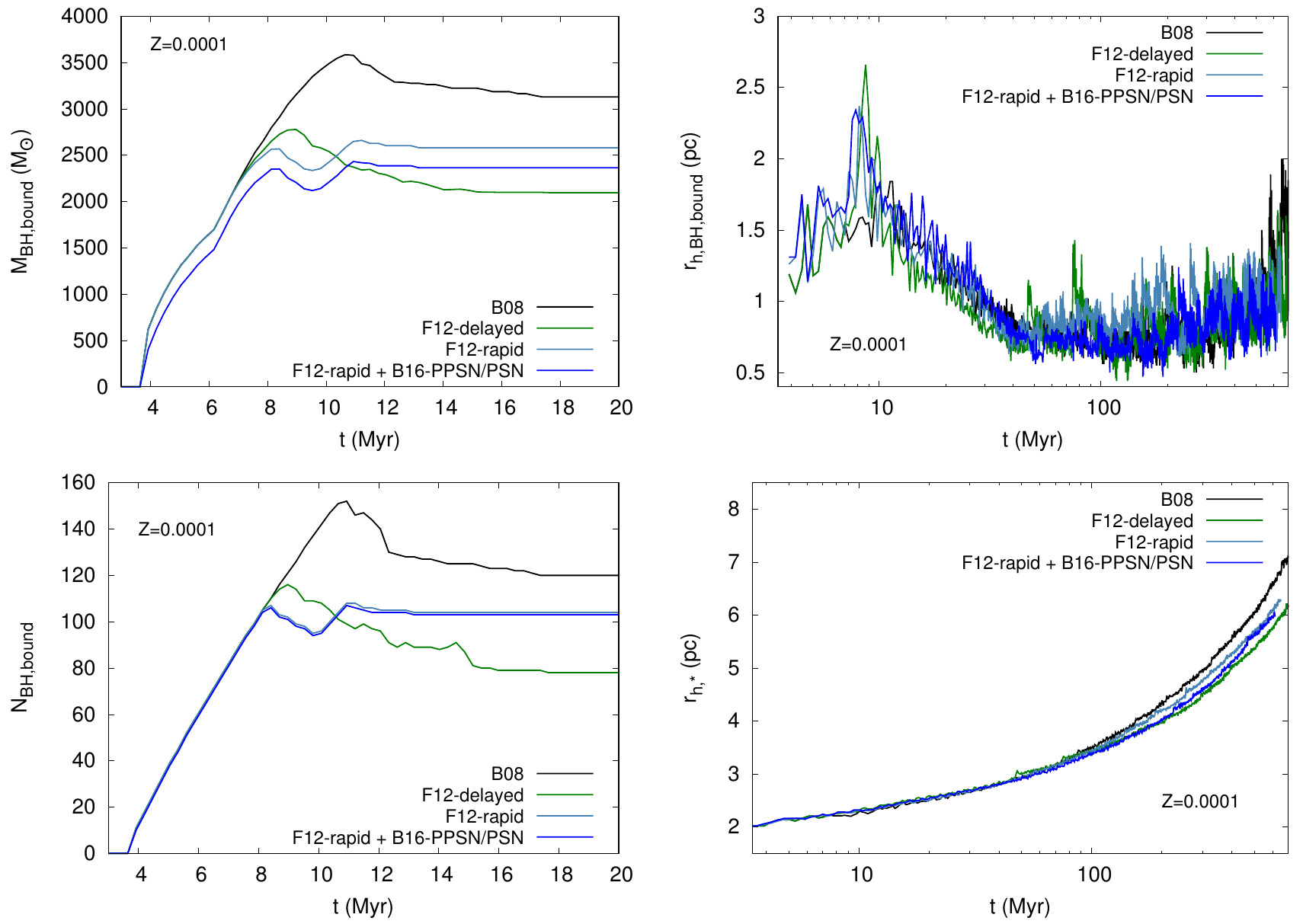}
\caption{The early (up to 20 Myr) evolution of the total mass, $\mbhbound$ (top left),
and number, $\nbhbound$ (bottom left), of the BHs
bound within the $\mcl(0)=5.0\times10^4\Ms$ (initially all single stars)
model cluster (see Sec.~\ref{nbcode}) for the various remnant-formation schemes (legend),
for the standard, fallback-controlled kick model as given by Eqn.~\ref{eq:vkick_std} ($Z=0.0001$ is assumed).
The corresponding evolutions of the half-mass radius of the cluster's BH subsystem, $\rhbhbound$ (top right),
and that of only
the luminous members, $\rhstar$ (bottom right), are also shown.}
\label{fig:bhmass_cmp1}
\end{figure*}

\begin{figure*}
\centering
\includegraphics[width=9.0cm]{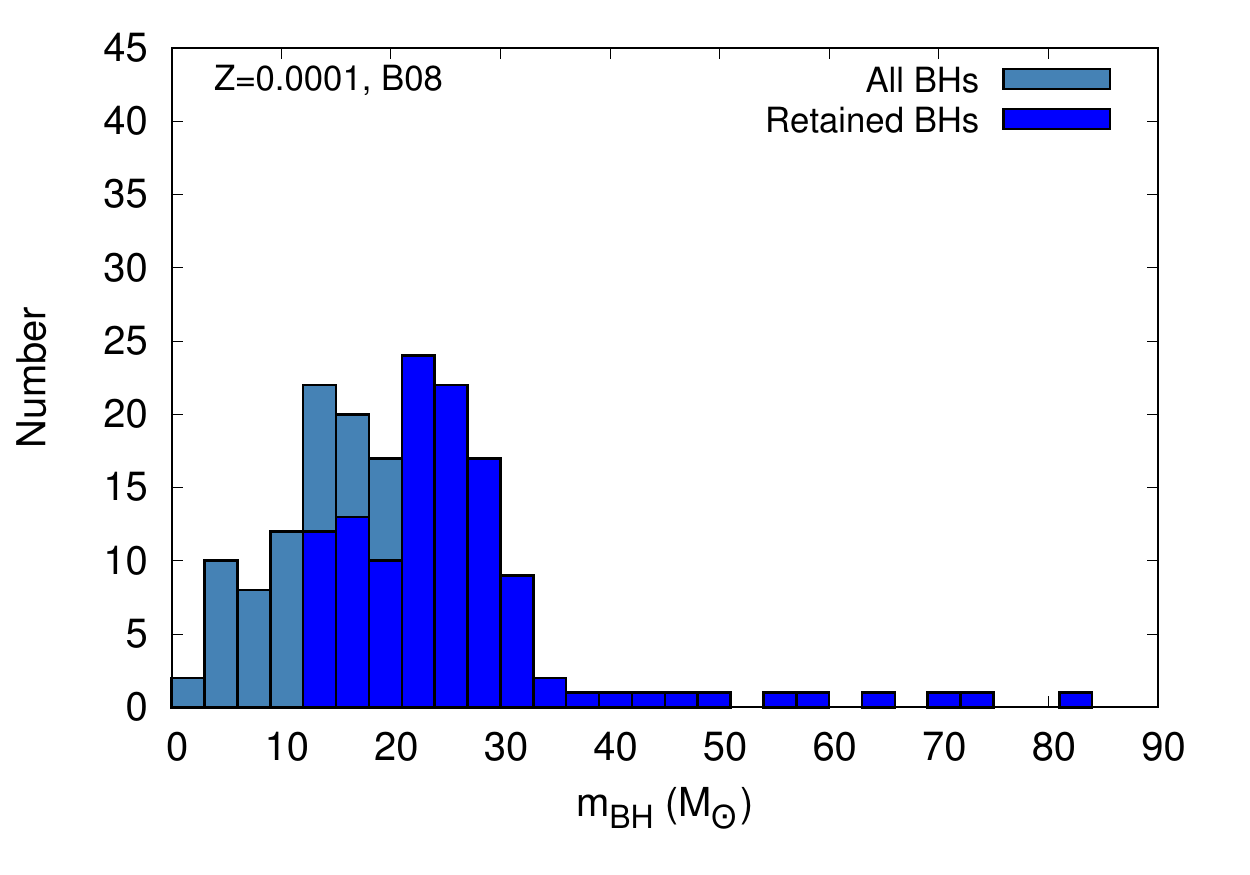}
\includegraphics[width=9.0cm]{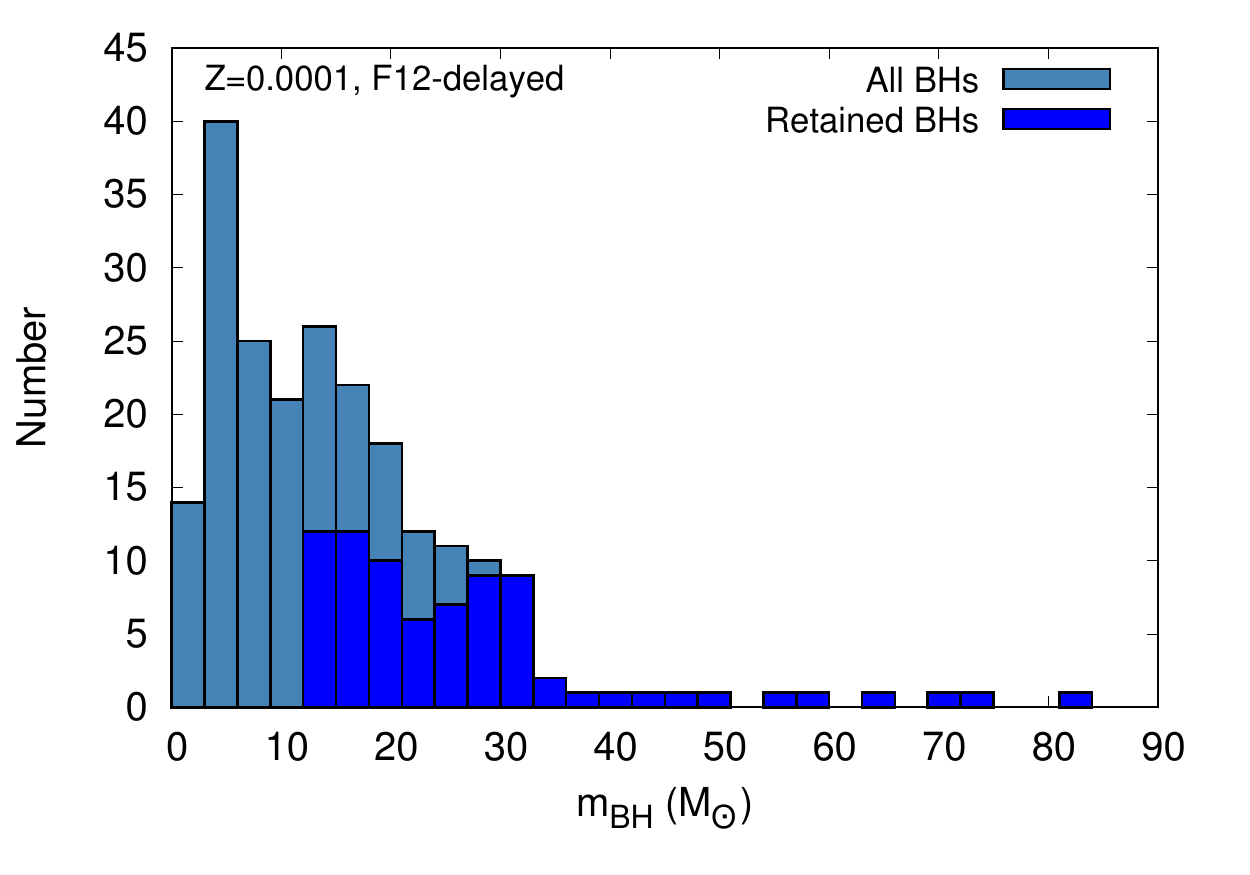}\\
\includegraphics[width=9.0cm]{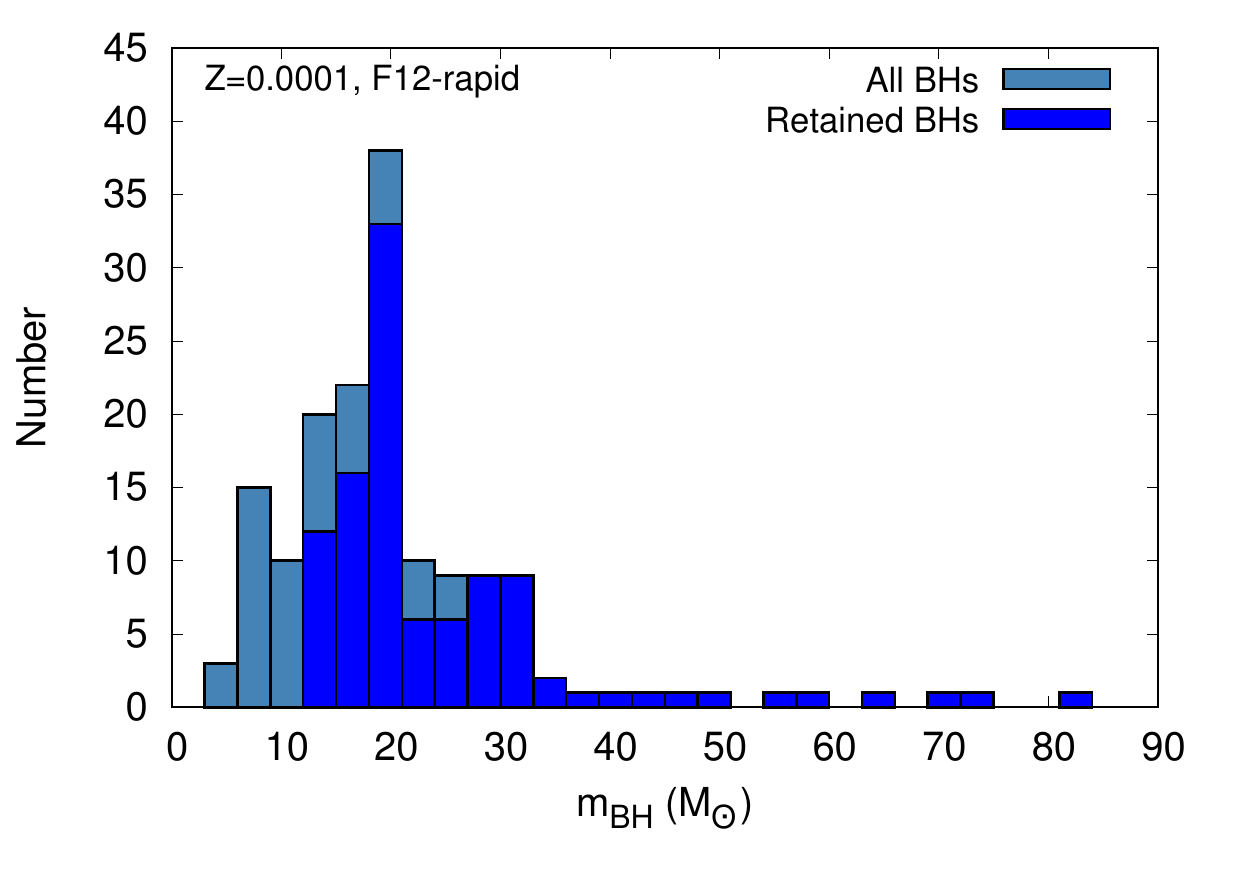}
\includegraphics[width=9.0cm]{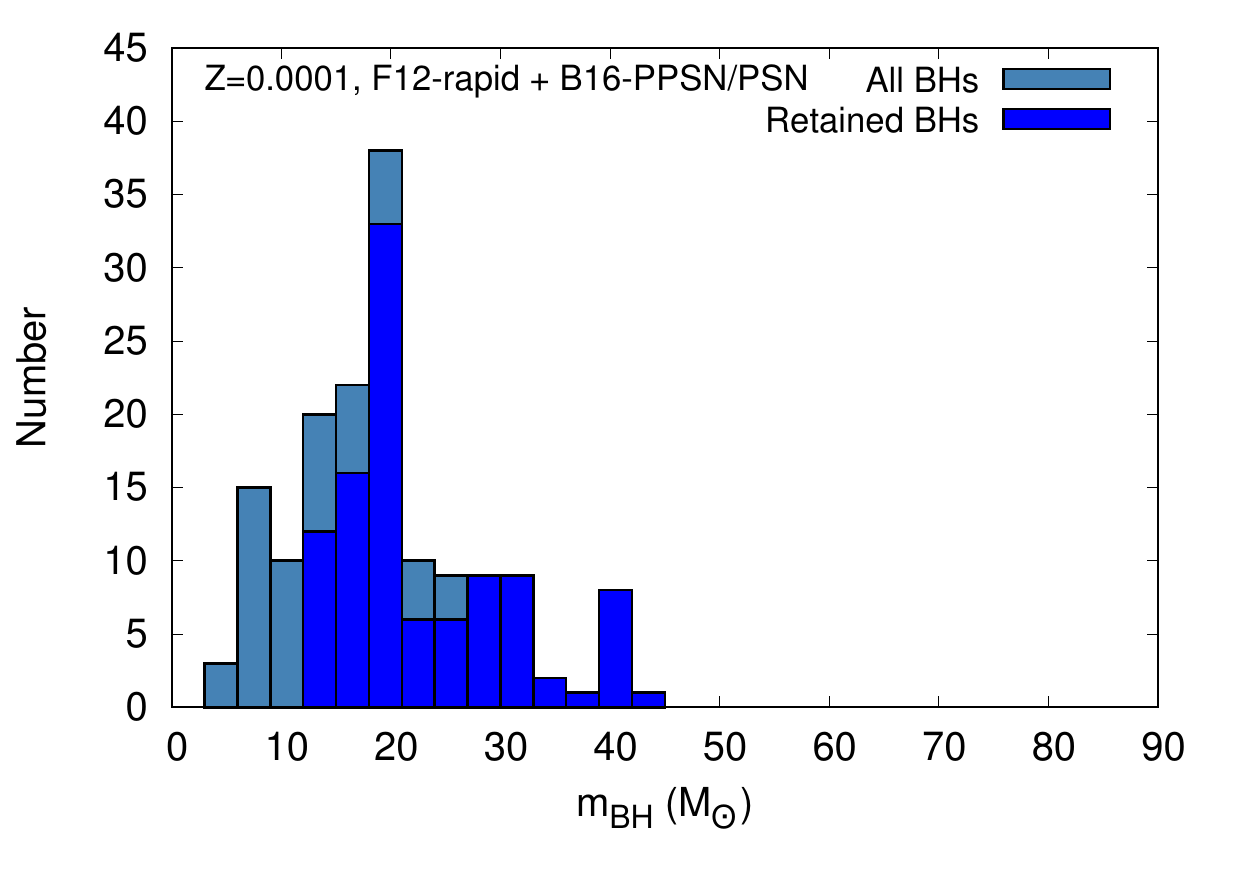}
\caption{The BH mass distributions obtained in the
$\mcl(0)=5.0\times10^4\Ms$ model clusters (initially all single stars; Sec.~\ref{nbcode}) for
four remnant-formation scenarios as indicated in the legends ($Z=0.0001$ taken).
On each panel, both the BHs' natal mass distribution and the distribution at $t\approx20$ Myr
cluster-evolutionary time are shown (the steel-blue and blue histograms respectively). The latter
distribution well represents the mass spectrum of those BHs which remain gravitationally
bound to a medium-mass (young massive or open; here taken to be of $\approx5\times10^4\Ms$)
cluster after their birth and which are, therefore, available for long-term
dynamical processing in the parent cluster (see Sec.~\ref{stdkick}). The retained
BH mass distributions (blue histograms), in these panels, are the outcomes of the
standard, fallback-controlled natal kick model (Eqn.~\ref{eq:vkick_std}).}
\label{fig:bhmass_cmp2}
\end{figure*}

\begin{figure*}
\centering
\includegraphics[width=10.5cm]{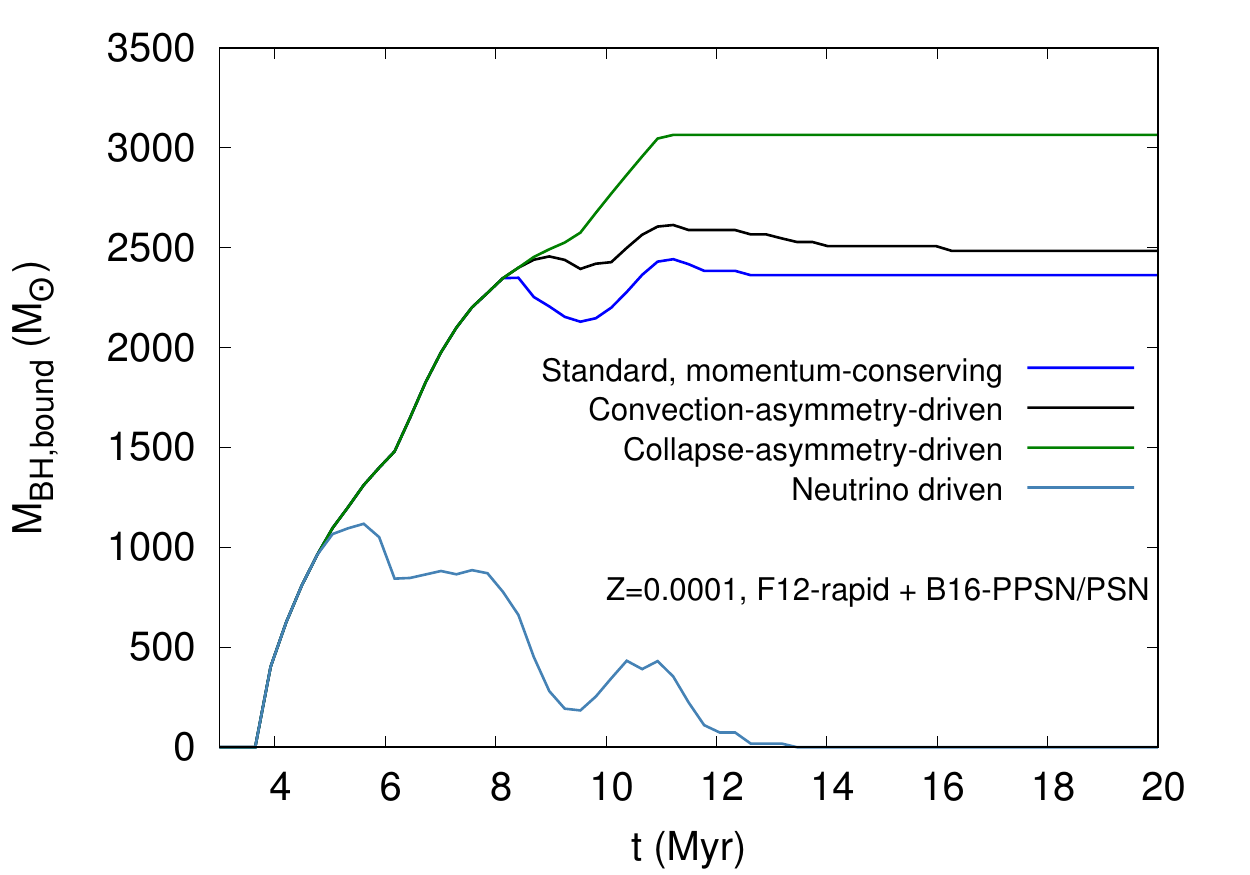}
\includegraphics[width=10.5cm]{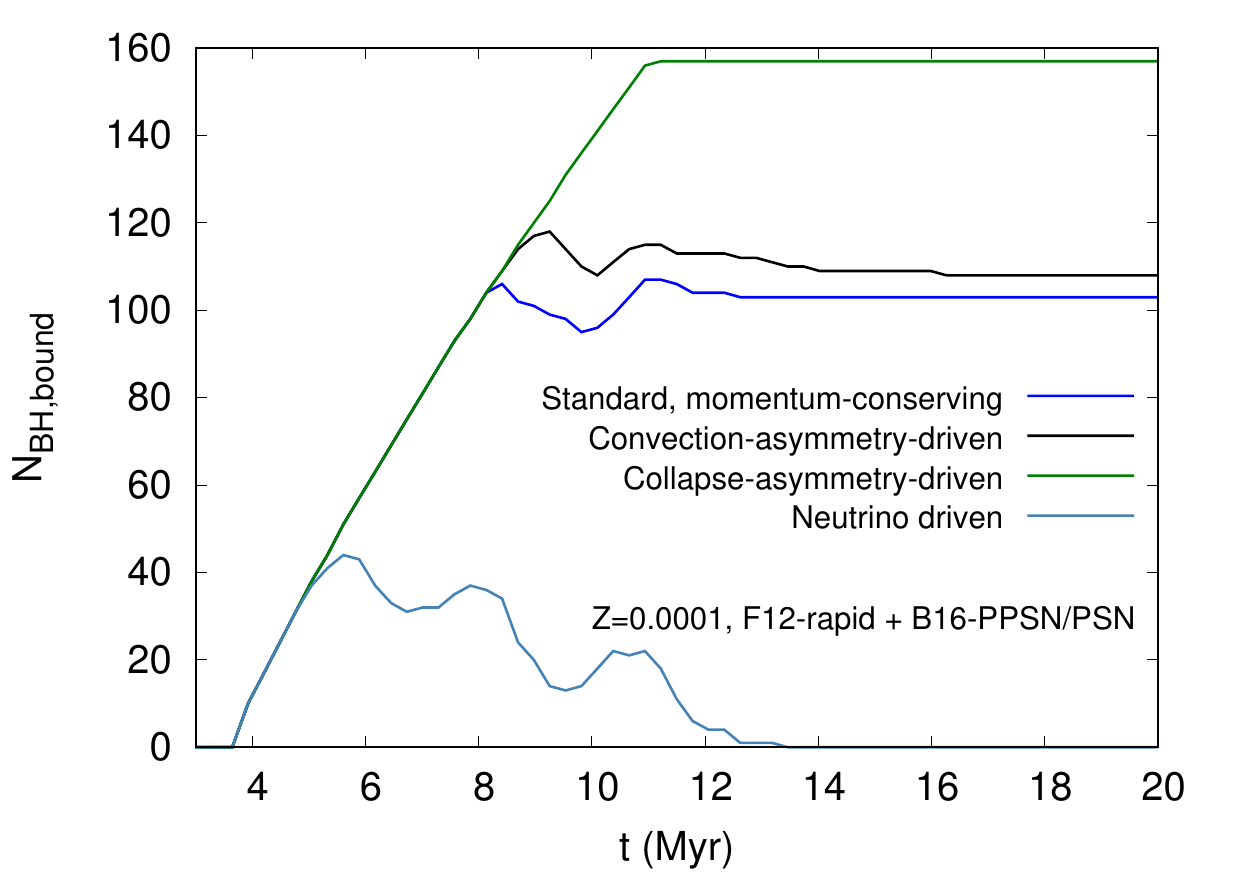}
\includegraphics[width=10.5cm]{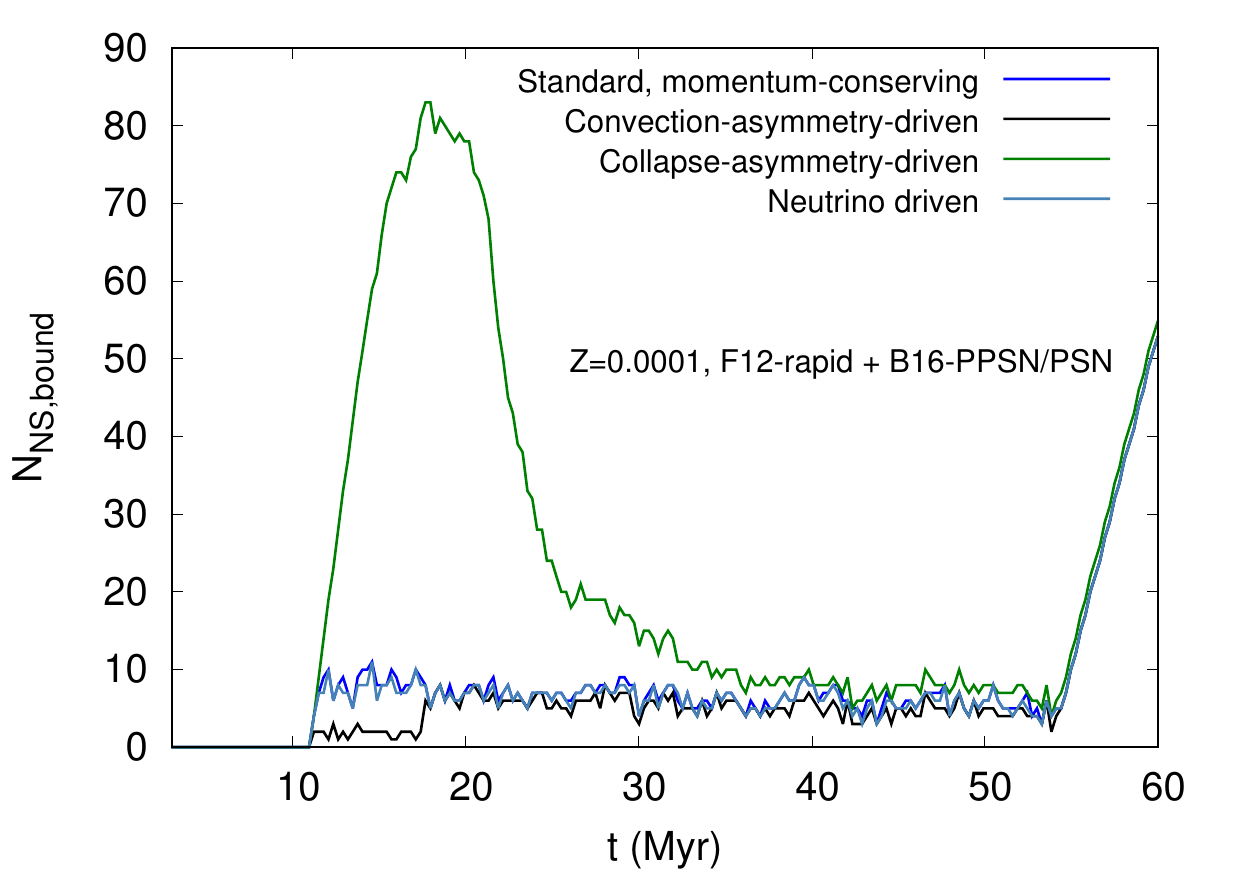}
\caption{The early evolution of the total mass, $\mbhbound$ (top panel),
and number, $\nbhbound$ (middle), of the BHs and the number, $\nnsbound$ (bottom),
of the NSs
bound within the $\mcl(0)=5.0\times10^4\Ms$ (initially all single stars)
model cluster (see Sec.~\ref{nbcode}) for the various natal-kick engines
(legends; see Secs.~\ref{stdkick} and \ref{altkick}),
for the F12-rapid+B16-PPSN/PSN remnant-mass model (Sec.~\ref{newrem}; $Z=0.0001$ is assumed).
The collapse-asymmetry-driven kick engine produces relatively low kicks also for some NSs
which marginally escape the cluster, taking time, resulting in
their buildup at $\approx20$ Myr followed by a slow decline (bottom panel). After $\approx55$ Myr
when the ECS-NSs begin to form, the latter's population
build up solely for all the kick-mechanism cases.}
\label{fig:kcomp_num}
\end{figure*}

\begin{figure*}
\centering
\includegraphics[width=9.0cm,angle=0]{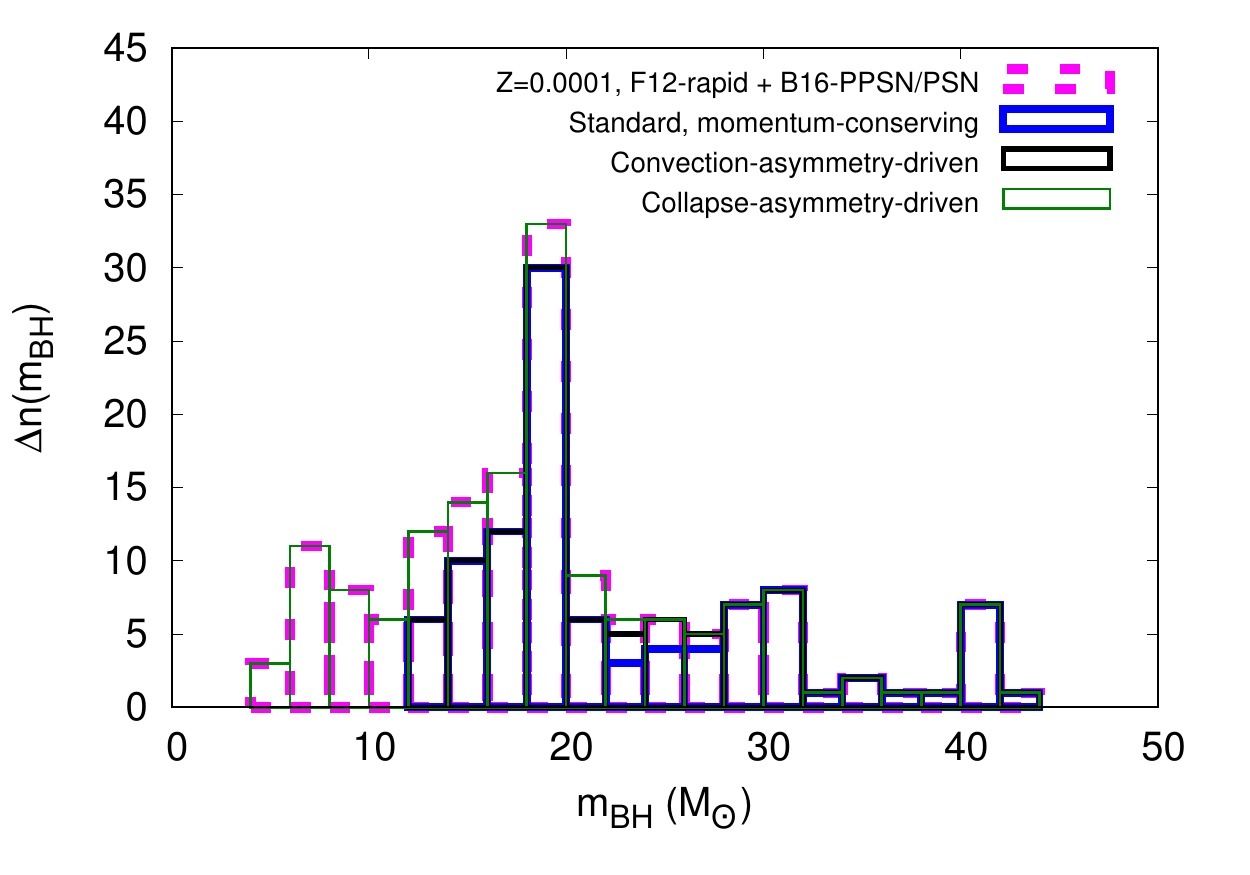}
\includegraphics[width=9.0cm,angle=0]{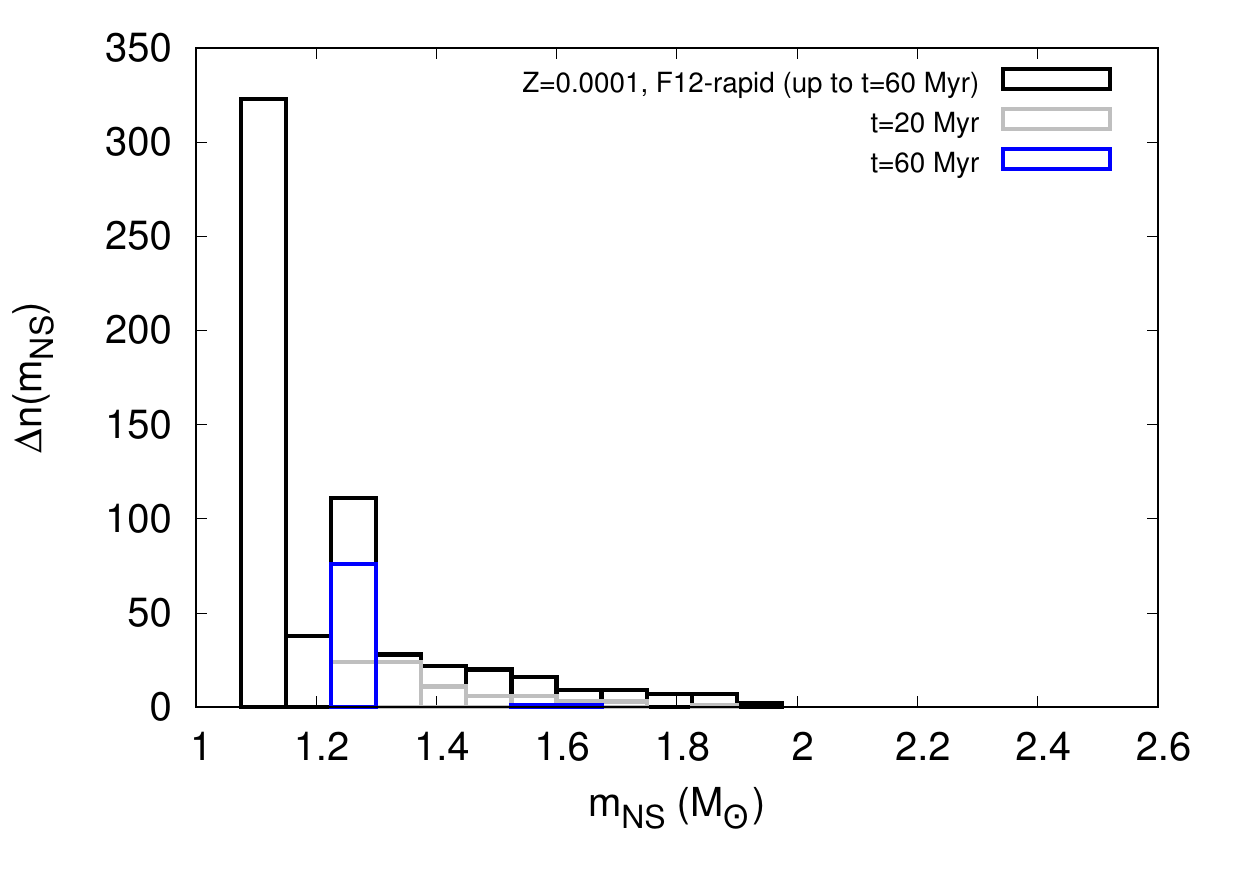}
\caption{{\bf Left:} Comparison between the retained BHs' mass distribution
(inside a $\mcl(0)\approx5\times10^4\Ms$, $\rh(0)\approx2$ pc,
initially single-only stars model cluster; see
also Fig.~\ref{fig:bhmass_cmp2}) for the various natal-kick
prescriptions considered in this work (legends; see Secs.~\ref{stdkick} and \ref{altkick}).
The F12-rapid+B16-PPSN/PSN
remnant-mass model (Sec.~\ref{newrem}) and $Z=0.0001$ is assumed.
The dashed histogram represents the BHs' natal mass distribution for
these remnant-mass model and $Z$. No BHs are retained in such clusters in the
neutrino-driven-kick case (Sec.~\ref{nknu}; Fig.~\ref{fig:kcomp_num}). In contrast,
nearly all BHs are retained if the kicks are collapse asymmetry driven. 
{\bf Right:} The NSs' \emph{natal} mass distribution for the F12-rapid(+B16-PPSN/PSN)
remnant-mass model (black histogram). The \emph{retained} NSs' mass distributions,
for the collapse-asymmetry-driven kick model, at $t\approx20$ Myr and
60 Myr are also shown (grey and blue histograms respectively).
The marginally-escaping NSs at 20 Myr (Sec.~\ref{nkmech}; Fig.~\ref{fig:kcomp_num},
bottom panel) are all depleted by 60 Myr
when only the low-/zero-kicked ECS-NSs retain in the cluster.}
\label{fig:kcomp_bhdist}
\end{figure*}

\begin{figure*}
\centering
\includegraphics[width=9.0cm,angle=0]{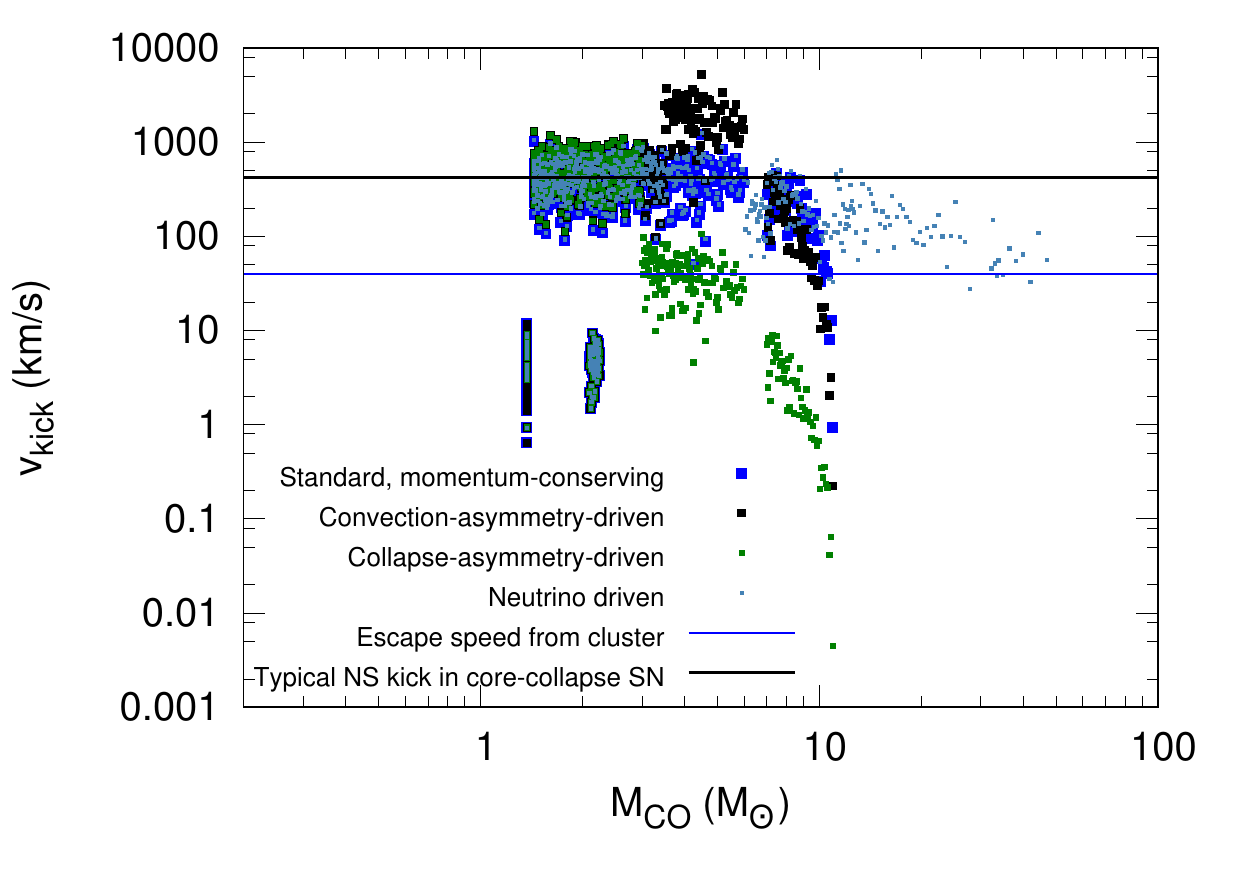}
\includegraphics[width=9.0cm,angle=0]{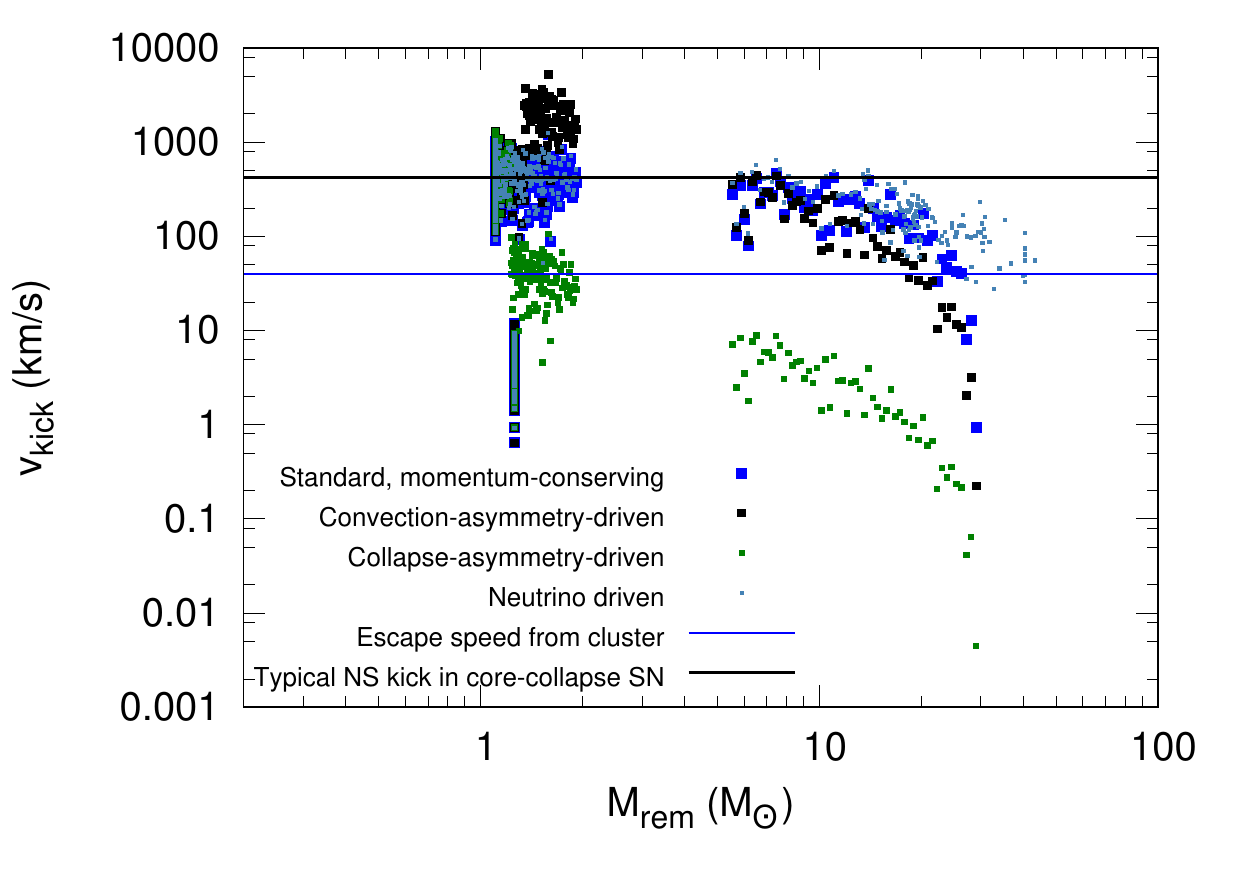}
\caption{The natal kicks, $\vkick$, as generated by $\nbseven$,
as a function of $\mco$ (left panel) and $\mrem$ (right panel),
for the different natal-kick recipes considered here.
The F12-rapid+B16-PPSN/PSN
remnant-mass model (Sec.~\ref{newrem}) and $Z=0.0001$ is assumed.
Due to the logarithmic vertical
axis, direct-collapse BHs with $\fbfac=1$ and $\vkick=0$ are not shown in
these panels; the sharp drop in $\vkick$ with increasing $\mco$ or $\mrem$
is the approach towards direct collapse. The typical $\vesc$ for
the $\mcl(0)\approx5.0\times10^4\Ms$, $\rh(0)\approx2$ pc clusters
considered here (blue, solid line) and the \citet{Hobbs_2005}
$\signs\approx265\kmps$ w.r.t. which all the kick models are
scaled (see Secs.~\ref{stdkick} and \ref{altkick}; black, solid line)
are indicated.}
\label{fig:kcomp_comass}
\end{figure*}

\begin{itemize}

\item A reimplementation of the B10 wind model (Sec.~\ref{newwind}) and
the adoption of newer remnant-mass prescriptions such as F12-rapid
and F12-delayed including the PPSN/PSN conditions of B16 (Sec.~\ref{newrem}) in
the widely-used population synthesis program $\bse$
reproduces the ZAMS mass-remnant mass relations
of $\startrack$ nearly perfectly over a wide range of ZAMS mass,
remnant-formation scheme, and metallicity.
(Sec.~\ref{comp1}, Figs.~\ref{fig:cmp1}, \ref{fig:cmp_psn}, \ref{fig:cmp_ext}).
Such agreements,
however, require smaller stellar-evolutionary time step parameters
(the $\ptsone$, $\ptstwo$, and $\ptsthree$ parameters) than
their default values while running $\bse$
(Sec.~\ref{tstep}, Fig.~\ref{fig:tstep}), that
slightly slows down the program.
The new $\bse$, therefore, now incorporates remnant-mass and stellar-wind
prescriptions, in terms of detail, flexibility and the outcomes,
at par with state-of-the-art population-synthesis programs such as
$\startrack$ and $\mobse$.

\item These new developments in $\bse$ naturally get ported into
the direct N-body code $\nbseven$ since the latter adopts
$\bse$ as its stellar-evolution and remnant-formation driver.
The adoption of the new $\bse$ routines (the functions
$\mlwind$ and $\hrdiag$) in $\nbseven$ yields a similar quality of
reproduction of the $\startrack$ ZAMS mass-remnant mass relations,
for NSs and BHs formed from single stars within model star clusters
(Sec.~\ref{nbcode}, Fig.~\ref{fig:cmp_nb}).

\item To deal with the retention of NSs and BHs in such model
clusters at their birth, an explicit fallback modulation
is implemented in the natal-kick generator of $\nbseven$,
which can be directly ported into the corresponding implementation
in the standalone $\bse$ program (their $\kick$ routines;
Sec.~\ref{stdkick}).
In addition to this standard, fallback-controlled natal kick
prescription, its variants describing convection-asymmetry-driven and
collapse-asymmetry-driven natal kicks are prescribed and as well
a neutrino-driven kick model (Sec.~\ref{altkick}; Fig.~\ref{fig:kcomp_comass}).

The above implementations have now been tested to work very stably
in an extensive set of long-term $\nbseven$ runs (Banerjee, S., in prep.).
These new stellar-evolutionary ingredients and natal-kick models
have now as well been ported to $\nbpp$ and tested to perform as stably (in prep.).

The various remnant and natal-kick prescriptions in the routines $\hrdiag$
and $\kick$ can, at present, be selected based on switches built in these routines
(for $\nbseven$, also the stellar-evolutionary time step parameters),
requiring recompilation of $\bse$ and $\nbseven$.  
In the near future, arrangements will be made to provide these switches
in the inputs, to make the options more accessible. Such a tested version
of standalone \bse, with all the newly-implemented prescriptions
described here, can be obtained freely.
\footnote{\url{https://tinyurl.com/v7uadf4}}

\item For a given cluster (and given IMF), the amount of BHs retaining (\ie,
remaining gravitationally bound to the cluster) due to low-enough natal kicks
to become available for long-term dynamical processing
depends on the remnant-formation mechanism (Fig.~\ref{fig:bhmass_cmp1}),
the natal-kick mechanism (Fig.~\ref{fig:kcomp_num}), and metallicity (Table~\ref{tab_retfrac}).  
For all remnant-formation and natal-kick schemes
(except for the collapse-asymmetry-driven case; see below),
the retained BHs are
generally top-heavier in mass spectrum compared to their
birth mass spectrum (Figs.~\ref{fig:bhmass_cmp2} \& \ref{fig:kcomp_bhdist}) due
to the preferential escape of lower-mass BHs as a result of their
higher natal kicks (or smaller fallback mass; Fig.~\ref{fig:kcomp_comass}). 
The F12-delayed remnant-formation model gives birth to the
largest number of BHs (as opposed to the resulting retained BHs where
B08 maximizes) but it also produces the largest number of escapees
ultimately resulting in the smallest number of retainers.

\item Table~\ref{tab_retfrac} gives the BH mass and number retention
fractions in a moderately-massive stellar cluster as a function of
remnant-formation scheme and metallicity, for the standard, fallback-controlled
natal kick.
The convection-asymmetry-driven
kick mechanism would result in nearly the same retention fractions,
the collapse-asymmetry-driven
mechanism would lead to all retention fractions $\approx1.0$ and the neutrino-driven
mechanism would lead to all retention fractions $\approx0.0$
(Fig.~\ref{fig:kcomp_num}, \ref{fig:kcomp_comass}). Hence,
if $\sim100$ BHs still retain in some present-day GCs as inferred when
their observed photometric and internal-kinematic structures are
compared with model computations
\citep[\eg,][]{Askar_2018,Kremer_2018,Kremer_2018b}, 
then the neutrino-driven mechanism can possibly be ruled out.
If forthcoming observations suggest the presence of $\lesssim10\Ms$
BHs in GCs and/or open clusters and/or young massive clusters,
then this would provide a support to the collapse-asymmetry
mechanism (Sec.~\ref{nkmech}, Fig.~\ref{fig:kcomp_bhdist}).
For the case of the latter kick mechanism, a population
of slow-escaping core-collapse NSs can also be expected
in the field of view of $\sim10-20$ Myr-aged young clusters
(Sec.~\ref{nkmech}, Fig.~\ref{fig:kcomp_num}),
which may be identified in the future through radio observations.
The BH retention fractions in Table~\ref{tab_retfrac}
represent those in a moderate-escape-velocity system, essentially
due to direct collapse or a substantial amount of material fallback alone:
for massive GCs, their higher escape velocity would additionally aid
retention, potentially resulting in somewhat higher retention fractions in all cases.
These inferences encourage future, more elaborate studies
of discriminatory signatures of natal-kick mechanisms
in stellar clusters, especially in more massive systems (Leveque, A., et al., in preparation). 

\item If BH-progenitor stars exist in high primordial-binary
fractions, as observed in young massive and open clusters,
then, depending on the mass loss
during star-star merger process, the BHs' natal and cluster-retained mass spectra would
become wider; at low metallicities, the BH mass can exceed well beyond $40\Ms$
even in the presence of PPSN,
forming BHs bound to a stellar cluster well within the
PSN mass gap (Sec.~\ref{primbin}, Fig.~\ref{fig:bhwbin}).
Since such BHs are still direct derivatives of stellar
progenitors, they would potentially have very low spins
(unless the star-star merger happens late in the evolution of both stars). A
high primordial-binary fraction among BH progenitors would
also lead to a moderately higher BH-retention fraction
in both mass and number (Table~\ref{tab_retfrac})
although the total number of BHs produced, under such circumstance, would be
somewhat smaller due to the mergers (Fig.~\ref{fig:bhwbin}).
The BH masses get altered due to pre-BH-formation stellar mergers
and/or other stellar-hydrodynamical processes (\eg, CE envelope ejection,
tidally-enhanced stellar wind).
In this work,
these modified BH masses are the outcomes of the simplistic treatment
of these complex physical processes in the binary-evolution engine of $\bse$
and as well of the single-star-like treatment while forming
remnant masses from stellar-merger products that would
have altered stellar structure, which
is why their values should be taken with caution.

The above effect, if real, would still preserve the (potentially-future)
LIGO-detected BBH mergers comprising $\gtrsim40\Ms$ members
as signatures of dynamically-triggered BBH mergers in clusters,
except that such inferences need not, anymore, depend on
low natal spins of BHs \citep[see, \eg,][]{Rodriguez_2018}.
If stellar progenitors indeed produce low-spinning BHs
irrespective of their own spin \citep{Belczynski_2020}, internal structure, and
chemical composition, then such mass-gap BBH mergers
could potentially exhibit low effective spin parameter. Whether
a dynamically-active environment such as that inside a
stellar cluster aid in inducing star-star mergers
that lead to low-spin, PSN-mass-gap BHs, will be investigated in
a separate study.
If the single-star remnant mass-ZAMS mass relation
is to be unaffected by massive primordial-binary mergers,
then a substantial, up to $\approx40$\%, mass loss would be necessary
in such mergers (Sec.~\ref{primbin}, Fig.~\ref{fig:bhwbin3}).
In this context, it would be worth bearing in mind
that the PPSN cutoff mass is not strictly constrained (see B16).
In particular, the recent measurement of a 
$\approx50\Ms$ primary in the BBH merger event GW170729
(\citealt{Abbott_GWTC1,Abbott_GWTC1_prop}; but see \citealt{Fishbach_2020})
may call for a moderate revision of the
PPSN BH mass limit in population-synthesis codes to
increase it by $\approx10\Ms$ as was already proposed
in \citet[][see their Fig. 3]{Belczynski_2020}. In fact,
considering the full mass spectrum of BBHs detected to date
by LIGO-Virgo \citep{Abbott_GWTC1},
GW170729 is statistically consistent with being a first-generation
merger \citep{Chatziioannou_2019,Kimball_2020}.
Also, a more elaborate study of
the influence of a dynamically-active environment on
massive-binary evolution will be taken up in the
near future.

\end{itemize}

\begin{acknowledgements}
We thank the anonymous referee for their useful comments and
suggestions which have improved the manuscript.
SB acknowledges the support from the Deutsche Forschungsgemeinschaft
(DFG; German Research Foundation) through the individual research grant
``The dynamics of stellar-mass black holes in dense stellar systems and their
role in gravitational-wave generation'' (BA 4281/6-1; PI: S. Banerjee).
SB acknowledges the pleasant hospitality of the
Nicolaus Copernicus Astronomical Centre of the Polish Academy of Sciences
(CAMK) where this work was initiated.  
SB acknowledges the generous support and efficient system maintenance
of the computing teams at the Argelander-Institut f\"ur Astronomie (AIfA)
and the CAMK where the computations have been performed.
KB acknowledges support from the Polish National Science Center (NCN) grant Maestro
(2018/30/A/ST9/00050).
The work of CF was done at Los Alamos National Laboratory under
project number 20190021DR.
This work was supported by the Deutsche Forschungsgemeinschaft
(DFG, German Research Foundation) – Project-ID 138713538 –
SFB 881 (``The Milky Way System'', subproject A06) and by the
Volkswagen Foundation under the Trilateral Partnerships grants
No. 90411 and 97778. 
The special GPU accelerated supercomputer Laohu at NAOC
has been used and we thank the Center of Information and 
Computing of NAOC for support. 
This work has been benefited from support by the International Space
Science Institute, Bern, Switzerland, through its International Team
program ref. no. 393 ``The Evolution of Rich Stellar Populations and
BH Binaries'' (2017–18). 
The work of PB was also partially supported under the special 
program of the NAS of Ukraine ``Support for the development of 
priority fields of scientific research'' (CPCEL 6541230).
RS and PB acknowledge support through the Silk Road Project at the
National Astronomical Observatories (NAOC),
Chinese Academy of Sciences, through the National Natural Science Foundation
of China (NSFC, grant No. 11673032), and through the Strategic Priority Research Program (Pilot B)
``Multi-wavelength gravitational wave universe'' of the Chinese Academy of Sciences (No. XDB23040100).
Visits of SB and KB have been supported by the Silk Road Project at NAOC.
JH has been supported by the Kavli Visitor Program of
Kavli Institute for Astronomy and Astrophysics, Peking University.
LW acknowledges the support from Alexander von Humboldt Foundation and JSPS
International Research Fellow (School of Science, The university of
Tokyo).
We thank Mirek Giersz and Diogo Belloni for pointing out
an issue with white dwarf formation and white dwarf masses
under our newly implemented schemes which has helped us to locate and fix an important bug
in one of the routines
in the updated $\bse$ code presented in this work.
The ``time step issue'' in $\bse$ (Sec.~\ref{tstep})
was hinted by Mirek Giersz during a lunch at CAMK.
\end{acknowledgements}

%
\bibliographystyle{aa}
\bibliography{bibliography/biblio.bib}

\begin{thebibliography}{112}
\expandafter\ifx\csname natexlab\endcsname\relax\def\natexlab#1{#1}\fi

\bibitem[{{Aarseth}(2003)}]{2003gnbs.book.....A}
{Aarseth}, S.~J. 2003, {Gravitational N-Body Simulations, Cambridge University
  Press, Cambridge, UK, pp.~430.~ISBN 0521432723}, 430

\bibitem[{{Aarseth}(2012)}]{2012MNRAS.422..841A}
{Aarseth}, S.~J. 2012, \mnras, 422, 841

\bibitem[{{Abadie} {et~al.}(2010){Abadie}, {Abbott}, {Abbott}, {Abernathy},
  {Accadia}, {Acernese}, {Adams}, {Adhikari}, {Ajith}, {Allen}, \&
  et~al.}]{Abadie_2010}
{Abadie}, J., {Abbott}, B.~P., {Abbott}, R., {et~al.} 2010, Classical and
  Quantum Gravity, 27, 173001

\bibitem[{{Abbott} {et~al.}(2016{\natexlab{a}}){Abbott}, {Abbott}, {Abbott},
  {Abernathy}, {Acernese}, {Ackley}, {Adams}, {Adams}, {Addesso}, {Adhikari},
  \& et~al.}]{2016ApJ...818L..22A}
{Abbott}, B.~P., {Abbott}, R., {Abbott}, T.~D., {et~al.} 2016{\natexlab{a}},
  \apjl, 818, L22

\bibitem[{{Abbott} {et~al.}(2016{\natexlab{b}}){Abbott}, {Abbott}, {Abbott},
  {Abernathy}, {Acernese}, {Ackley}, {Adams}, {Adams}, {Addesso}, {Adhikari},
  \& et~al.}]{2016PhRvL.116f1102A}
{Abbott}, B.~P., {Abbott}, R., {Abbott}, T.~D., {et~al.} 2016{\natexlab{b}},
  Physical Review Letters, 116, 061102

\bibitem[{{Abbott} {et~al.}(2019{\natexlab{a}}){Abbott}, {Abbott}, {Abbott},
  {Abraham}, {Acernese}, {Ackley}, {Adams}, {Adhikari}, {Adya}, {Affeldt}, \&
  et~al.}]{Abbott_GWTC1_prop}
{Abbott}, B.~P., {Abbott}, R., {Abbott}, T.~D., {et~al.} 2019{\natexlab{a}},
  \apjl, 882, L24

\bibitem[{{Abbott} {et~al.}(2019{\natexlab{b}}){Abbott}, {Abbott}, {Abbott},
  {Abraham}, {Acernese}, {Ackley}, {Adams}, {Adhikari}, {Adya}, {Affeldt}, \&
  et~al.}]{Abbott_GWTC1}
{Abbott}, B.~P., {Abbott}, R., {Abbott}, T.~D., {et~al.} 2019{\natexlab{b}},
  Physical Review X, 9, 031040

\bibitem[{Abbott {et~al.}(2017{\natexlab{a}})Abbott, Abbott, Abbott, Acernese,
  Ackley, Adams, Adams, Addesso, Adhikari, Adya, \& et~al.}]{Abbott_GW170104}
Abbott, B.~P., Abbott, R., Abbott, T.~D., {et~al.} 2017{\natexlab{a}}, Physical
  Review Letters, 118, 221101

\bibitem[{Abbott {et~al.}(2017{\natexlab{b}})Abbott, Abbott, Abbott, Acernese,
  Ackley, Adams, Adams, Addesso, Adhikari, Adya, \& et~al.}]{Abbott_GW170814}
Abbott, B.~P., Abbott, R., Abbott, T.~D., {et~al.} 2017{\natexlab{b}}, Physical
  Review Letters, 119, 141101

\bibitem[{{Arca Sedda} {et~al.}(2018){Arca Sedda}, {Askar}, \&
  {Giersz}}]{ArcaSedda_2018}
{Arca Sedda}, M., {Askar}, A., \& {Giersz}, M. 2018, \mnras, 479, 4652

\bibitem[{{Askar} {et~al.}(2018){Askar}, {Arca Sedda}, \&
  {Giersz}}]{Askar_2018}
{Askar}, A., {Arca Sedda}, M., \& {Giersz}, M. 2018, \mnras, 478, 1844

\bibitem[{{Askar} {et~al.}(2017){Askar}, {Szkudlarek}, {Gondek-Rosi{\'n}ska},
  {Giersz}, \& {Bulik}}]{Askar_2016}
{Askar}, A., {Szkudlarek}, M., {Gondek-Rosi{\'n}ska}, D., {Giersz}, M., \&
  {Bulik}, T. 2017, \mnras, 464, L36

\bibitem[{{Banerjee}(2017)}]{Banerjee_2017}
{Banerjee}, S. 2017, \mnras, 467, 524

\bibitem[{{Banerjee}(2018{\natexlab{a}})}]{Banerjee_2018}
{Banerjee}, S. 2018{\natexlab{a}}, \mnras, 481, 5123

\bibitem[{{Banerjee}(2018{\natexlab{b}})}]{Banerjee_2017b}
{Banerjee}, S. 2018{\natexlab{b}}, \mnras, 473, 909

\bibitem[{{Banerjee} {et~al.}(2010){Banerjee}, {Baumgardt}, \&
  {Kroupa}}]{Banerjee_2010}
{Banerjee}, S., {Baumgardt}, H., \& {Kroupa}, P. 2010, \mnras, 402, 371

\bibitem[{{Banerjee} {et~al.}(2012){Banerjee}, {Kroupa}, \&
  {Oh}}]{Banerjee_2012}
{Banerjee}, S., {Kroupa}, P., \& {Oh}, S. 2012, \mnras, 426, 1416

\bibitem[{{Belczynski} {et~al.}(2010){Belczynski}, {Benacquista}, \&
  {Bulik}}]{Belczynski_2010c}
{Belczynski}, K., {Benacquista}, M., \& {Bulik}, T. 2010, \apj, 725, 816

\bibitem[{Belczynski {et~al.}(2010)Belczynski, Bulik, Fryer, Ruiter, Valsecchi,
  Vink, \& Hurley}]{Belczynski_2010}
Belczynski, K., Bulik, T., Fryer, C.~L., {et~al.} 2010, The Astrophysical
  Journal, 714, 1217

\bibitem[{{Belczynski} {et~al.}(2016{\natexlab{a}}){Belczynski}, {Heger},
  {Gladysz}, {Ruiter}, {Woosley}, {Wiktorowicz}, {Chen}, {Bulik},
  {O'Shaughnessy}, {Holz}, {Fryer}, \& {Berti}}]{Belczynski_2016a}
{Belczynski}, K., {Heger}, A., {Gladysz}, W., {et~al.} 2016{\natexlab{a}},
  \aap, 594, A97

\bibitem[{{Belczynski} {et~al.}(2016{\natexlab{b}}){Belczynski}, {Holz},
  {Bulik}, \& {O'Shaughnessy}}]{Belczynski_2016}
{Belczynski}, K., {Holz}, D.~E., {Bulik}, T., \& {O'Shaughnessy}, R.
  2016{\natexlab{b}}, \nat, 534, 512

\bibitem[{{Belczynski} {et~al.}(2002){Belczynski}, {Kalogera}, \&
  {Bulik}}]{Belczynski_2002}
{Belczynski}, K., {Kalogera}, V., \& {Bulik}, T. 2002, \apj, 572, 407

\bibitem[{Belczynski {et~al.}(2008)Belczynski, Kalogera, Rasio, Taam, Zezas,
  Bulik, Maccarone, \& Ivanova}]{Belczynski_2008}
Belczynski, K., Kalogera, V., Rasio, F.~A., {et~al.} 2008, The Astrophysical
  Journal Supplement Series, 174, 223

\bibitem[{{Belczynski} {et~al.}(2017{\natexlab{a}}){Belczynski}, {Klencki},
  {Fields}, {Olejak}, {Berti}, {Meynet}, {Fryer}, {Holz}, {O'Shaughnessy},
  {Brown}, {Bulik}, {Leung}, {Nomoto}, {Madau}, {Hirschi}, {Jones}, {Mondal},
  {Chruslinska}, {Drozda}, {Gerosa}, {Doctor}, {Giersz}, {Ekstrom}, {Georgy},
  {Askar}, {Wysocki}, {Natan}, {Farr}, {Wiktorowicz}, {Miller}, {Farr}, \&
  {Lasota}}]{Belczynski_2020}
{Belczynski}, K., {Klencki}, J., {Fields}, C.~E., {et~al.} 2017{\natexlab{a}},
  arXiv e-prints, arXiv:1706.07053

\bibitem[{{Belczynski} {et~al.}(2017{\natexlab{b}}){Belczynski}, {Klencki},
  {Meynet}, {Fryer}, {Brown}, {Chruslinska}, {Gladysz}, {O'Shaughnessy},
  {Bulik}, {Berti}, {Holz}, {Gerosa}, {Giersz}, {Ekstrom}, {Georgy}, {Askar},
  \& {Lasota}}]{2017arXiv170607053B}
{Belczynski}, K., {Klencki}, J., {Meynet}, G., {et~al.} 2017{\natexlab{b}},
  ArXiv e-prints (arXiv:1706.07053) [\eprint[arXiv]{1706.07053}]

\bibitem[{{Belczynski} {et~al.}(2016{\natexlab{c}}){Belczynski}, {Repetto},
  {Holz}, {O'Shaughnessy}, {Bulik}, {Berti}, {Fryer}, \&
  {Dominik}}]{Belczynski_2016b}
{Belczynski}, K., {Repetto}, S., {Holz}, D.~E., {et~al.} 2016{\natexlab{c}},
  \apj, 819, 108

\bibitem[{{Burrows} \& {Hayes}(1996)}]{Burrows_1996}
{Burrows}, A. \& {Hayes}, J. 1996, \prl, 76, 352

\bibitem[{{Chatterjee} {et~al.}(2017){Chatterjee}, {Rodriguez}, \&
  {Rasio}}]{2016arXiv160300884C}
{Chatterjee}, S., {Rodriguez}, C.~L., \& {Rasio}, F.~A. 2017, \apj, 834, 68

\bibitem[{{Chatziioannou} {et~al.}(2019){Chatziioannou}, {Cotesta}, {Ghonge},
  {Lange}, {Ng}, {Calder{\'o}n Bustillo}, {Clark}, {Haster}, {Khan},
  {P{\"u}rrer}, {Raymond}, {Vitale}, {Afshari}, {Babak}, {Barkett}, {Blackman},
  {Boh{\'e}}, {Boyle}, {Buonanno}, {Campanelli}, {Carullo}, {Chu}, {Flynn},
  {Fong}, {Garcia}, {Giesler}, {Haney}, {Hannam}, {Harry}, {Healy},
  {Hemberger}, {Hinder}, {Jani}, {Khamersa}, {Kidder}, {Kumar}, {Laguna},
  {Lousto}, {Lovelace}, {Littenberg}, {London}, {Millhouse}, {Nuttall}, {Ohme},
  {O'Shaughnessy}, {Ossokine}, {Pannarale}, {Schmidt}, {Pfeiffer}, {Scheel},
  {Shao}, {Shoemaker}, {Szilagyi}, {Taracchini}, {Teukolsky}, \&
  {Zlochower}}]{Chatziioannou_2019}
{Chatziioannou}, K., {Cotesta}, R., {Ghonge}, S., {et~al.} 2019, \prd, 100,
  104015

\bibitem[{{Claret}(2007)}]{Claret_2007}
{Claret}, A. 2007, \aap, 467, 1389

\bibitem[{{de Mink} {et~al.}(2013){de Mink}, {Langer}, {Izzard}, {Sana}, \& {de
  Koter}}]{deMink_2013}
{de Mink}, S.~E., {Langer}, N., {Izzard}, R.~G., {Sana}, H., \& {de Koter}, A.
  2013, \apj, 764, 166

\bibitem[{{Di Carlo} {et~al.}(2019){Di Carlo}, {Giacobbo}, {Mapelli},
  {Pasquato}, {Spera}, {Wang}, \& {Haardt}}]{DiCarlo_2019}
{Di Carlo}, U.~N., {Giacobbo}, N., {Mapelli}, M., {et~al.} 2019, arXiv
  e-prints, arXiv:1901.00863

\bibitem[{{Duquennoy} \& {Mayor}(1991)}]{Duq_1991}
{Duquennoy}, A. \& {Mayor}, M. 1991, \aap, 248, 485

\bibitem[{{Eldridge} \& {Tout}(2004)}]{Eldridge_2004}
{Eldridge}, J.~J. \& {Tout}, C.~A. 2004, \mnras, 353, 87

\bibitem[{{Farr} {et~al.}(2017){Farr}, {Stevenson}, {Miller}, {Mandel}, {Farr},
  \& {Vecchio}}]{Farr_2017}
{Farr}, W.~M., {Stevenson}, S., {Miller}, M.~C., {et~al.} 2017, \nat, 548, 426

\bibitem[{{Fishbach} {et~al.}(2019){Fishbach}, {Farr}, \&
  {Holz}}]{Fishbach_2020}
{Fishbach}, M., {Farr}, W.~M., \& {Holz}, D.~E. 2019, arXiv e-prints,
  arXiv:1911.05882

\bibitem[{{Fragos} {et~al.}(2009){Fragos}, {Willems}, {Kalogera}, {Ivanova},
  {Rockefeller}, {Fryer}, \& {Young}}]{Fragos_2009}
{Fragos}, T., {Willems}, B., {Kalogera}, V., {et~al.} 2009, \apj, 697, 1057

\bibitem[{Fryer(1999)}]{Fryer_1999}
Fryer, C.~L. 1999, The Astrophysical Journal, 522, 413

\bibitem[{{Fryer}(2004)}]{Fryer_2004}
{Fryer}, C.~L. 2004, \apj, 601, L175

\bibitem[{{Fryer} {et~al.}(2012){Fryer}, {Belczynski}, {Wiktorowicz},
  {Dominik}, {Kalogera}, \& {Holz}}]{Fryer_2012}
{Fryer}, C.~L., {Belczynski}, K., {Wiktorowicz}, G., {et~al.} 2012, \apj, 749,
  91

\bibitem[{Fryer \& Kalogera(2001)}]{Fryer_2001}
Fryer, C.~L. \& Kalogera, V. 2001, The Astrophysical Journal, 554, 548

\bibitem[{{Fryer} \& {Kusenko}(2006)}]{Fryer_2006}
{Fryer}, C.~L. \& {Kusenko}, A. 2006, \apjs, 163, 335

\bibitem[{{Fryer} \& {Young}(2007)}]{Fryer_2007}
{Fryer}, C.~L. \& {Young}, P.~A. 2007, \apj, 659, 1438

\bibitem[{{Fujii} \& {Portegies Zwart}(2013)}]{Fujii_2013}
{Fujii}, M.~S. \& {Portegies Zwart}, S. 2013, \mnras, 430, 1018

\bibitem[{{Fuller} {et~al.}(2003){Fuller}, {Kusenko}, {Mocioiu}, \&
  {Pascoli}}]{Fuller_2003}
{Fuller}, G.~M., {Kusenko}, A., {Mocioiu}, I., \& {Pascoli}, S. 2003, \prd, 68,
  103002

\bibitem[{{Fuller} {et~al.}(2019){Fuller}, {Piro}, \& {Jermyn}}]{Fuller_2019}
{Fuller}, J., {Piro}, A.~L., \& {Jermyn}, A.~S. 2019, \mnras, 485, 3661

\bibitem[{{Gaburov} {et~al.}(2008){Gaburov}, {Lombardi}, \& {Portegies
  Zwart}}]{Gaburov_2008}
{Gaburov}, E., {Lombardi}, J.~C., \& {Portegies Zwart}, S. 2008, \mnras, 383,
  L5

\bibitem[{{Geller} {et~al.}(2019){Geller}, {Leigh}, {Giersz}, {Kremer}, \&
  {Rasio}}]{Geller_2019}
{Geller}, A.~M., {Leigh}, N. W.~C., {Giersz}, M., {Kremer}, K., \& {Rasio},
  F.~A. 2019, arXiv e-prints, arXiv:1902.00019

\bibitem[{{Gessner} \& {Janka}(2018)}]{Gessner_2018}
{Gessner}, A. \& {Janka}, H.-T. 2018, \apj, 865, 61

\bibitem[{{Giacobbo} {et~al.}(2018){Giacobbo}, {Mapelli}, \&
  {Spera}}]{Giacobbo_2018}
{Giacobbo}, N., {Mapelli}, M., \& {Spera}, M. 2018, \mnras, 474, 2959

\bibitem[{{Giersz} {et~al.}(2013){Giersz}, {Heggie}, {Hurley}, \&
  {Hypki}}]{Giersz_2013}
{Giersz}, M., {Heggie}, D.~C., {Hurley}, J.~R., \& {Hypki}, A. 2013, \mnras,
  431, 2184

\bibitem[{{Giesers} {et~al.}(2018){Giesers}, {Dreizler}, {Husser}, {Kamann},
  {Anglada Escud{\'e}}, {Brinchmann}, {Carollo}, {Roth}, {Weilbacher}, \&
  {Wisotzki}}]{Giesers_2018}
{Giesers}, B., {Dreizler}, S., {Husser}, T.-O., {et~al.} 2018, \mnras, 475, L15

\bibitem[{{Glebbeek} {et~al.}(2009){Glebbeek}, {Gaburov}, {de Mink}, {Pols}, \&
  {Portegies Zwart}}]{Glebbeek_2009}
{Glebbeek}, E., {Gaburov}, E., {de Mink}, S.~E., {Pols}, O.~R., \& {Portegies
  Zwart}, S.~F. 2009, \aap, 497, 255

\bibitem[{{Hamann} \& {Koesterke}(1998)}]{Hamann_1998}
{Hamann}, W.-R. \& {Koesterke}, L. 1998, \aap, 335, 1003

\bibitem[{{Hobbs} {et~al.}(2005){Hobbs}, {Lorimer}, {Lyne}, \&
  {Kramer}}]{Hobbs_2005}
{Hobbs}, G., {Lorimer}, D.~R., {Lyne}, A.~G., \& {Kramer}, M. 2005, \mnras,
  360, 974

\bibitem[{{Humphreys} \& {Davidson}(1994)}]{Humphreys_1994}
{Humphreys}, R.~M. \& {Davidson}, K. 1994, \pasp, 106, 1025

\bibitem[{Hurley {et~al.}(2000)Hurley, Pols, \& Tout}]{Hurley_2000}
Hurley, J.~R., Pols, O.~R., \& Tout, C.~A. 2000, Monthly Notices of the Royal
  Astronomical Society, 315, 543

\bibitem[{Hurley {et~al.}(2002)Hurley, Tout, \& Pols}]{Hurley_2002}
Hurley, J.~R., Tout, C.~A., \& Pols, O.~R. 2002, Monthly Notices of the Royal
  Astronomical Society, 329, 897

\bibitem[{{Hypki} \& {Giersz}(2013)}]{Hypki_2013}
{Hypki}, A. \& {Giersz}, M. 2013, \mnras, 429, 1221

\bibitem[{{Ivanova} {et~al.}(2013){Ivanova}, {Justham}, {Chen}, {De Marco},
  {Fryer}, {Gaburov}, {Ge}, {Glebbeek}, {Han}, {Li}, {Lu}, {Marsh},
  {Podsiadlowski}, {Potter}, {Soker}, {Taam}, {Tauris}, {van den Heuvel}, \&
  {Webbink}}]{Ivanova_2013}
{Ivanova}, N., {Justham}, S., {Chen}, X., {et~al.} 2013, \aapr, 21, 59

\bibitem[{{Janka}(2013)}]{Janka_2013}
{Janka}, H.-T. 2013, \mnras, 434, 1355

\bibitem[{{Joshi} {et~al.}(2000){Joshi}, {Rasio}, \& {Portegies
  Zwart}}]{Joshi_2000}
{Joshi}, K.~J., {Rasio}, F.~A., \& {Portegies Zwart}, S. 2000, \apj, 540, 969

\bibitem[{{Kimball} {et~al.}(2020){Kimball}, {Berry}, \&
  {Kalogera}}]{Kimball_2020}
{Kimball}, C., {Berry}, C., \& {Kalogera}, V. 2020, Research Notes of the
  American Astronomical Society, 4, 2

\bibitem[{{Kremer} {et~al.}(2018{\natexlab{a}}){Kremer}, {Chatterjee}, {Ye},
  {Rodriguez}, \& {Rasio}}]{Kremer_2018b}
{Kremer}, K., {Chatterjee}, S., {Ye}, C.~S., {Rodriguez}, C.~L., \& {Rasio},
  F.~A. 2018{\natexlab{a}}, ArXiv e-prints [\eprint[arXiv]{1808.02204}]

\bibitem[{{Kremer} {et~al.}(2018{\natexlab{b}}){Kremer}, {Ye}, {Chatterjee},
  {Rodriguez}, \& {Rasio}}]{Kremer_2018}
{Kremer}, K., {Ye}, C.~S., {Chatterjee}, S., {Rodriguez}, C.~L., \& {Rasio},
  F.~A. 2018{\natexlab{b}}, \apjl, 855, L15

\bibitem[{{Kroupa}(1995)}]{Kroupa_1995a}
{Kroupa}, P. 1995, \mnras, 277, 1491

\bibitem[{{Kroupa}(2001)}]{Kroupa_2001}
{Kroupa}, P. 2001, \mnras, 322, 231

\bibitem[{{Kruckow} {et~al.}(2018){Kruckow}, {Tauris}, {Langer}, {Kramer}, \&
  {Izzard}}]{Kruckow_2018}
{Kruckow}, M.~U., {Tauris}, T.~M., {Langer}, N., {Kramer}, M., \& {Izzard},
  R.~G. 2018, \mnras, 481, 1908

\bibitem[{{Kudritzki} \& {Reimers}(1978)}]{Kudritzki_1978}
{Kudritzki}, R.~P. \& {Reimers}, D. 1978, \aap, 70, 227

\bibitem[{{Lattimer} \& {Yahil}(1989)}]{Lattimer_1989}
{Lattimer}, J.~M. \& {Yahil}, A. 1989, \apj, 340, 426

\bibitem[{{Leigh} \& {Geller}(2013)}]{Leigh_2013}
{Leigh}, N.~W.~C. \& {Geller}, A.~M. 2013, \mnras, 432, 2474

\bibitem[{{Lombardi} {et~al.}(2002){Lombardi}, {Warren}, {Rasio}, {Sills}, \&
  {Warren}}]{Lombardi_2002}
{Lombardi}, James~C., J., {Warren}, J.~S., {Rasio}, F.~A., {Sills}, A., \&
  {Warren}, A.~R. 2002, \apj, 568, 939

\bibitem[{{Mandel}(2016)}]{Mandel_2016}
{Mandel}, I. 2016, \mnras, 456, 578

\bibitem[{{Mandel} \& {Farmer}(2017)}]{Mandel_2017}
{Mandel}, I. \& {Farmer}, A. 2017, \nat, 547, 284

\bibitem[{{Mapelli}(2016)}]{Mapelli_2016}
{Mapelli}, M. 2016, \mnras, 459, 3432

\bibitem[{{Mapelli} {et~al.}(2017){Mapelli}, {Giacobbo}, {Ripamonti}, \&
  {Spera}}]{Mapelli_2017}
{Mapelli}, M., {Giacobbo}, N., {Ripamonti}, E., \& {Spera}, M. 2017, \mnras,
  472, 2422

\bibitem[{{Marchant} {et~al.}(2016){Marchant}, {Langer}, {Podsiadlowski},
  {Tauris}, \& {Moriya}}]{Marchant_2016}
{Marchant}, P., {Langer}, N., {Podsiadlowski}, P., {Tauris}, T.~M., \&
  {Moriya}, T.~J. 2016, \aap, 588, A50

\bibitem[{{Meakin} \& {Arnett}(2006)}]{Meakin_2006}
{Meakin}, C.~A. \& {Arnett}, D. 2006, \apjl, 637, L53

\bibitem[{{Meakin} \& {Arnett}(2007)}]{Meakin_2007}
{Meakin}, C.~A. \& {Arnett}, D. 2007, \apj, 665, 690

\bibitem[{Morscher {et~al.}(2015)Morscher, Pattabiraman, Rodriguez, Rasio, \&
  Umbreit}]{Morscher_2015}
Morscher, M., Pattabiraman, B., Rodriguez, C., Rasio, F.~A., \& Umbreit, S.
  2015, The Astrophysical Journal, 800, 9

\bibitem[{{Nieuwenhuijzen} \& {de Jager}(1990)}]{Nieu_1990}
{Nieuwenhuijzen}, H. \& {de Jager}, C. 1990, \aap, 231, 134

\bibitem[{Nitadori \& Aarseth(2012)}]{Nitadori_2012}
Nitadori, K. \& Aarseth, S.~J. 2012, Monthly Notices of the Royal Astronomical
  Society, 424, 545

\bibitem[{{Park} {et~al.}(2017){Park}, {Kim}, {Lee}, {Bae}, \&
  {Belczynski}}]{Park_2017}
{Park}, D., {Kim}, C., {Lee}, H.~M., {Bae}, Y.-B., \& {Belczynski}, K. 2017,
  \mnras, 469, 4665

\bibitem[{{Plummer}(1911)}]{Plummer_1911}
{Plummer}, H.~C. 1911, \mnras, 71, 460

\bibitem[{Podsiadlowski {et~al.}(2004)Podsiadlowski, Langer, Poelarends,
  Rappaport, Heger, \& Pfahl}]{Podsiadlowski_2004}
Podsiadlowski, P., Langer, N., Poelarends, A. J.~T., {et~al.} 2004, The
  Astrophysical Journal, 612, 1044

\bibitem[{{Portegies Zwart} \& {McMillan}(2018)}]{PortegiesZwart2018}
{Portegies Zwart}, S. \& {McMillan}, S. 2018, {Astrophysical Recipes; The art
  of AMUSE}

\bibitem[{{Repetto} {et~al.}(2012){Repetto}, {Davies}, \&
  {Sigurdsson}}]{Repetto_2012}
{Repetto}, S., {Davies}, M.~B., \& {Sigurdsson}, S. 2012, \mnras, 425, 2799

\bibitem[{{Repetto} {et~al.}(2017){Repetto}, {Igoshev}, \&
  {Nelemans}}]{Repetto_2017}
{Repetto}, S., {Igoshev}, A.~P., \& {Nelemans}, G. 2017, \mnras, 467, 298

\bibitem[{{Repetto} \& {Nelemans}(2015)}]{Repetto_2015}
{Repetto}, S. \& {Nelemans}, G. 2015, \mnras, 453, 3341

\bibitem[{Rodriguez {et~al.}(2018)Rodriguez, Amaro-Seoane, Chatterjee, \&
  Rasio}]{Rodriguez_2018}
Rodriguez, C.~L., Amaro-Seoane, P., Chatterjee, S., \& Rasio, F.~A. 2018, Phys.
  Rev. Lett., 120, 151101

\bibitem[{Rodriguez {et~al.}(2016)Rodriguez, Chatterjee, \&
  Rasio}]{Rodriguez_2016}
Rodriguez, C.~L., Chatterjee, S., \& Rasio, F.~A. 2016, Physical Review D, 93

\bibitem[{{Samsing}(2018)}]{Samsing_2018}
{Samsing}, J. 2018, \prd, 97, 103014

\bibitem[{{Sana} {et~al.}(2013){Sana}, {de Koter}, {de Mink}, {Dunstall},
  {Evans}, {H{\'e}nault-Brunet}, {Ma{\'{\i}}z Apell{\'a}niz},
  {Ram{\'{\i}}rez-Agudelo}, {Taylor}, {Walborn}, {Clark}, {Crowther},
  {Herrero}, {Gieles}, {Langer}, {Lennon}, \& {Vink}}]{Sana_2013}
{Sana}, H., {de Koter}, A., {de Mink}, S.~E., {et~al.} 2013, \aap, 550, A107

\bibitem[{{Sana} \& {Evans}(2011)}]{Sana_2011}
{Sana}, H. \& {Evans}, C.~J. 2011, in IAU Symposium, Vol. 272, Active OB Stars:
  Structure, Evolution, Mass Loss, and Critical Limits, ed. C.~{Neiner},
  G.~{Wade}, G.~{Meynet}, \& G.~{Peters}, 474--485

\bibitem[{{Scheck} {et~al.}(2008){Scheck}, {Janka}, {Foglizzo}, \&
  {Kifonidis}}]{Scheck_2008}
{Scheck}, L., {Janka}, H.~T., {Foglizzo}, T., \& {Kifonidis}, K. 2008, \aap,
  477, 931

\bibitem[{{Scheck} {et~al.}(2004){Scheck}, {Plewa}, {Janka}, {Kifonidis}, \&
  {M{\"u}ller}}]{Scheck_2004}
{Scheck}, L., {Plewa}, T., {Janka}, H.~T., {Kifonidis}, K., \& {M{\"u}ller}, E.
  2004, \prl, 92, 011103

\bibitem[{{Spera} {et~al.}(2019){Spera}, {Mapelli}, {Giacobbo}, {Trani},
  {Bressan}, \& {Costa}}]{Spera_2019}
{Spera}, M., {Mapelli}, M., {Giacobbo}, N., {et~al.} 2019, \mnras, 485, 889

\bibitem[{{Spitzer}(1987)}]{1987degc.book.....S}
{Spitzer}, L. 1987, {Dynamical evolution of globular clusters, Princeton
  University Press, Princeton, NJ, 191 p.}

\bibitem[{{Spruit}(2002)}]{Spruit_2002}
{Spruit}, H.~C. 2002, \aap, 381, 923

\bibitem[{{Stevenson} {et~al.}(2017){Stevenson}, {Vigna-G{\'o}mez}, {Mandel},
  {Barrett}, {Neijssel}, {Perkins}, \& {de Mink}}]{Mandel_2017a}
{Stevenson}, S., {Vigna-G{\'o}mez}, A., {Mandel}, I., {et~al.} 2017, Nature
  Communications, 8, 14906

\bibitem[{{Timmes} {et~al.}(1996){Timmes}, {Woosley}, \&
  {Weaver}}]{Timmes_1996}
{Timmes}, F.~X., {Woosley}, S.~E., \& {Weaver}, T.~A. 1996, \apj, 457, 834

\bibitem[{{Toonen} {et~al.}(2014){Toonen}, {Claeys}, {Mennekens}, \&
  {Ruiter}}]{Toonen_2014}
{Toonen}, S., {Claeys}, J.~S.~W., {Mennekens}, N., \& {Ruiter}, A.~J. 2014,
  \aap, 562, A14

\bibitem[{{Vassiliadis} \& {Wood}(1993)}]{Vassiliadis_1993}
{Vassiliadis}, E. \& {Wood}, P.~R. 1993, \apj, 413, 641

\bibitem[{Vink \& de~Koter(2005)}]{Vink_2005}
Vink, J.~S. \& de~Koter, A. 2005, Astronomy and Astrophysics, 442, 587

\bibitem[{Vink {et~al.}(2001)Vink, de~Koter, \& Lamers}]{Vink_2001}
Vink, J.~S., de~Koter, A., \& Lamers, H. J. G. L.~M. 2001, Astronomy and
  Astrophysics, 369, 574

\bibitem[{{Voss} \& {Tauris}(2003)}]{Voss_2003}
{Voss}, R. \& {Tauris}, T.~M. 2003, \mnras, 342, 1169

\bibitem[{{Wang} {et~al.}(2020){Wang}, {Kroupa}, {Takahashi}, \&
  {Jerabkova}}]{Wang_2020}
{Wang}, L., {Kroupa}, P., {Takahashi}, K., \& {Jerabkova}, T. 2020, \mnras,
  491, 440

\bibitem[{Wang {et~al.}(2016)Wang, Spurzem, Aarseth, Giersz, Askar, Berczik,
  Naab, Schadow, \& Kouwenhoven}]{Wang_2016}
Wang, L., Spurzem, R., Aarseth, S., {et~al.} 2016, Mon. Not. R. Astron. Soc.,
  458, 1450

\bibitem[{Wang {et~al.}(2015)Wang, Spurzem, Aarseth, Nitadori, Berczik,
  Kouwenhoven, \& Naab}]{Wang_2015}
Wang, L., Spurzem, R., Aarseth, S., {et~al.} 2015, Monthly Notices of the Royal
  Astronomical Society, 450, 4070

\bibitem[{{Willems} {et~al.}(2005){Willems}, {Henninger}, {Levin}, {Ivanova},
  {Kalogera}, {McGhee}, {Timmes}, \& {Fryer}}]{Willems_2005}
{Willems}, B., {Henninger}, M., {Levin}, T., {et~al.} 2005, \apj, 625, 324

\bibitem[{{Woosley}(2017)}]{Woosley_2017}
{Woosley}, S.~E. 2017, \apj, 836, 244

\bibitem[{{Ziosi} {et~al.}(2014){Ziosi}, {Mapelli}, {Branchesi}, \&
  {Tormen}}]{Ziosi_2014}
{Ziosi}, B.~M., {Mapelli}, M., {Branchesi}, M., \& {Tormen}, G. 2014, \mnras,
  441, 3703

\end{thebibliography}

\begin{table*}
\caption{The formation and retention of stellar-remnant BHs in
a $\mcl(0)\approx5.0\times10^4\Ms$ model cluster (Sec.~\ref{nbcode}) as a function
of metallicity (column 1) and SN/remnant-formation scheme (column 2;
Sec.~\ref{newrem}; ``+MB'' implies the cluster model includes a population
of massive primordial binaries as described in Sec.~\ref{primbin}). The columns
3, 4, 5, and 6 give the fraction, $\fbhm$, of the cluster's initial stellar mass
that is converted
into BHs, the fraction, $\fretbhm$, of them that are retained in the cluster due
to low-enough natal kicks, the fraction, $\fbhn$, of the cluster's initial
number of stars that is converted into BHs, and the fraction, $\fretbhn$,
of them that are retained in the cluster, respectively. In this table,
$\fretbhm$ and $\fretbhn$ correspond to the standard, fallback-controlled natal kicks
(Sec.~\ref{stdkick}) --- Sec.~\ref{altkick} discusses how the retentions are affected
by alternative natal-kick prescriptions, namely, the convection-asymmetry-driven
kick results in nearly the same retention fractions, the collapse-asymmetry-driven
kick leads to all retention fractions $\approx1.0$ and the neutrino-driven kick
leads to all retention fractions $\approx0.0$. Such BH retention fractions
are representatives of those in a typical young massive or an open cluster
or a low-mass GC 
in the Milky Way or a local-group galaxy.}
\label{tab_retfrac}
\centering
\begin{tabular}{llcccc}
\hline
Metallicity $Z$ & Remnant scheme & $\fbhm/10^{-2}$ & \fretbhm & $\fbhn/10^{-3}$ & \fretbhn\\
\hline
0.0001          & F12-rapid+B16-PPSN/PSN & 6.130  & 0.771   & 1.833   & 0.656   \\
0.0001          & F12-rapid              & 6.562  & 0.786   & 1.844   & 0.658   \\
0.0001          & F12-delayed            & 6.973  & 0.601   & 2.580  & 0.353   \\
0.0001          & B08                    & 7.493  & 0.835   & 2.055  & 0.682   \\
\hline
0.002          & F12-rapid+B16-PPSN/PSN & 4.999  & 0.777  & 1.751  & 0.633    \\
0.002          & F12-rapid              & 5.018  & 0.777  & 1.751  & 0.633   \\
0.002          & F12-delayed            & 5.334  & 0.566  & 2.533  & 0.318   \\
0.002          & B08                    & 5.604  & 0.808  & 1.938  & 0.675   \\
\hline
0.02          & F12-rapid              & 2.117  & 0.594  & 1.564  & 0.530   \\
0.02          & F12-delayed            & 1.944  & 0.185  & 2.055  & 0.074   \\
0.02          & B08                    & 2.220  & 0.538  & 1.716  & 0.374   \\
\hline
0.0001        & F12-rapid+B16-PPSN/PSN (+MB) & 6.152 & 0.793  & 1.167  & 0.700   \\
0.0001        & F12-rapid (+MB)              & 7.418 & 0.813  & 1.249 & 0.710   \\
0.0001        & F12-delayed (+MB)            & 7.563 & 0.736  & 1.447  & 0.564   \\
0.0001        & B08 (+MB)                    & 7.992  & 0.881  & 1.296 & 0.793   \\
\hline
\end{tabular}
\end{table*}

\begin{table*}
\caption{Remnant outcomes of selected massive binaries when evolved with
$\startrack$ (column 5) and updated $\bse$ (column 6). Columns 1-4 give, respectively,
the binaries' primary (the more massive member) ZAMS mass, $\mzamsa$,
secondary ZAMS mass, $\mzamsb$, initial semi-major-axis, $a_0$, and initial eccentricity, $e_0$.
When the outcome of the binary evolution is two compact remnants, their
masses are indicated across a `$+$' symbol in columns 5 and 6, the left and the right
side mass being the derivative of the primary and the secondary stellar member respectively.
If the two remnants remain bound, then the semi-major-axis ($\Rs$) and eccentricity of
the double-compact binary are also indicated at selected evolutionary times (indicated
in parentheses). If the binary becomes detached in the course of evolution (due to, \eg,
SN natal kick), then only the remnant mass pair is indicated. If the binary evolution
leads to only one remnant (due to star-star or star-remnant merger), then just
this remnant's mass is indicated. Due to the adopted NS upper mass limit,
a remnant of $>2.5\Ms$ ($\leq2.5\Ms$) is a BH (NS). These $\bse$ and $\startrack$
evolutions are done for the F12-rapid+B16-PPSN/PSN remnant-mass
model with the same choices of evolutionary parameters, the
values of the most relevant ones being $Z=0.0001$, $\ace=1.0$, and
$\fmrg=0.5$.}
\label{tab_binevol}
\centering
\begin{tabular}{ccccll}
\hline
$\mzamsa/\Ms$ & $\mzamsb/\Ms$ & $a_0/\Rs$ & $e_0$ & \startrack evolution & \bse evolution\\
\hline
118.5 & 25.5 & 73.6    & 0.73 & $40.5\Ms$        & $40.5\Ms$ \\
\hline
138.5 & 20.0 & 3806.2  & 0.28 & PSN (no remnant) & $41.0\Ms+1.6\Ms$ \\
\hline
82.3  & 80.0 & 38.8    & 0.58 & $40.5\Ms$        & $40.5\Ms$ \\
\hline
48.8  & 38.9 & 48.7    & 0.00 & $40.9\Ms$        & $51.2\Ms$ \\
\hline
39.4  & 17.5 & 26.5    & 0.46 & $17.8\Ms$        & $1.7\Ms + 18.3\Ms$ \\
\hline
27.2  & 17.0 & 288.1   & 0.57 & $29.3\Ms$        & $6.3\Ms + 29.8\Ms$ \\
\hline
60.6  & 28.0 & 63.0    & 0.60 & $29.3\Ms$        & $22.2\Ms + 23.4\Ms$, $19.1\Rs$, $0.06$ ($t=13{\rm~Gyr}$) \\
\hline
33.9  & 24.5 & 270.7   & 0.00 & $17.0\Ms$        & $10.8\Ms + 15.8\Ms$, $8.5\Rs$, $0.07$ ($t=9.8{\rm~Myr}$) \\
	&        &           &      &                  & BBH merger ($t=193.3{\rm~Myr}$) \\
\hline
68.7  & 23.1 & 54.2    & 0.78 & $33.3\Ms$        & $31.8\Ms$ \\
\hline
96.6  & 21.3 & 4126.7  & 0.21 & $39.6\Ms + 1.28\Ms$ & $37.6\Ms + 1.8\Ms$ \\
\hline
32.4  & 19.6 & 67.1    & 0.00 & $22.7\Ms$        & $8.5\Ms + 14.4\Ms$ \\
\hline
84.7  & 57.3 & 56.1    & 0.33 & $59.2\Ms$        & $32.2\Ms + 40.5\Ms$, $8.4\Rs$, $0.06$ ($t=5.8{\rm~Myr}$) \\
	&        &           &      &                  & BBH merger ($t=14.1{\rm~Myr}$) \\ 
\hline
\end{tabular}
\end{table*}

\begin{figure*}
\vspace{-0.6 cm}
\centering
\includegraphics[width=7.5cm,angle=0]{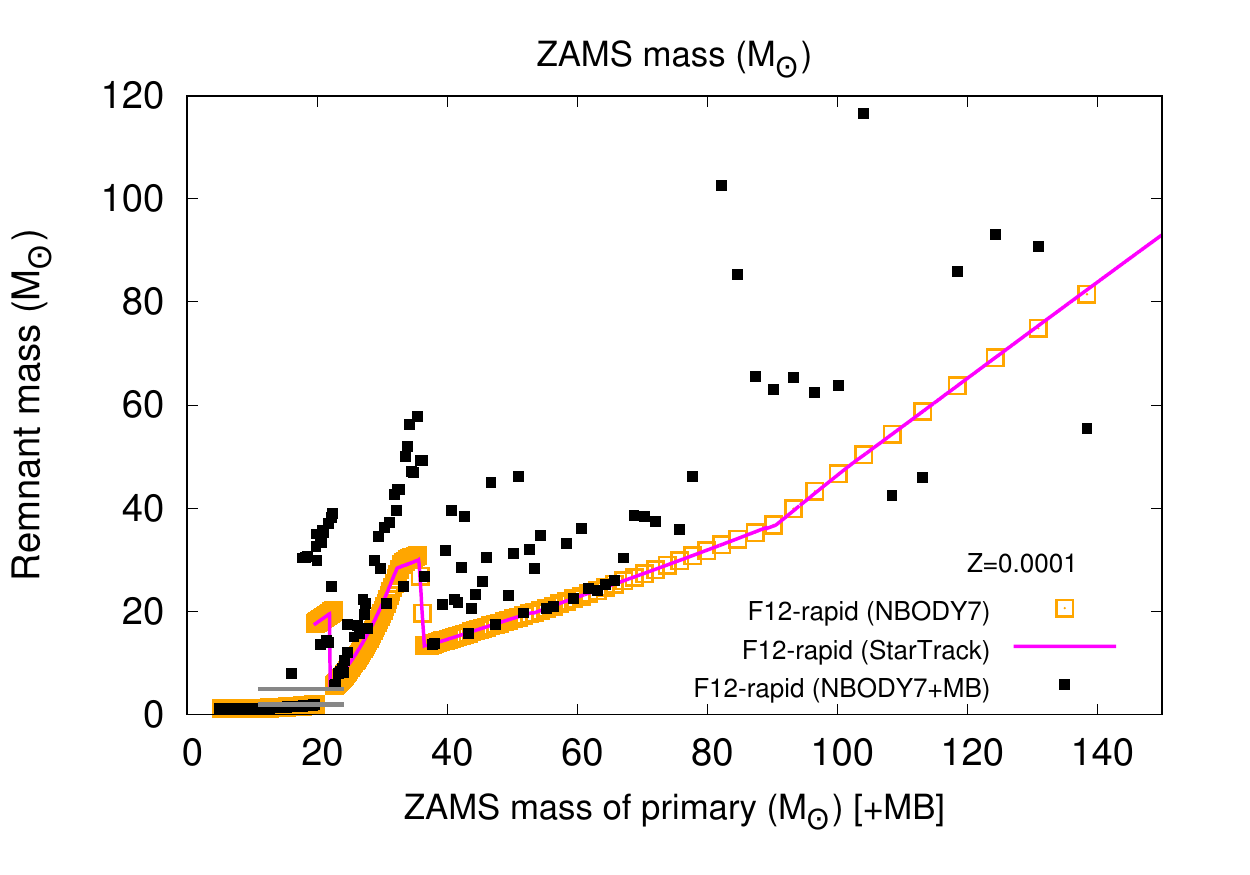}
\includegraphics[width=7.5cm,angle=0]{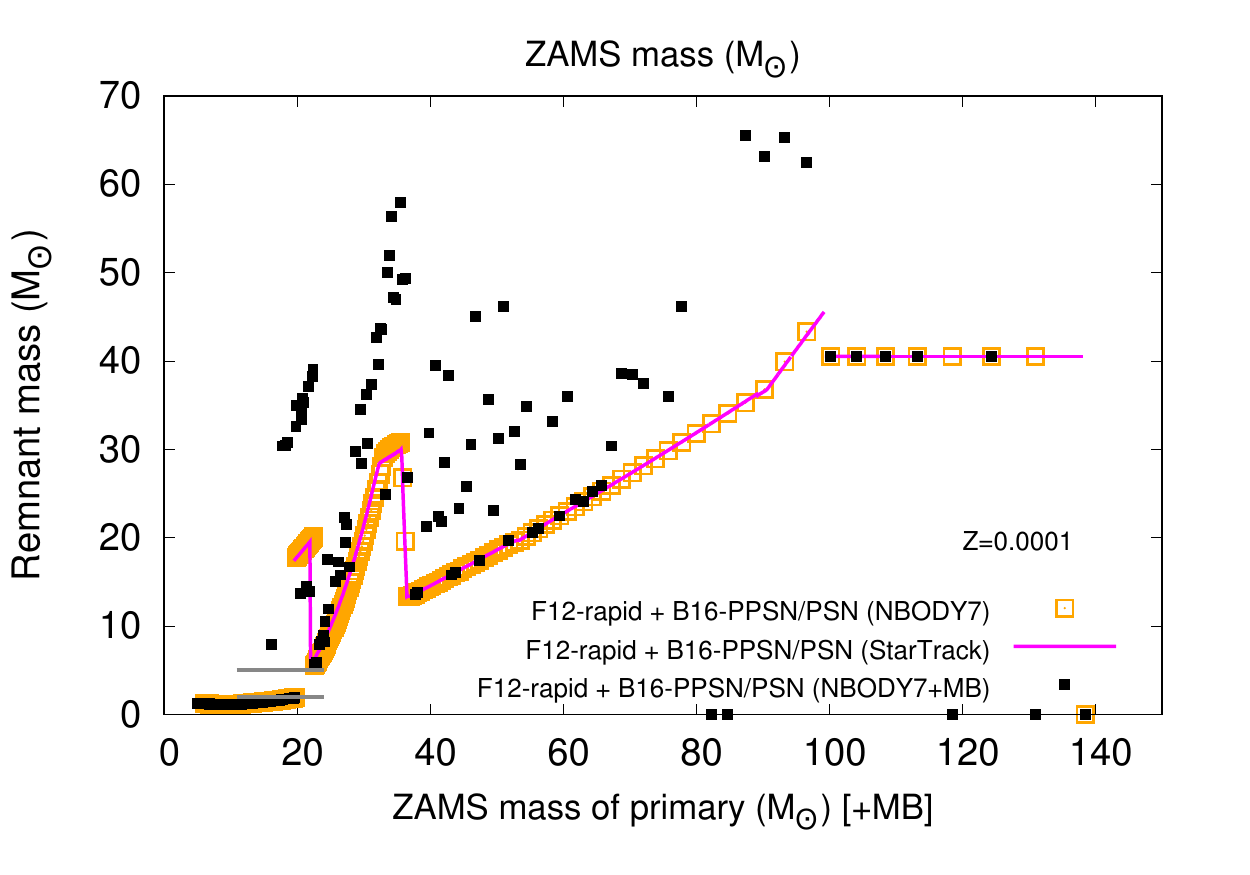}\\
\includegraphics[width=7.5cm,angle=0]{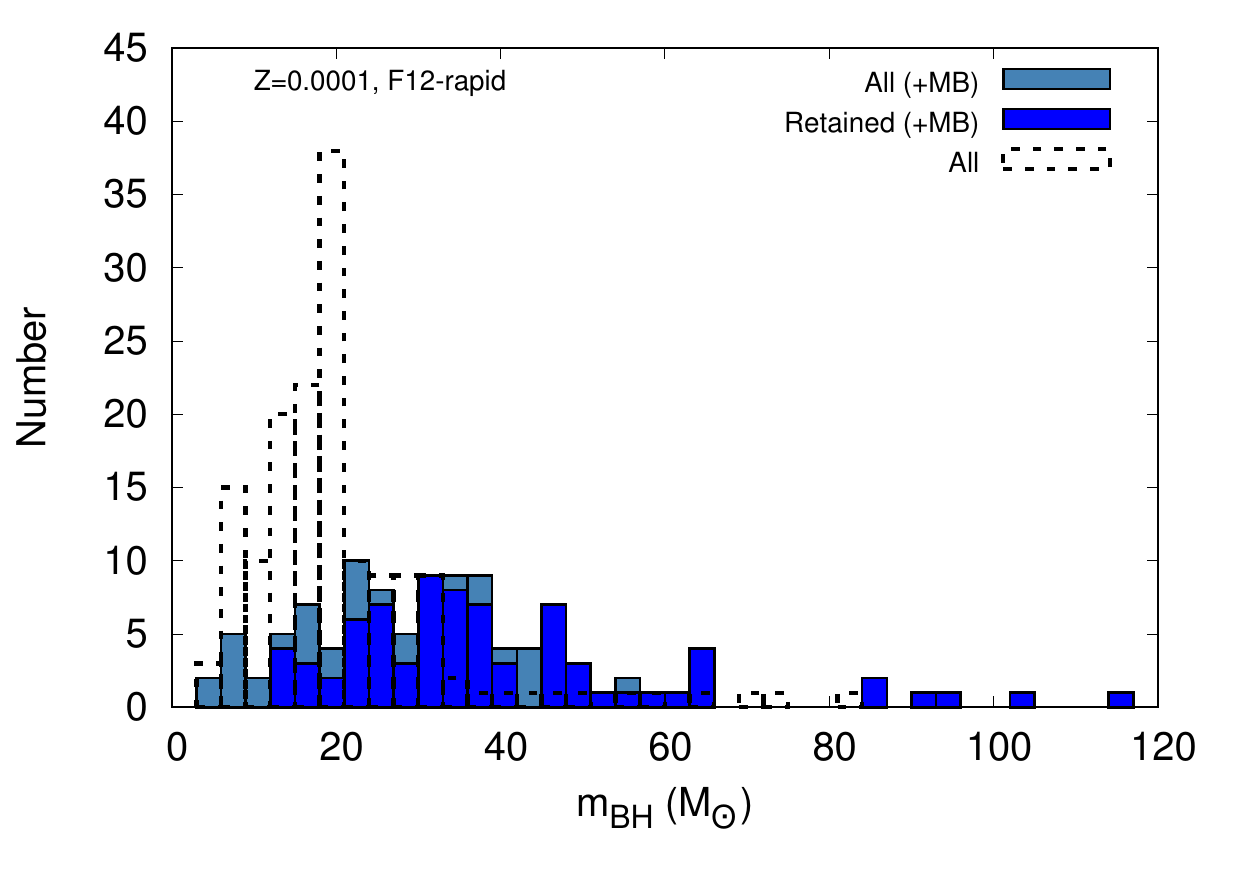}
\includegraphics[width=7.5cm,angle=0]{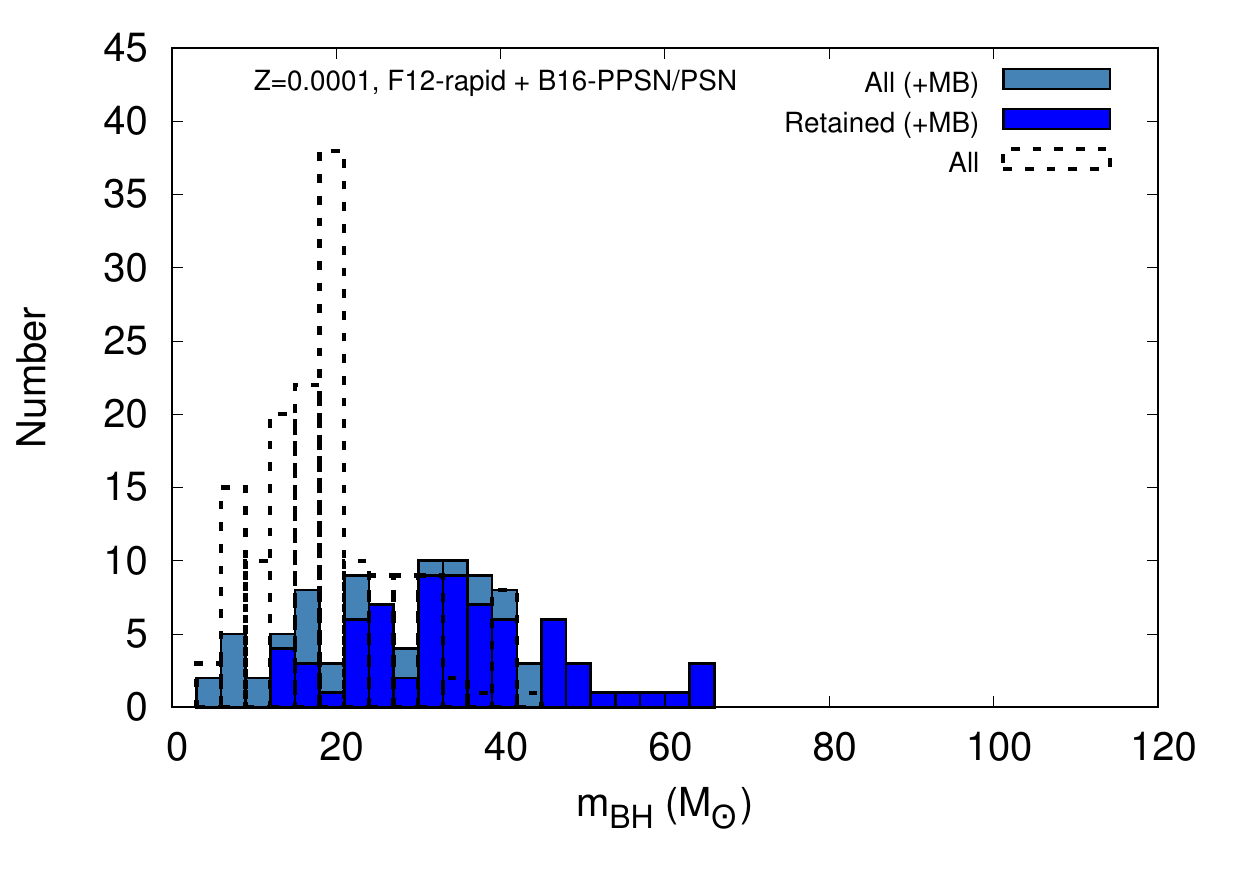}\\
\includegraphics[width=7.5cm,angle=0]{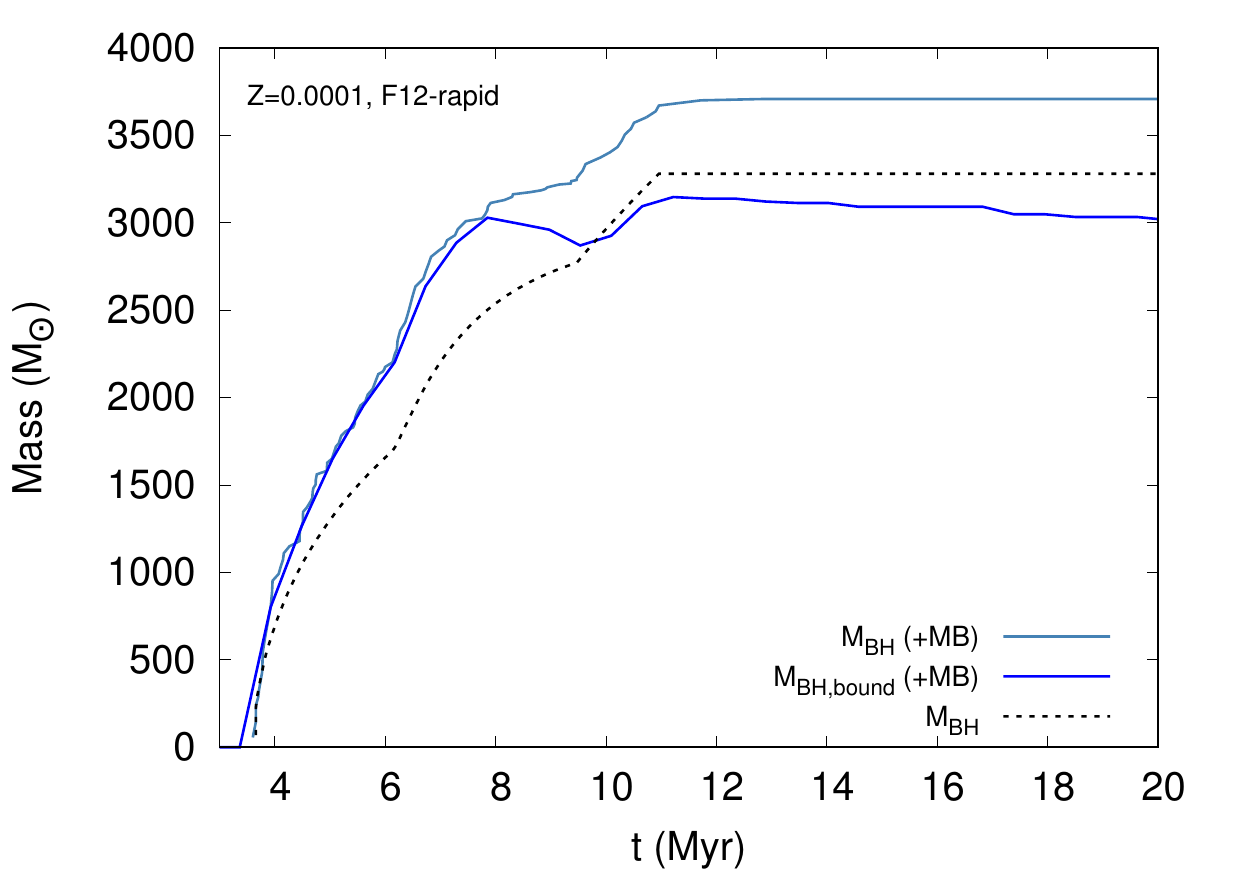}
\includegraphics[width=7.5cm,angle=0]{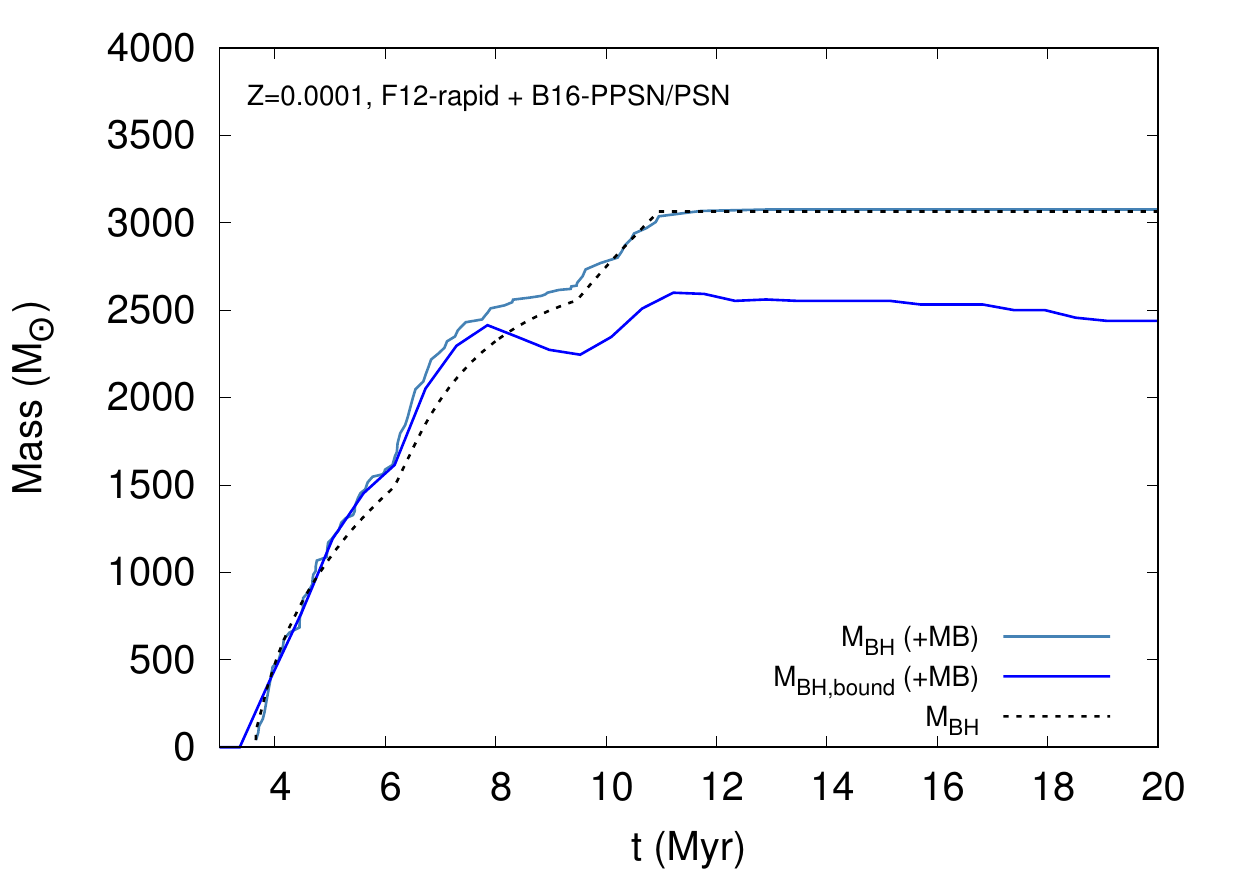}\\
\includegraphics[width=7.5cm,angle=0]{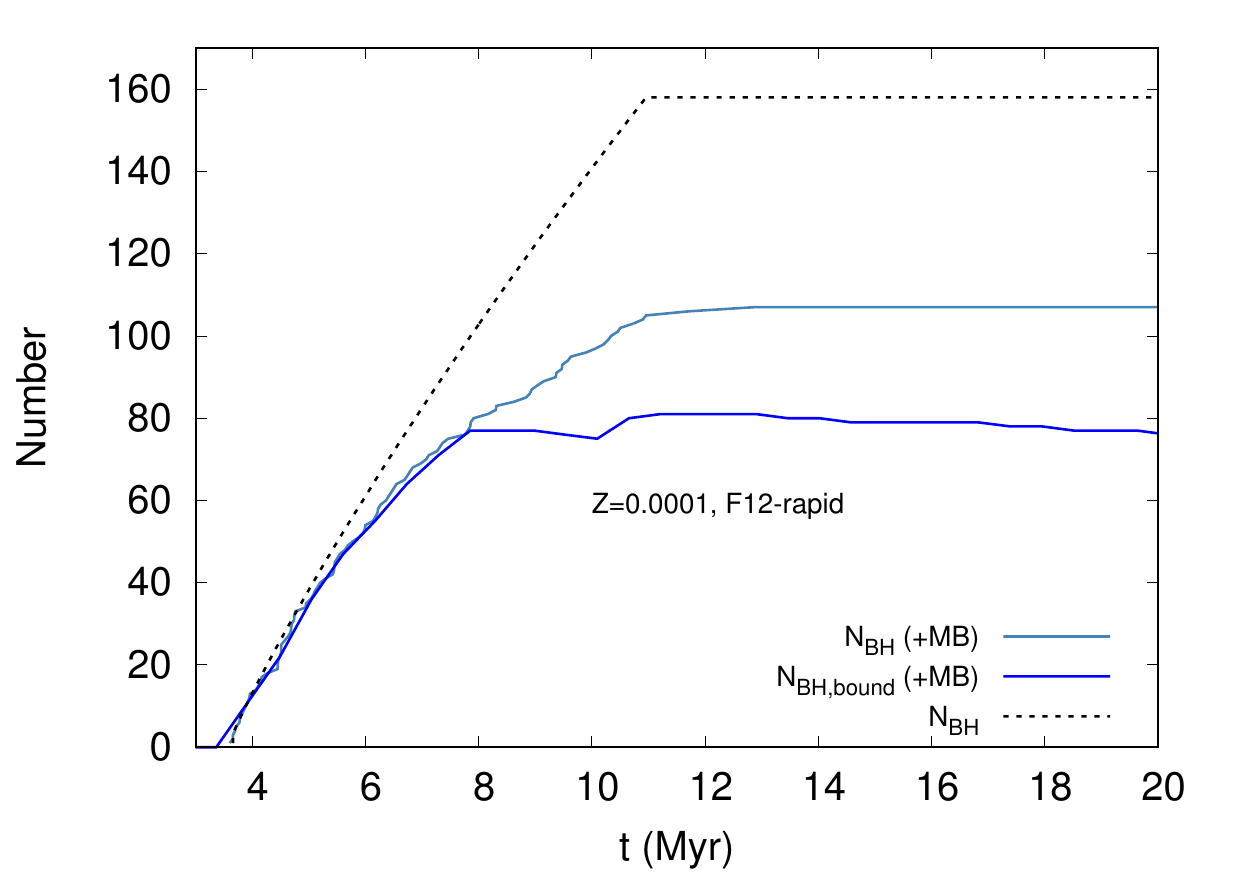}
\includegraphics[width=7.5cm,angle=0]{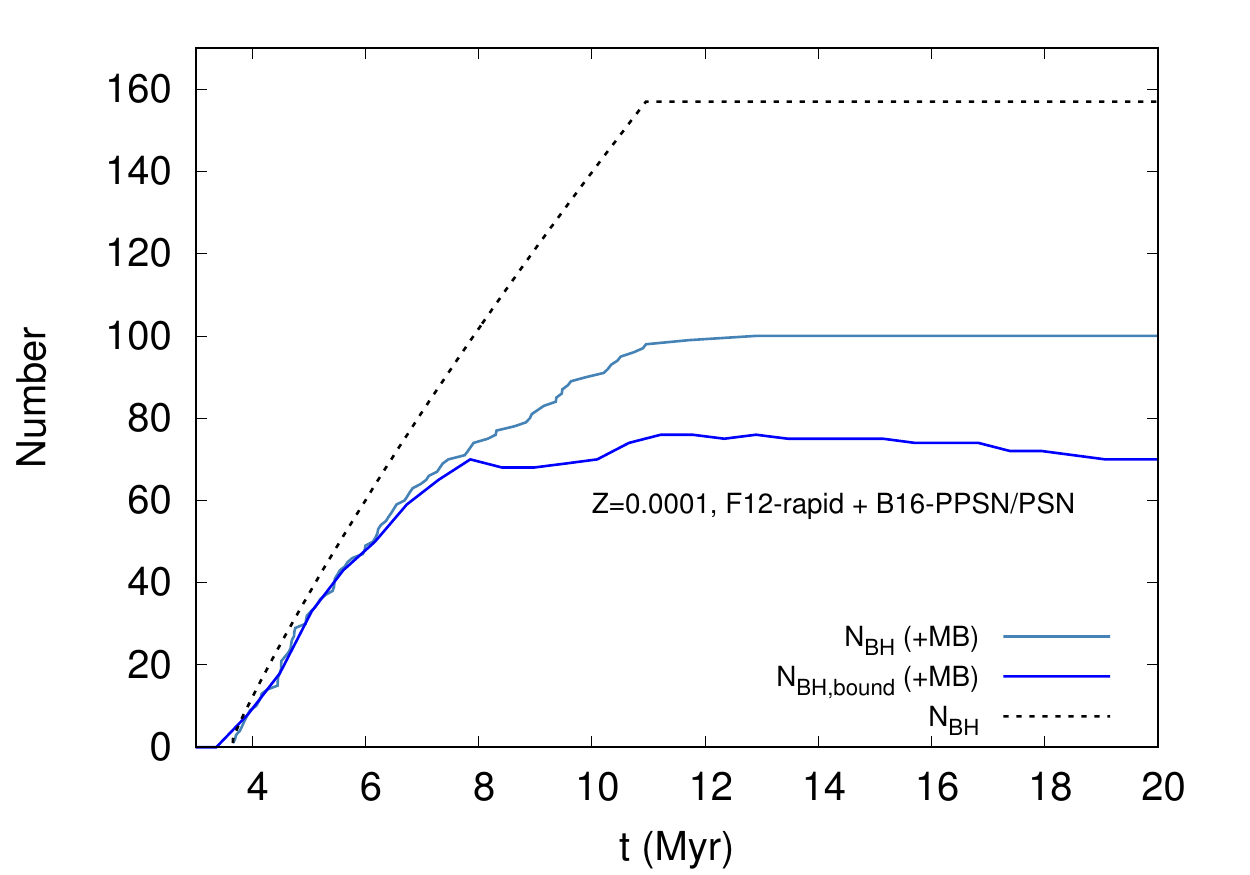}
\vspace{-0.2 cm}
\caption{The remnant production and retention in the $\nbseven$-computed, $\mcl(0)\approx5.0\times10^4\Ms$,
$\rh(0)\approx2$ pc model cluster with all BH-progenitor stars paired in primordial binaries
(see Sec.~\ref{primbin}).
In all the panels, ``+MB'' indicates the presence of such a population of massive primordial
binaries; the absence of this symbol in a legend implies the case where
all stars are initially single. The F12-rapid remnant-formation prescription and $Z=0.0001$
is assumed without (left column) and with (right column) PPSN/PSN.
{\bf Top row:} ZAMS mass vs. remnant mass (black, filled squares) compared with the corresponding
$\startrack$ outcomes from pure single-star evolution (blue, solid lines;
\cf Fig.~\ref{fig:cmp_nb}). For the
BH progenitors that have undergone a star-star merger before the BH formation, the ZAMS mass
of the primary (the more massive of the members participating in the star-star merger, at the
time of the merger) is plotted in the abscissa. For reference, the ZAMS mass-remnant mass
points (orange, empty squares) from an identical model cluster but with initially only
single stars (as in Sec.~\ref{nbcode}; \cf Fig.~\ref{fig:cmp_nb}) are also shown.
{\bf Second row:} the BHs' natal and retained mass distributions (steel blue and
blue filled histograms respectively; assuming the standard,
fallback-controlled natal kicks). They are compared with the corresponding
BH natal mass distribution without
any massive primordial binaries (dashed, empty histogram \cf Fig.~\ref{fig:bhmass_cmp2}).
{\bf Third row:} the cumulative time development of the total BH mass, $M_{\rm BH}$,
as they are born (steel blue, solid line) and the total BH mass, $\mbhbound$, that
is bound to the cluster as a function of time (blue, solid line). The $M_{\rm BH}(t)$
corresponding to initially only single stars is also shown (black, dashed line).
{\bf Fourth row:} the same as the third row but for the BH numbers $N_{\rm BH}$
and $\nbhbound$.}
\label{fig:bhwbin}
\end{figure*}

\begin{figure*}
\centering
\includegraphics[width=9.0cm,angle=0]{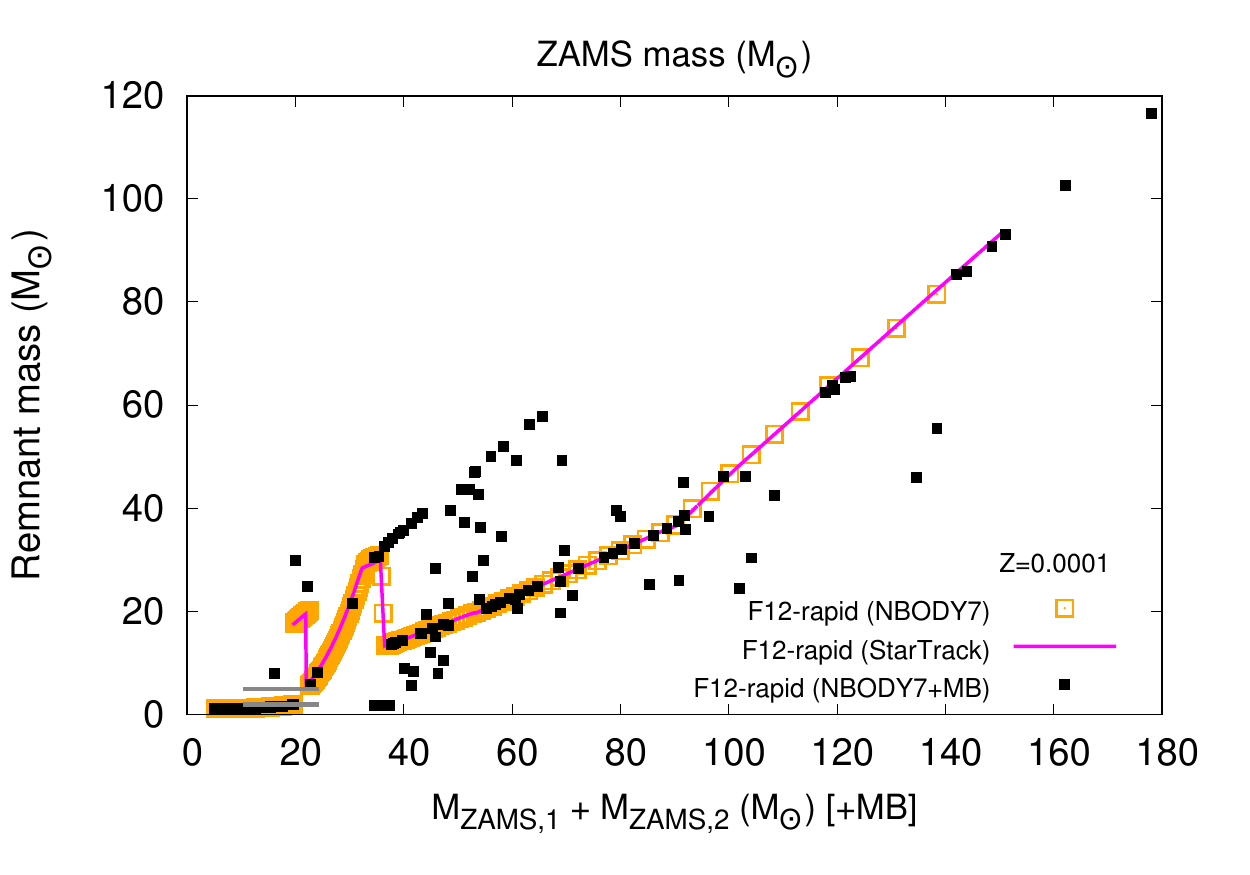}
\includegraphics[width=9.0cm,angle=0]{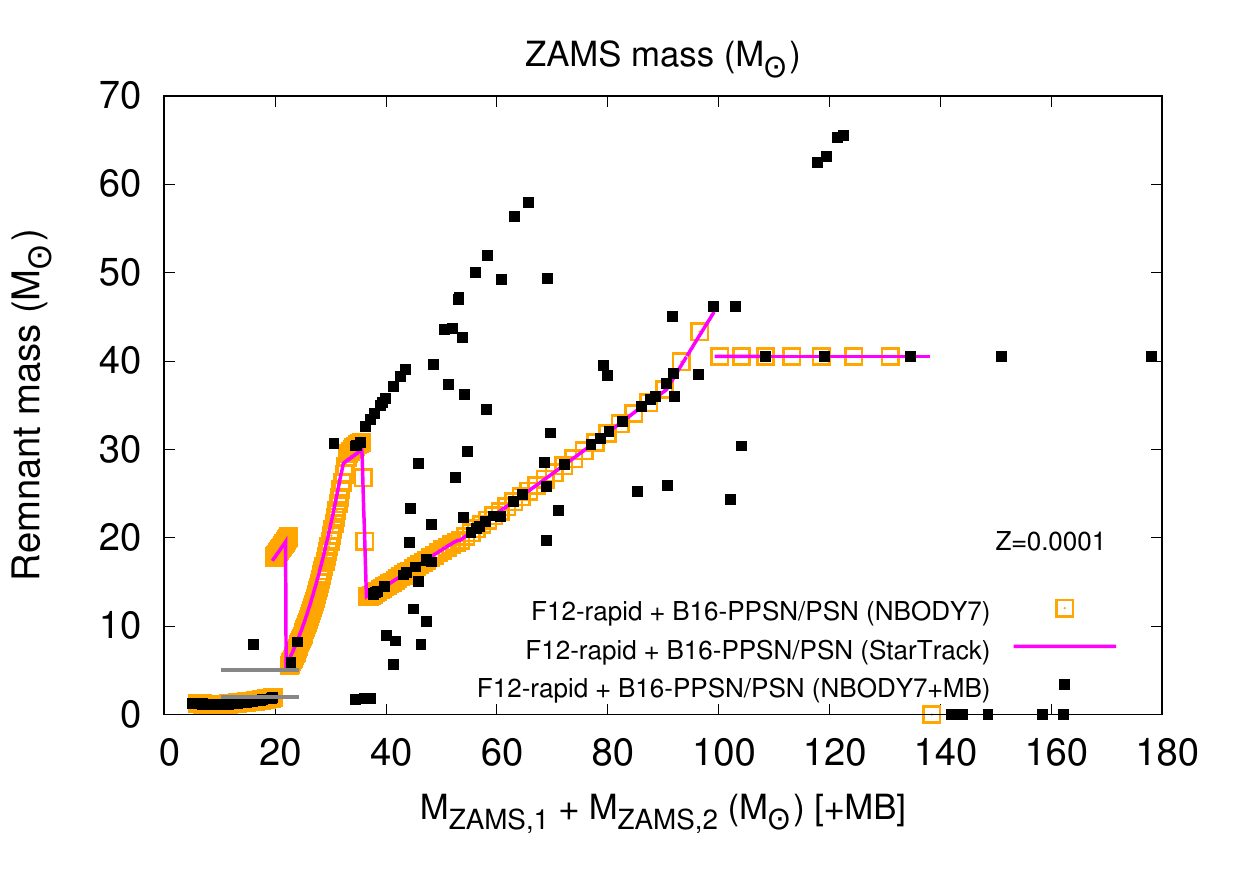}
\caption{The same as in the top row of Fig.~\ref{fig:bhwbin} except that for the model
cluster with massive primordial binaries (see Sec.~\ref{primbin}), the ZAMS
mass of the primary plus the ZAMS mass of the secondary of the pre-remnant-formation, star-star
mergers are plotted along the abscissa. The legends are the same as in Fig.~\ref{fig:bhwbin}.}
\label{fig:bhwbin2}
\end{figure*}

\begin{figure*}
\centering
\includegraphics[width=6.4cm,angle=0]{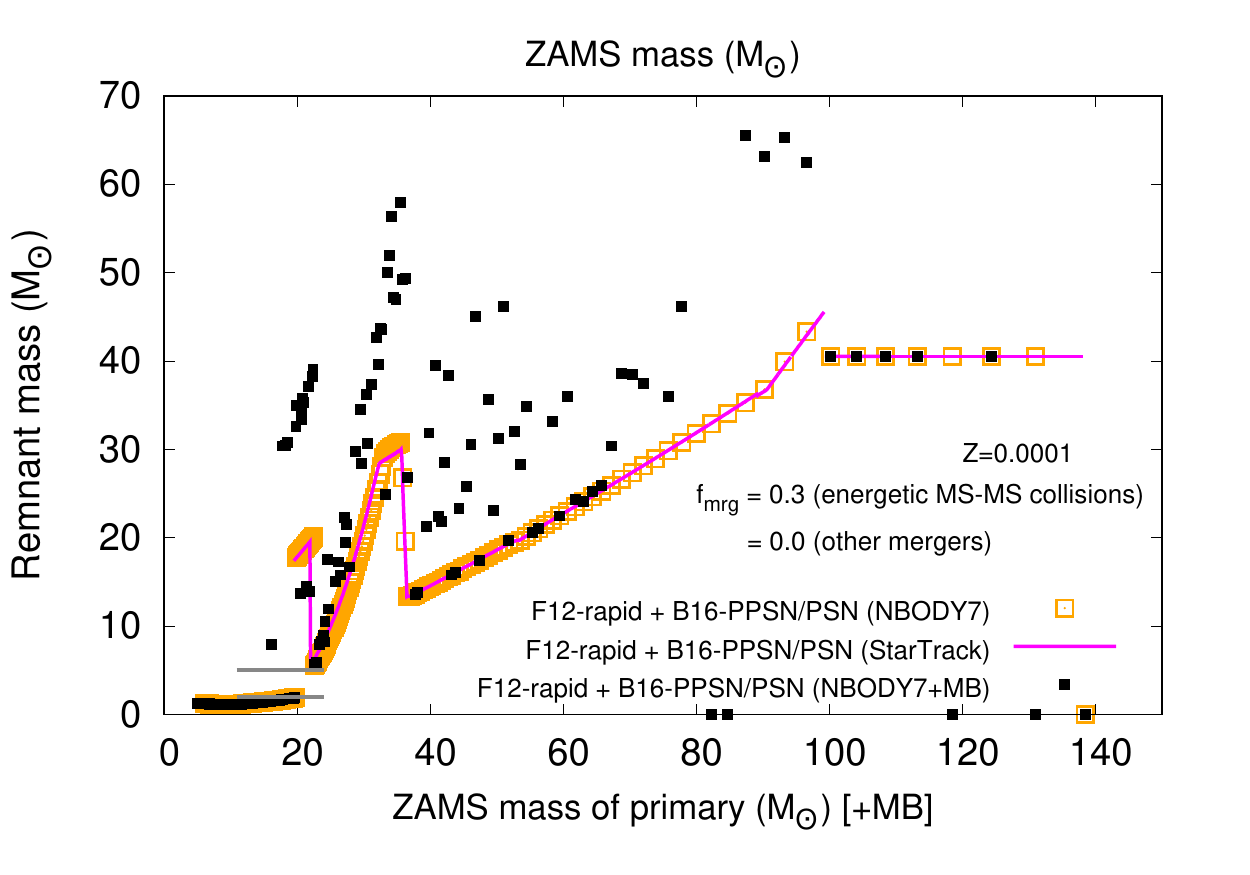}
\hspace{-0.6 cm}
\includegraphics[width=6.4cm,angle=0]{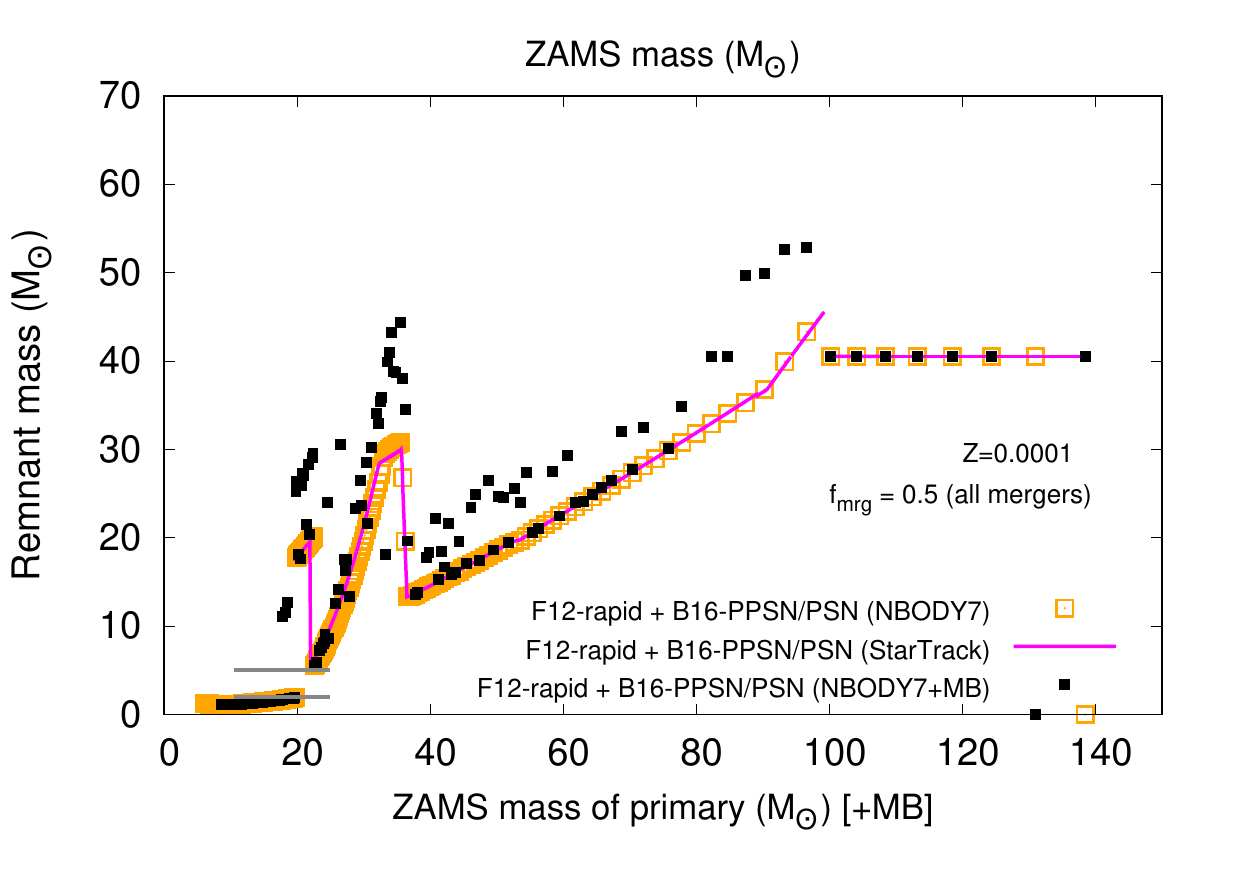}
\hspace{-0.6 cm}
\includegraphics[width=6.4cm,angle=0]{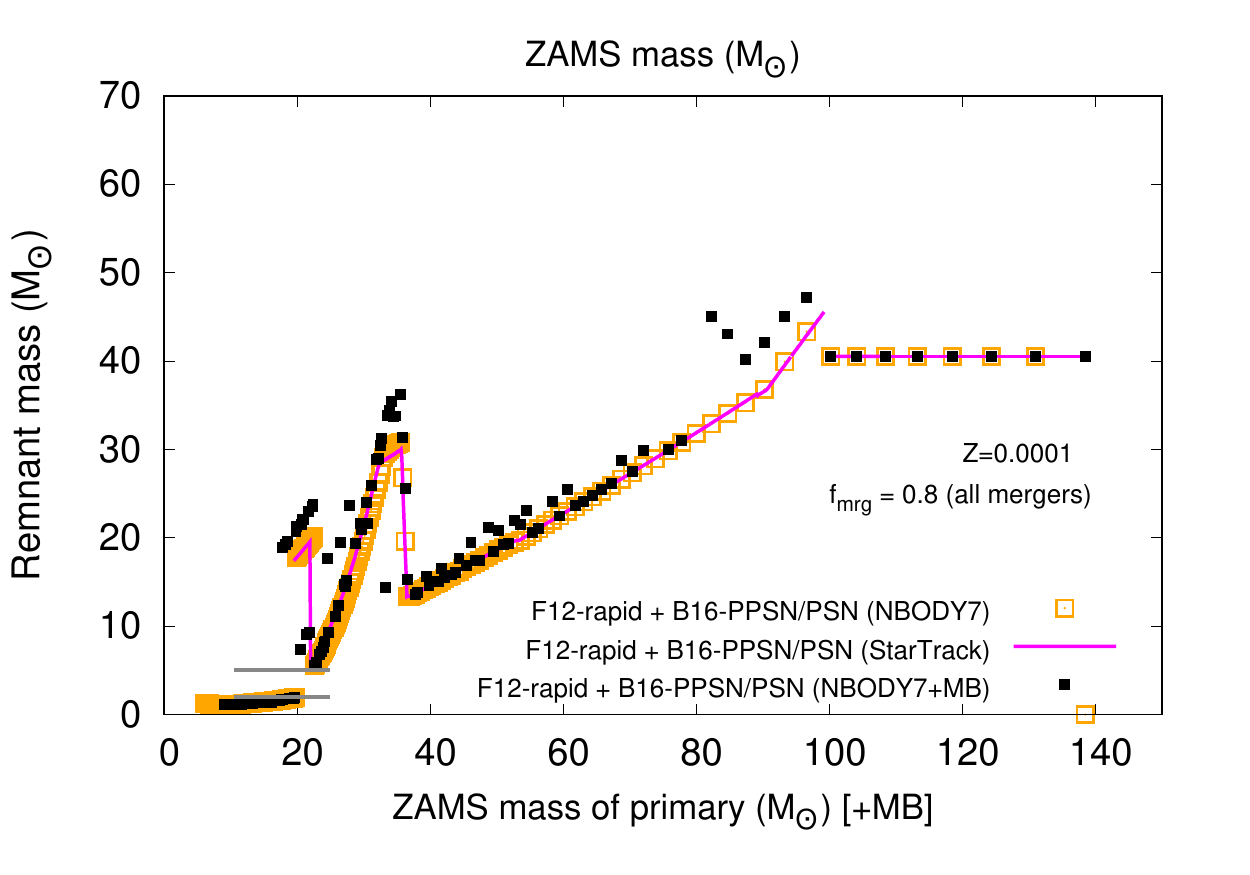}
\caption{The remnant production and retention in the $\nbseven$-computed, $\mcl(0)\approx5.0\times10^4\Ms$,
$\rh(0)\approx2$ pc model cluster with all BH-progenitor stars paired in primordial binaries
(see Sec.~\ref{primbin}). The legends are the same as in the top panels of Fig.~\ref{fig:bhwbin}.
The panels show different computations that adopt different extents of mass loss during a star-star merger
process: the left panel is for the case that is default in $\nbseven/\bse$, namely,
mass loss fraction of $\fmrg=0.3$ from the secondary for energetic MS-MS collisions only
(see text) and zero otherwise (the data in this panel are identical to those in the top-right
panel of Fig.~\ref{fig:bhwbin}), the middle panel is for the \nbseven computation where a mass loss fraction
of $\fmrg=0.5$ from the secondary is implemented for all star-star mergers, and the right
panel is for the computation where a mass loss fraction
of $\fmrg=0.8$ from the secondary is implemented for all star-star mergers. With
increasing extent of mass loss in mergers (left to right panel), the single-star ZAMS mass-remnant mass
relation (orange, empty squares) gets affected by massive-stellar mergers (black, filled squares)
to a diminishing extent. The F12-rapid remnant-formation prescription, including PPSN/PSN,
and $Z=0.0001$ is assumed in this demonstration.}
\label{fig:bhwbin3}
\end{figure*}

\begin{figure*}
\centering
\includegraphics[width=13.0cm,angle=0]{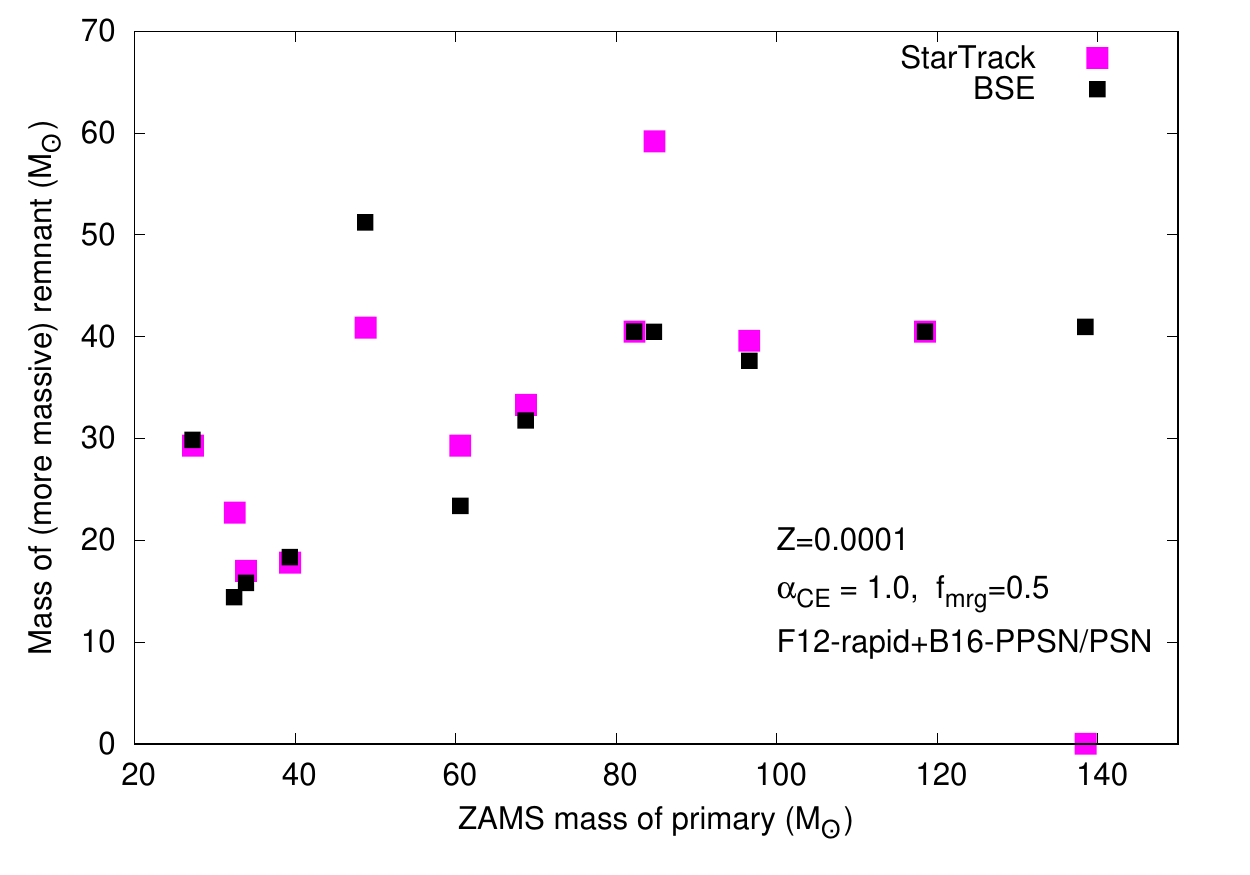}
\caption{The remnant outcomes of 12 massive binaries, with initial
parameters as given in Table~\ref{tab_binevol}, when evolved
with updated $\bse$ (black, filled squares) and $\startrack$ (magenta, filled squares).
Plotted along the X-axis are the binaries' primary (the more massive member) ZAMS masses
and the masses of the single remnant or of the heavier of the two remnants of
the binary evolutions are plotted along the Y-axis.
Due to the adopted NS upper mass limit,
a remnant of $>2.5\Ms$ ($\leq2.5\Ms$) is a BH (NS).
These $\bse$ and $\startrack$
evolutionary calculations are done for the F12-rapid+B16-PPSN/PSN remnant-mass
model with the same choices of evolutionary parameters, the
values of the most relevant ones being $Z=0.0001$, $\ace=1.0$, and
$\fmrg=0.5$.}
\label{fig:bevcomp}
\end{figure*}

\appendix

\onecolumn 

\section{Summary of the changes implemented in the updated $\bse$ code}\label{changes}

Below, we summarize the changes in the various subroutines of the
updated $\bse$ described in this work
with respect to the current publicly-available version of this program.

\subsection*{\tt mlwind.f}

Significant reorganization of the original {\tt mlwind} function is done
to implement the B10 wind recipe correctly; see Sec.~\ref{newwind}
for the details. The original \citet{Hurley_2000} wind and
various variants of both B10 and \citet{Hurley_2000} wind recipes can
be activated by appropriate choice of the {\tt mdflag} flag
variable. {\tt mdflag}$=3$ selects the B10 wind which
option is now recommended.

\subsection*{\tt hrdiag.f}

In the subroutine {\tt hrdiag}, the ``rapid'' and ``delayed'' remnant-mass prescriptions
of F12 are implemented utilizing the relevant
analytical formulae; see Sec.~\ref{newrem} for the details.
These formulae are inserted in the same parts of the code where
the final baryonic remnant mass is evaluated
from the older prescriptions. All the older remnant
prescriptions are preserved and can be selected via
choices of the input flag variable {\tt nsflag}: currently
{\tt nsflag}$=3$ (4) selects the F12-rapid (delayed) prescriptions.

The decision for undergoing PSN (giving a remnant of zero mass)
or PPSN truncation according to the criteria in B16 is
taken before the baryonic remnant mass evaluation and
is implemented as an extension of the existing conditional
statement that handles the cases of no remnant formation. PSN/PPSN
can be enabled (disabled) with the choice of the input flag variable
{\tt psflag}$=1(0)$.

Care is taken that these extensions are analogously coded in
both the hydrogen rich-star and helium-star sections  
of {\tt hrdiag}, as for the other remnant prescriptions.

Apart from these, two additional {\tt COMMON BLOCK}s are
introduced. One is\\
{\tt COMMON /FBACK/ FBFAC,FBTOT,MCO,ECS},\\
which enables the transport of the fallback fraction, fallback
mass, CO core mass, and an indicator for the
occurrence of an ECS to other subroutines where they would be
useful such as {\tt kick} (see below). The other is\\
{\tt COMMON /FLAGS2/ psflag,kmech,ecflag}\\
to port relevant new input flags ({\tt ecflag}
for enabling ECS is now read from input just for symmetry
and convenience).

Note that the {\tt ECS} indicator, which is set to unity directly in the conditional
statement in {\tt hrdiag} that assigns the ECS-NS mass, is
implemented to override the default approach of sensing
ECS events (outside {\tt hrdiag}) via $1.26\Ms$ NSs. This is an arrangement
to address the fact that in the newly-implemented F12-rapid
scheme, the core-collapse NSs' mass can go down to $\approx1.1\Ms$
so that a $1.26\Ms$ NS can as well be a core-collapse NS. 

\subsection*{\tt kick.f}

The analytical formulae for
modelling the various core-collapse SN natal kick mechanisms,
as explored in this work
(see Secs.~\ref{stdkick} \& \ref{altkick}),
are all implemented in the subroutine {\tt kick}.
To evaluate the kick-velocity magnitude,
the SN fallback fraction and the CO core mass are
needed which are supplied by the subroutine {\tt hrdiag}
through the {\tt FBACK} common block (see above).
{\tt kick} also uses the {\tt FLAGS2} common block  
to receive the value of the input flag variable
{\tt kmech} that toggles among the kick models.

Note that unlike {\tt mlwind} and {\tt hrdiag},
the {\tt kick} subroutine is structurally quite different in
$\nbseven$, which also has to keep track of
the identities of the members receiving the kick.
The natal kick formulae have been analogously
implemented in $\nbseven$'s version of {\tt kick}.
This $\nbseven$ subroutine is currently private and will be
made public elsewhere.

\subsection*{\tt star.f}

The subroutine {\tt star} differs slightly from its public
$\bse$ version and is now identical with {\tt star.f} that
is supplied with the public version of $\nbseven$.

\subsection*{\tt sse.f}

An extra {\tt READ} statement is added in
this main routine to read in the
choices of the input flags {\tt psflag}, {\tt kmech},
and {\tt ecflag} (see above).

\subsection*{\tt bse.f}

The same as in {\tt sse.f}.

\subsection*{\tt const\_bse.h}

In this header file, the common block\\
{\tt COMMON /FLAGS2/ psflag,kmech,ecflag}\\
is added to carry the user-input values of
{\tt psflag}, {\tt kmech}, and {\tt ecflag} to other subroutines.

\subsection*{\tt README\_NEW}

The descriptions of the new ingredients, including references,
are elaborated in this companion manual. The functionalities
of the new input flags (see above) and the amended formats
of the runtime input files are also described here.

\subsection*{Stellar mass limit}

Apart from the changes described above,
the imposed $100\Ms$ stellar mass limit is, by default, commented out in
{\tt star.f}, {\tt hrdiag.f}, {\tt mix.f}, and {\tt evolv2.f}.
This facilitates the user to explore the very-massive-star regime
where PPSN and PSN would occur and also the
outcome of massive-stellar mergers. However, it should
be borne in mind that extending beyond $100\Ms$ is
a rather large extrapolation of the underlying
stellar structure employed in $\bse$.

The stellar-mass ceiling is by default suppressed
in the public $\nbseven$.

\twocolumn

\end{document}